\documentclass[aps,prb,twocolumn,superscriptaddress]{revtex4-2}

\usepackage{amsmath}    

\usepackage{amsfonts}
\usepackage{amssymb}
\usepackage[colorlinks=true,citecolor=blue,linkcolor=red]{hyperref}
\usepackage{graphicx}
\usepackage[dvipsnames]{xcolor}
\usepackage{dsfont}

\usepackage{bm} 
\usepackage{dcolumn}
\usepackage{mathtools}

\usepackage{array,booktabs,ragged2e}	
\usepackage{soul}						
\usepackage{cancel}						
\usepackage{tabularx}
\usepackage{multirow}
\usepackage{hhline}
\usepackage{adjustbox}
\usepackage{sidecap}
\usepackage{threeparttable}
\usepackage{colortbl} 
\usepackage{tabularx}
\usepackage[utf8]{inputenc}
\usepackage[T1]{fontenc}
\usepackage{quiver}
\usepackage{booktabs}

\def\bra#1{\left<{#1}\right|}					
\def\ket#1{\left|{#1}\right>}					
\def\eps{\varepsilon}

\newcolumntype{P}[1]{>{\raggedleft\arraybackslash}p{#1}}
\newcolumntype{R}[1]{>{\centering\arraybackslash}p{#1}}

\usepackage{bm,color,amsmath,amssymb,mathrsfs,latexsym,graphicx,psfrag,mathtools}
\usepackage{color}
\usepackage[dvipsnames]{xcolor}

\usepackage{empheq}

\newcommand{\bsl}[1]{\boldsymbol{#1}}

\renewcommand{\mod}{\mathrm{mod}}

\newcommand{\utext}[1]{_{\text{#1}}}

\newcommand{\ii}{\mathrm{i}}

\newcommand{\dsN}{\mathbb{N}}
\newcommand{\dsR}{\mathbb{R}}

\newcommand{\Tr}{\mathop{\mathrm{Tr}}}
\renewcommand{\Re}{\mathop{\mathrm{Re}}}
\renewcommand{\Im}{\mathop{\mathrm{Im}}}

\newcommand{\refcite}[1]{Ref.\,\cite{#1}}

\newcommand{\mat}[1]{\left(\begin{matrix}#1\end{matrix}\right)}

\newcommand{\eq}[1]{\begin{equation} #1 \end{equation}}

\newcommand{\eqa}[1]{\begin{align}\begin{split} #1 \end{split}\end{align}}

\let\oldAA\AA
\renewcommand{\AA}{\text{\normalfont\oldAA}}

\newcommand{\ie}{{\emph{i.e.}}}
\newcommand{\eg}{{\emph{e.g.}}}

\newcommand{\G}{\mathcal{G}}

\newcommand{\V}{\mathcal{V}}
\renewcommand{\L}{\mathcal{L}}

\newcommand{\diag}{\text{diag}}

\newcommand{\K}{\text{K}}

\newcommand{\BZ}{\text{1BZ}}

\newcommand{\HP}[1]{%
  \ifmmode
    {\color{orange}{#1}}%
  \else
    {\color{orange}(HP: {#1})}%
  \fi
}

\usepackage{comment}

\usepackage{cleveref}
\crefname{appendix}{App.}{Apps.}
\crefname{equation}{Eq.}{Eqs.}
\crefname{figure}{Fig.}{Figs.}
\Crefname{figure}{Figure}{Figures}
\crefname{table}{Tab.}{Tabs.}
\crefname{section}{Sec.}{Secs.}
\creflabelformat{appendix}{[#2#1#3]}

\usepackage{amsmath,amssymb,amsthm}
\theoremstyle{definition}

\begin{document}

\title{ Representability-Aware Neural Networks for Reduced Density Matrices: Application to Fractional Chern Insulators}

\author{Justin B. Hart}
\affiliation{Department of Physics, University of Florida, Gainesville, FL, USA}

\author{Awwab A. Azam}
\affiliation{Department of Electrical and Computer Engineering, University of Florida, Gainesville, FL, USA}

\author{Thomas Li}
\affiliation{Palo Alto High School, Palo Alto, CA, USA }

\author{Yunxuan Li}
\affiliation{Google, Mountain View, CA, USA}

\author{Ye Bi}
\affiliation{Department of Animal Science, Iowa State University, Ames, IA, USA}

\author{Haining Pan}
\affiliation{Department of Physics and Astronomy, Center for Materials Theory, Rutgers University, Piscataway, NJ, USA}
\affiliation{Department of Physics, University of Florida, Gainesville, FL, USA}

\author{Jiabin Yu}
\email{yujiabin@ufl.edu}
\affiliation{Department of Physics, University of Florida, Gainesville, FL, USA}
\affiliation{Quantum Theory Project, University of Florida, Gainesville, FL, USA}

\begin{abstract}
Artificial neural networks (NNs) have shown potential for studying strongly correlated quantum systems, often as variational many-body wavefunction ansätze. However, many physical properties, including spontaneous symmetry breaking and the energy of two-body Hamiltonians, can be computed directly from reduced density matrices rather than the full wavefunction. Here, we develop a representability-aware and interpolable NN framework for predicting two-particle reduced density matrices (2-RDMs). The NN incorporates a subset of representability conditions through its architecture and loss function, and can operate on different momentum meshes, enabling evaluating the representability conditions across multiple meshes, which we call interpolated representability condition. The framework can be used either to predict 2-RDMs on large momentum meshes by interpolating exact results from small meshes, or as a variational 2-RDM ansatz optimized by energy minimization on arbitrary meshes. We apply this approach to the fractional Chern insulator in the one-band projected model of twisted bilayer MoTe$_2$ at twist angle $3.89^\circ$ and hole filling $2/3$. Trained on exact-diagonalization (ED) 2-RDMs from meshes with $12$ or $18$ momentum points using six different NN architectures, the best NN is the residual multilayer perceptron, which predicts the $6\times6$ 2-RDM with $97.07\%-98.18\%$  accuracy relative to the ED 2-RDM but predicts an energy $77.353$ meV above ED ground-state energy.
We then variationally optimize the NN on several meshes including  $6\times6$, predicting a $6\times 6$ energy of just $0.104$ meV below ED while maintaining $98.94\%-98.96\%$ accuracy. Compared with the conventional boundary-point semidefinite programming, which gives an energy $5.560$ meV below ED with $96.40\%-98.94\%$ accuracy, the NN achieves a more accurate energy and similar accuracy while using only less than 1/20 as many parameters. Eventually, we add a symmetric mesh of $48$ momentum points to the variational optimization of the NN, and provide a prediction of the many-body ground-state energy and the many-body quantum metric on that mesh.
\end{abstract}

\maketitle

\section{Introduction}

Artificial neural networks (NNs) have been increasingly explored as tools for studying strongly correlated quantum systems. A prominent direction is the construction of NN many-body wavefunction ansätze (or neural quantum state), whose parameters are optimized by minimizing the energy within variational quantum Monte Carlo~\cite{carleoSolvingQuantumManybody2017,vaswani2023attentionneed,chngMachineLearningPhases2017,Nomura2017BoltzmannStronglyCorrelated,glasserNeuralNetworkQuantumStates2018,chooSymmetriesManyBodyExcitations2018,Lou_2019_NN_Wavefunction,chooTwodimensionalFrustratedJ1J22019,sharirDeepAutoregressiveModels2020,irikuraNeuralnetworkQuantumStates2020,hibat-allahRecurrentNeuralNetwork2020,Pfau_2020_NN_Wavefunction,hermannDeepneuralnetworkSolutionElectronic2020,scherbelaSolvingElectronicSchrodinger2021,liFermionicNeuralNetwork2021,MacDonald_2021_CNN_Ansatze,leeNeuralnetworkVariationalQuantum2021,adamsVariationalMonteCarlo2021,inuiDeterminantfreeFermionicWave2021,luoGaugeEquivariantNeural2021,gutierrezRealTimeEvolution2022,vivasNeuralNetworkQuantumStates2022,luoGaugeEquivariantNeural2022,chenSimulating2+1DLattice2022,pesciaNeuralnetworkQuantumStates2022,liDeeplearningDensityFunctional2022,robledomorenoFermionicWaveFunctions2022,Li_2022_NNWavefunction,vonglehnSelfAttentionAnsatzAbinitio2023,cassellaDiscoveringQuantumPhase2023,kimNeuralnetworkQuantumStates2023,luoPairingbasedGraphNeural2023,renGroundStateMolecules2023,wilsonNeuralNetworkAnsatz2023,martynVariationalNeuralNetworkAnsatz2023,MacDonald_2023_NN_VMC,chenANTNBridgingAutoregressive2024,Li2024nnMonteCarlo,Lou_2024_NN_SF,Pescia_2024_NN_ansatz_electron_gas,goldshlagerKaczmarzinspiredApproachAccelerate2024,chenExactEfficientRepresentation2025a,tengSolvingVisualizingFractional2025,romeroSpectroscopyTwodimensionalInteracting2025,liDeepLearningSheds2025,maTransformerBasedNeuralNetworks2025,fuMinimalUniversalRepresentation2025a,ibarra-garcia-padillaAutoregressiveNeuralQuantum2025,Geier_2025_ML_QMC,Luo_2025_ML_QMC_tMoTe2,Qian_2025_NN_LL_mixing,Gu_2025_Hubbard_NNansatz,Roth_2025_NN_ansatze,Annabelle_2025_t-J_NN_ansatze,Roth_2025_SC_Hubbard_NN_ansatze,abouelkomsanTopologicalOrderDeep2025a,Valenti_2025_RMG_Wiger_NQS,zaklamaAttentionBasedFoundationModel2025,fadonExtractingAnyonStatistics2025,langeSimulatingSuperconductivityMixeddimensional2026,leeProjectorNeuralTensorNetwork2026,Fu_2026_FermiSets,solinasNeuralQuantumStates2026,viterittiApproachingThermodynamicLimit2026a,Abouelkomsan_2026_FirstPrinciplesAIFQHCrystallization,Geier_2026_PredictingMagnetismAI,solankiNeuralQuantumStates2026,Zaklama_2026_LargeElectronModel,linteauVarianceReductionForces2026,Zhu_2026_DisorderAwareFQHNQS,liSpinadaptedNeuralNetwork2026,Viteritti_2026_BeyondVariationalBiasHubbard,Nazaryan_2026_QERNEL,Graham_2026_QuantumElectronQuasicrystal,hulNeuralNetworkQuantum2026,meraliParallelScanRecurrent2026}.
In practice, a complete representation of the many-body wavefunction is not always necessary---many experimentally and theoretically relevant observables are directly determined by $n$-particle reduced density matrices ($n$-RDMs). For instance, the order parameters for spontaneous symmetry breaking are often the 1-RDM, while the many-body ground-state energy of a system with two-body interactions is determined by the 2-RDM.
 
This observation motivates an alternative route for studying strongly correlated quantum matter: instead of learning or optimizing the full wavefunction, one may directly determine the $n$-RDMs. 
Conventionally, this is formulated as an optimization problem over RDMs subject to certain physical constraints, often referred to as semidefinite programming or the many-body bootstrap
\cite{Coleman_1963_RDM,garrodReductionNParticleVariational1964,kummerNRepresentabilityProblemReduced1967a,erdahlTwoAlgorithmsLower1979,vandenbergheSemidefiniteProgramming1996,colemanReducedDensityMatrices2000,nakataVariationalCalculationsFermion2001,mazziottiVariationalMinimizationAtomic2002,zhaoReducedDensityMatrix2004,Mazziotti2004sdpQuantumChem,mazziottiVariationalTwoelectronReduced2005,Hammond2006Hubbard,gidofalviComputationQuantumPhase2006,Liu_2007_NRep_NP_Complete,mazziottiReducedDensityMatrixMechanicsApplication2007,erdahlLowerBoundMethod2007,schwerdtfegerConvexSetDescriptionQuantum2009,shenviActiveSpaceNRepresentabilityConstraints2010,Mazziotti_2011_BPSDP,mazziottiTwoelectronReducedDensity2012,baumgratzLowerBoundsGround2012,barthelSolvingCondensedMatterGroundState2012,verstichelVariationalTwoParticleDensity2012,Mazziotti_2012_N_rep,mazziottiEnhancedConstraintsAccurate2016,rubio-garciaVariationalReducedDensity2019,benavides-riverosReducedDensityMatrix2020,head-marsdenActiveSpacePairTwoElectron2020,han2020quantummanybodybootstrap,mazziottiDualconeVariationalCalculation2020,castilloEffectiveSolutionConvex2021,schillingEnsembleReducedDensity2021,liChallengesVariationalReducedDensityMatrix2021,knightReducedDensityMatrix2022,SagerSmith_2022_MLTwoElectrons,Mazziotti_2023_N_rep_QMB,wangCertifyingGroundstateProperties2023,deprinceVariationalDeterminationTwoelectron2024,scheerHamiltonianBootstrap2024,Gao2025QuantumSDP,DelgadoGranados_2025_ML2RDM,gaoBootstrappingFlatbandSuperconductors2025,DelgadoGranados_2026_SemidefiniteML2RDM,Martinez_2026_ML2RDM,paul2026bootstrappinggroundstateproperties}.
Here the physical constraints are called $N$-representability conditions, where the full implementation of them ensure that the $n$-RDMs arise from a physical $N$-particle density matrix. 
Nevertheless, the full characterization of the complete $N$-representability conditions is nondeterministic polynomial-time complete (NP-complete)~\cite{Liu_2007_NRep_NP_Complete,mazziottiTwoelectronReducedDensity2012,Mazziotti_2012_N_rep}, and thus only partial sets of constraints are imposed in practice.
Explicitly, the $N$-representability conditions of $n$-RDMs are classified into various subsets, labeled as $(n,p)$ with $n \leq p\leq N$, where the $(n,p)$ conditions intuitively capture the conditions imposed by the reduction from $p$-RDM to $n$-RDM \cite{zhaoReducedDensityMatrix2004,mazziottiVariationalTwoelectronReduced2005,mazziottiTwoelectronReducedDensity2012,Mazziotti_2012_N_rep,Mazziotti_2023_N_rep_QMB}.
Semidefinite programming methods have been extensively applied in quantum chemistry \cite{nakataVariationalCalculationsFermion2001,mazziottiVariationalMinimizationAtomic2002,zhaoReducedDensityMatrix2004,Mazziotti2004sdpQuantumChem,mazziottiVariationalTwoelectronReduced2005,erdahlLowerBoundMethod2007,Mazziotti_2011_BPSDP,mazziottiEnhancedConstraintsAccurate2016,mazziottiDualconeVariationalCalculation2020,Mazziotti_2023_N_rep_QMB,deprinceVariationalDeterminationTwoelectron2024}, and more recently to strongly correlated lattice and continuum systems \cite{gidofalviComputationQuantumPhase2006,schwerdtfegerConvexSetDescriptionQuantum2009,shenviActiveSpaceNRepresentabilityConstraints2010,baumgratzLowerBoundsGround2012,barthelSolvingCondensedMatterGroundState2012,verstichelVariationalTwoParticleDensity2012,rubio-garciaVariationalReducedDensity2019,benavides-riverosReducedDensityMatrix2020,head-marsdenActiveSpacePairTwoElectron2020,han2020quantummanybodybootstrap,castilloEffectiveSolutionConvex2021,schillingEnsembleReducedDensity2021,liChallengesVariationalReducedDensityMatrix2021,knightReducedDensityMatrix2022,wangCertifyingGroundstateProperties2023,Gao2025QuantumSDP,scheerHamiltonianBootstrap2024,gaoBootstrappingFlatbandSuperconductors2025}, including the Hubbard model~\cite{Hammond2006Hubbard,verstichelVariationalTwoParticleDensity2012,han2020quantummanybodybootstrap,scheerHamiltonianBootstrap2024} and (fractional) quantum Hall systems~\cite{Gao2025QuantumSDP}.
However, their broader application in strongly correlated systems remains largely unexplored, for example in fractional Chern insulators (FCIs).
 
In this work, we design NNs for calculating 2-RDMs of strongly correlated systems.
Unlike the previous work that used the NN to only interpolate $n$-RDMs from small momentum meshes to large meshes~\cite{Azam2025mlRDM}, we incorporate part of the $N$-representability conditions into the NNs. 
Specifically, we build the NNs to include the Hermiticity, fermionic antisymmetry, and the complete $(2,2)$ conditions (\ie, the positive semidefiniteness and mutual relations of various versions of 2-RDMs) either explicitly or into the loss functions.
Then, we use the NN in two ways.
In the first implementation, the NN is trained on 2-RDMs obtained from exact diagonalization (ED) on small system sizes and then used to predict the 2-RDM on larger momentum meshes.
In the second implementation, the NN is used directly as a variational ansatz for the 2-RDM, with its parameters (or weights) optimized by minimizing the many-body energy as well as the loss for the $(2,2)$ conditions that are not explicitly enforced.
A key advantage of the NN representation is that, during the variational optimization on a target momentum mesh, such as $3\times 4$, $3\times 6$, $6\times 6$, etc, the same network can also be evaluated on auxiliary meshes, such as $2\times 6$, etc.
We can therefore minimize the energies and the $(2,2)$ loss on the target mesh, while simultaneously penalizing violations of the $(2,2)$ representability conditions on multiple auxiliary meshes.
We refer to these cross-mesh constraints as interpolated $(2,2)$ conditions, which have no direct analogue in conventional semidefinite programming formulations.

As a demonstration, we apply the NNs to FCIs~\cite{neupert2011fractional,sheng,regnault2011fractional,tang2011high,Sun2011} in the one-band projected model of twisted MoTe$_2$ ($t$MoTe$_2$)~\cite{wu_topological_2019,yu2020giant,pan_band_2020,zhang_electronic_2021,li2021spontaneous,devakul_magic_2021,morales2023pressure,wang_fractional_2024,reddy_fractional_2023,qiu_interaction-driven_2023,dong2023composite,wang_topological_2023,goldman2023zero,morales2024magic,liu2024gate,xu_maximally_2024,reddy_toward_2023,yu_fractional_2024,abouelkomsan_band_2024,li2024electrically,jia_moire_2024,mao2024transfer,zhang_polarization-driven_2024,wang_topology_2023,li2024contrasting,sheng2024quantum,reddy_NonAbelian_2024,ahn_NonAbelian_2024,wang_Higher_2024,shen2024stabilizing,wang2024phase,xu_Multiple_2024,song2024phase,wu2024time,Kwan_2024_FTI,PhysRevResearch.6.L032063,zaklama2025structure,Zhang2024UniversalMoireModel,Wu_Fengcheng_2025_Topological_magnons_tMoTe2,Goncalves_2025_Quasiparticle_tMoTe2,Wu_Fengcheng_2025_tMoTe2_quasi_particles} at $3.89^\circ$ and hole filling $2/3$, where the model we use is constructed directly from density functional theory calculations without any continuous parameter fitting~\cite{Zhang2024UniversalMoireModel}.
We choose to study $t$MoTe$_2$ because FCIs were first experimentally observed there~\cite{cai2023signatures,zeng2023integer,park2023observation,Xu2023FCItMoTe2,Ji2024LocalProbetMoTe2,Young2024MagtMoTe2,Kang2024_tMoTe2_2.13,xu2024interplaytopologycorrelationssecond,An_2025_FM_tMoTe2,Jia_2025_SC_tMoTe2,Xu_2025_Interplay_SecondMoireBand_tMoTe2,Kang_2025_TRbreaking_FQSHE,Park_2025_tMoTe2_gap,Xu_2025_SC_tMoTe2,Mak_2025_SC_tWSe2,Park_2026_vanhovesingularitydriventopological}.
We first examine the ability of six NN architectures---a multi-layer perceptron (MLP) \cite{Hornik_1989_MLP}, a residual MLP \cite{He2015Residual,Lee-Thorp2022FNet}, a Kolmogorov-Arnold Network (KAN) \cite{Liu_2025_KAN,Yu_2024_MLP_vs_KAN}, a Sinusoidal Representation Network (SIREN) \cite{Sitzmann2020SIREN}, a Flexible spectral-bias tuning in Implicit Neural Representation (FINER) \cite{Liu2024FINER,Essakine2025ImplictNeural}, and a Transformer-Based Network (TBN) \cite{vaswani2023attentionneed,Katharopoulos2020LinearAttention}---to predict a 2-RDM on a $6\times 6$ mesh when trained on smaller meshes of 18 or 12 momentum points.  We find that the residual MLP with a depth of 10 residual blocks gives the best overall performance with an accuracy of $97.07\%-98.18\%$, but provides a predicted many-body energy about $77.353$ meV higher than the ED ground-state energy, much higher than the neutral excitation gap of about $2$ meV. To resolve this issue, we further variationally optimize it to minimize the energy on various target meshes from $3\times 4$ to $6 \times 6$ simultaneously, together with the interpolated $(2,2)$ condition. We find that our NN is able to give an energy prediction that is only $0.104$ meV lower than ED energy, in addition to a $2$-RDM prediction accuracy of $98.94\%-98.96\%$.
The NN result outperforms the conventional boundary-point semidefinite programming (BPSDP) method~\cite{Mazziotti_2011_BPSDP} with the (2,2), $T1$, and $T2$ conditions ($T1$ and $T2$ are part of $(2,3)$), which gives an energy prediction that is $5.560$ meV below the ED energy while maintaining a $2$-RDM prediction accuracy of about $96.40\%-98.94\%$. 
In particular, the parameters needed in our NN are less than $1/20$ as many as those in BPSDP with $(2,2)+T1+T2$ conditions.
Eventually, we perform the further variational optimization by adding a tilted $12\times 4$ mesh into the list of target meshes, and provide a prediction of the many-body ground-state energy and the many-body quantum metric.
These results establish representability-aware NNs as a compelling method for the study of strongly correlated phases. 

\section{$N$-representability Conditions for $n$-RDMs}
\label{sec:rep_conditions_RDMs}

We start with a brief review of the $n$-RDMs and their $N$-representability conditions, mainly following \cite{Mazziotti_2012_N_rep}.
Given a generic fermionic $N$-particle density matrix (with pure states included as a special case), its $n$-RDM is defined as
\eq{
\label{eq:physical_n-RDM}
\prescript{n}{}{D}^{\rm true}_{i_1 i_2 \cdots i_n\, j_n j_{n-1} \cdots j_1}
=
\left\langle
c^\dagger_{i_1}c^\dagger_{i_2}\cdots c^\dagger_{i_n}
c_{j_n}c_{j_{n-1}}\cdots c_{j_1}
\right\rangle ,
}
where $\langle \cdots \rangle$ denotes the expectation value with respect to the density matrix, and $c^\dagger_i$ is the fermion creation operator with $i$ labeling physical degrees of freedom such as the Bloch momentum.
The superscript ``true'' indicates that the tensor is a physical $n$-RDM generated by an $N$-particle density matrix.
Clearly, $\prescript{n}{}{D}^{\rm true}$ is antisymmetric under any odd permutation of the indices $i_1,\ldots,i_n$ or $j_1,\ldots,j_n$, is Hermitian, and is positive semidefinite (PSD):
\eq{
\sum_{i_1\cdots i_n,j_1\cdots j_n}
\chi_{i_1\cdots i_n}
\prescript{n}{}{D}^{\rm true}_{i_1\cdots i_n\,j_n\cdots j_1}
\chi^*_{j_1\cdots j_n}
\geq 0
}
for any complex tensor $\chi$.
However, a generic tensor $\prescript{n}{}{D}$ (with the same shape as $\prescript{n}{}{D}^{\rm true}$) may not correspond to any physical $n$-RDM of the form in \cref{eq:physical_n-RDM}, even if it is antisymmetric, Hermitian, and PSD.
The conditions that $\prescript{n}{}{D}$ needs to satisfy in order to be generated by an $N$-particle density matrix are called the $N$-representability conditions.

For a Hamiltonian with at most two-body interactions, the exact many-body ground-state energy can be computed directly from the exact ground-state $2$-RDM.
Therefore, we focus on the $2$-RDM in this work.
A classical approach to variational $2$-RDM optimization is BPSDP.
The key idea is to impose a tractable subset of the $N$-representability conditions as linear and semidefinite constraints, since enforcing the complete set of $N$-representability conditions makes the optimization NP-complete~\cite{Liu_2007_NRep_NP_Complete}.

More specifically, the $N$-representability conditions of the $2$-RDM can be formulated as follows.
A $(2,p)$ condition is obtained from the positive semidefiniteness of an operator
\eq{
\label{eq:C_2p_condition}
\hat C = \sum_l \hat T_l \hat T_l^\dagger ,
}
where each $\hat T_l$ is a polynomial of degree $p$ in the fermionic creation and annihilation operators, and $\hat C$ contains only two-body, one-body, and constant terms after normal ordering.
Then, $\langle \hat C \rangle$ consists of only linear combinations of components of $\prescript{2}{}{D}^{\rm true}$ and a constant, and replacing $\prescript{2}{}{D}^{\rm true}$ by a candidate $\prescript{2}{}{D}$ in
$
    \langle \hat C \rangle \geq 0
$
gives a $(2,p)$ condition on $\prescript{2}{}{D}$.
The complete $(2,p)$ condition is given by enforcing positive semidefiniteness for all such operators in \cref{eq:C_2p_condition}.
The complete $(2,N)$ condition, together with extra linear constraints such as the normalization $\sum_{ij}\prescript{2}{}{D}_{ijji} = N(N-1)$, fermionic antisymmetry and Hermiticity, give the complete $N$-representability condition.

In BPSDP calculations done in this work, we at most consider the complete $(2,2)$ and partial $(2,3)$ conditions, in addition to the normalization, fermionic antisymmetry and Hermiticity.
It turns out that the complete $(2,2)$ condition can be re-expressed as the positive semidefiniteness of 
\eqa{
& \prescript{2}{}{D}_{ijkl}^{\rm true} = \left\langle c_i^\dagger c_j^\dagger c_k c_l 
    \right\rangle\\
& \prescript{2}{}{Q}_{ijkl}^{\rm true} = \left\langle c_i c_j c_k^\dagger c_l ^\dagger \right\rangle \\
& \prescript{2}{}{G}_{ijkl}^{\rm true} = \left\langle c_i^\dagger c_j c_k^\dagger c_l \right\rangle,
}
along with the relations between them, via the fermion anticommutator.
The partial $(2,3)$ conditions we include are the $T1$ and $T2$ conditions~\cite{Erdahl1978T1T2,Zhao2007T1T2,Mazziotti_2012_N_rep}. 
Explicitly, the $T1$ condition corresponds to the following form of $\hat C$ in \cref{eq:C_2p_condition}:
\eq{
\hat{C} = \frac{1}{2}\left(\hat{C}_{T1,1}\hat{C}_{T1,1}^\dagger + \hat{C}_{T1,2}\hat{C}_{T1,2}^\dagger \right),
}
where $
\hat{C}_{T1,1} = \sum_{jkl} a_{jkl} \hat{c}_j^\dagger \hat{c}_k^\dagger \hat{c}_l^\dagger $, 
$\hat{C}_{T1,2} = \sum_{jkl} a_{jkl}^* \hat{c}_j \hat{c}_k \hat{c}_l $, 
and $a_{jkl}\in\mathbb{C}$ is an arbitrary complex tensor. Meanwhile, the $T2$ condition corresponds to the following form of $\hat C$ in \cref{eq:C_2p_condition}:
\eq{
\hat{C}  = \frac{1}{2}\left(\hat{C}_{T2,1}\hat{C}_{T2,1}^\dagger + \hat{C}_{T2,2}\hat{C}_{T2,2}^\dagger \right),
}
where 
$
\hat{C}_{T2,1}  = \sum_{jkl} b_{jkl} \hat{c}_j^\dagger \hat{c}_k^\dagger \hat{c}_l
$
,
$
\hat{C}_{T2,2}  = \sum_{jkl} b_{jkl}^* \hat{c}_j \hat{c}_k \hat{c}_l^\dagger 
$,
and $b_{jkl}\in\mathbb{C}$ is an arbitrary complex tensor.
A more detailed review of the $N$-representability conditions is given in \cref{app:n_part_RDM_n_rep}.

\section{Representability-Aware NNs}
\label{sec:NN_with_rep_conditions}

In this section, we discuss the representability-aware NN pipeline for the prediction of 2-RDMs. In total, we use six NN architectures: MLP \cite{Hornik_1989_MLP}, residual MLP \cite{He2015Residual,Lee-Thorp2022FNet}, KAN \cite{Liu_2025_KAN,Yu_2024_MLP_vs_KAN}, SIREN \cite{Sitzmann2020SIREN}, FINER \cite{Liu2024FINER,Essakine2025ImplictNeural}, and TBN \cite{vaswani2023attentionneed,Katharopoulos2020LinearAttention}.
The architectures differ in their inductive biases, expressiveness, and training, impacting their ability to represent and predict RDMs. 
Yet, regardless of the choice of the architecture, the NN pipeline is the same.
Therefore, we will focus on the pipeline and refer the readers to \cref{app:Architectures} for full descriptions of each architecture, including their hyperparameters and formal definitions.

\begin{figure}[t]
  \centering
  \includegraphics[width=\columnwidth]{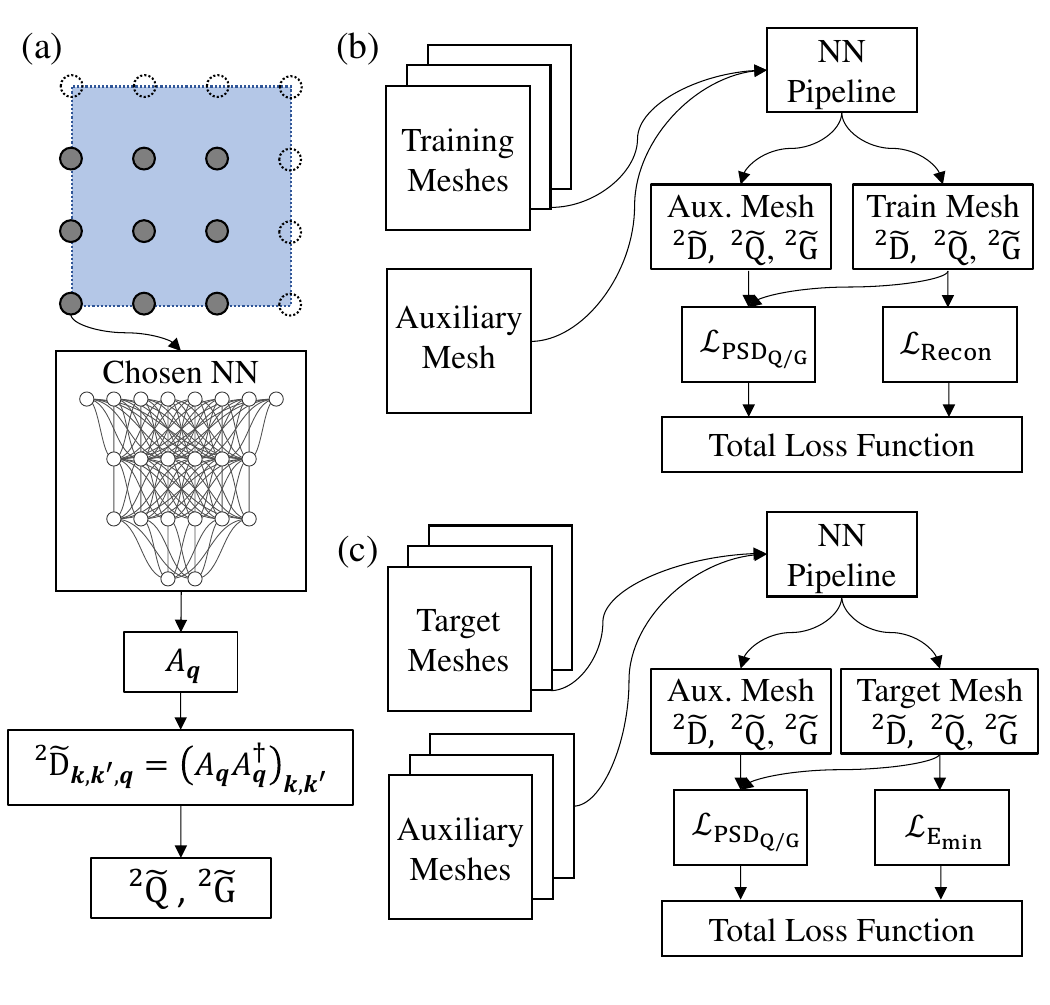}
  \caption{\textbf{Schematic of Representability-Aware NN } 
  (a) The input of the NN involves the Bloch momenta on certain momentum meshes. The input is passed through our chosen NN architecture, and the outputs are aggregated to form a tensor $A_{\bsl{q}}$ in \cref{eqn:tilde_D}, which is in turn used to construct $\prescript{2}{}{\tilde{D}}$, $\prescript{2}{}{\tilde{Q}}$ and $\prescript{2}{}{\tilde{G}}$. This pipeline supports prediction on any $L_1\times L_2$ conventional or tilted mesh in any convention as the NN can be evaluated at arbitrarily dense or arbitrarily shaped momentum meshes in the first Brillouin zone (\BZ). 
  (b) Loss function for the interpolation training. 
  The NNs are trained on 2-RDMs on small meshes, where we provide true $\tilde D$, $\tilde Q$, and $\tilde G$ through the ED calculation.
  (c) Loss function for variational optimization. The 2-RDM is predicted on  desired meshes by minimizing the energy and the interpolated representability condition on the desired meshes and any auxiliary meshes. 
  }
  \label{fig:nn_diagram}
\end{figure}

The pipeline is summarized in \cref{fig:nn_diagram}(a).
Formally, each NN is a mapping 
\begin{equation}
    \Phi: \mathbb{R}^9\to \mathbb{R}^2,
\end{equation}
which is trained to minimize a loss function $\mathcal{L}(\theta) \in \mathbb{R}_{\geq 0}$  over tunable parameters (weights) $\theta\in\mathbb{R}^{n_{\mathrm{params}}}$ with $n_{\mathrm{params}}$ denoting the number of learnable parameters in the NN.
To explain the physical meaning of the map $\Phi$ and the loss function, let us be concrete, and henceforth consider the one-band projected many-body problem with Bloch momentum conservation, focusing on the $2$-RDM given by states with fixed many-body momentum, unless specified otherwise.
The index $i$ in $c^\dagger_{i}$ then only labels the Bloch momentum, and the momentum conservation, Hermiticity and positive semidefiniteness allows us to parametrize $\prescript{2}{}{D}$ as
\eq{
\prescript{2}{}{D}_{-\bsl{k}+\bsl{q},\bsl{k},\bsl{k}', -\bsl{k}'+\bsl{q}} = \prescript{2}{}{\tilde{D}}_{\bsl{k},\bsl{k'},\bsl{q}} = (A_{\bsl{q}}A_{\bsl{q}}^\dagger)_{\bsl{k},\bsl{k'}} 
    \label{eqn:tilde_D}\ ,
}
where $\bsl{k}$, $\bsl{k}'$ and $\bsl{q}$ take values from a momentum mesh in the Brillouin zone.
In this work, we will always consider a 2D $L_1\times L_2$ mesh with the form $\{  \bsl{k} = k_1/L_1 \bsl{G}_1 + k_2/L_2 \bsl{G}_2 | k_i = 0,1,2,...,L_i-1\} $ with $L_1,L_2$ positive integers and $\bsl{G}_1,\bsl{G}_2$ two reciprocal lattice vectors.
If $\bsl{G}_1,\bsl{G}_2$ are primitive, we say that the mesh is conventional, otherwise, we say that it is tilted (see \cref{app:basics} for details). 
$N_k = L_1 L_2$ is the number of Bloch momentum points in the mesh, and thus $A_{\bsl{q}} \in \mathbb{C}^{N_k\times N_k}$ is a $N_k \times N_k$ matrix.

The input of the NN $\Phi$ contains information about the Bloch momentum $\bsl{q}$, the matrix index of $A_{\bsl{q}}$ (labeled as $\bsl{k},\bsl{p}$), the size of the momentum mesh ($L_1, L_2$), and $N$, forming a real vector $\bsl{x}\in \dsR^9$.
The output of the NN is the value of the corresponding element of $A_{\bsl{q}}$, which is a complex number and can therefore be represented by two real numbers, $\ie$, as an element of $\dsR^2$.
Explicitly, we have
\eq{
\Phi(\bsl{x}) = \left[A_{\bsl{q}}\right]_{\bsl{k}\bsl{p}}
}
with
\eq{
\bsl{x} = (v(\bsl{k}),v(\bsl{p}),v(\bsl{q}), 1/L_1, 1/L_2, 1/N)\ ,
}
\begin{align}
    v(\bsl{k}) =  \left(\frac{1}{\pi}(\bsl{a}_{1}\cdot\bsl{k}\ \bmod 2\pi)-1, \frac{1}{\pi}(\bsl{a}_{2}\cdot\bsl{k}\ \bmod 2\pi)-1\right) \  ,
\end{align} 
and $\bsl{a}_{1}$ and $\bsl{a}_{2}$ the primitive lattice vectors.
Here, $v(\bsl{k})$ acts as a feature encoding that projects the Bloch momentum onto the primitive lattice vectors and rescales the result from $[0,2\pi)$ to $[-1,1)$, so that the NN input is mesh-independent and normalized.
Importantly, as we can see, changing the momentum mesh keeps the dimension of $\bsl{x}$ invariant (although the values of $\bsl{x}$ itself can change).

The output of the NN is then anti-symmetrized by 
\eq{
A_{\bsl{q}} \rightarrow ( A_{\bsl{q}} - R_{\bsl{q}} A_{\bsl{q}})/2 
}
where $[R_{\bsl{q}}]_{\bsl{k},\bsl{k'}}=\delta_{-\bsl{k}+\bsl{q}\,\mod\,\bsl{G},\bsl{k'}}$ (where $\bsl{G}$ is the reciprocal lattice vector) represents flipping the first two creation operators in $\prescript{2}{}{D}$. 
The output $A_{\bsl{q}}$ is then used to construct $\prescript{2}{}{\tilde{D}}$ in \cref{eqn:tilde_D}, which is then normalized to $N(N-1)$.
The constructed $\prescript{2}{}{\tilde{D}}$ can then be used to construct $\prescript{2}{}{\tilde{Q}}$ and $\prescript{2}{}{\tilde{G}}$, via the relation between $\prescript{2}{}{Q}^{\rm true}$, $ \prescript{2}{}{G}^{\rm true}$ and $\prescript{2}{}{D}^{\rm true}$ and
\eqa{
& \prescript{2}{}{Q}^{}_{-\bsl{k}+\bsl{q},\, \bsl{k},\, \bsl{k}',\, -\bsl{k}'+\bsl{q}} = \prescript{2}{}{\tilde{Q}}_{\bsl{k},\bsl{k}',\bsl{q}}  \\
     & \prescript{2}{}{G}^{}_{\bsl{k}+\bsl{q},\, \bsl{k},\, \bsl{k}',\, \bsl{k}'+\bsl{q}}  = \prescript{2}{}{\tilde{G}}_{\bsl{k},\bsl{k}',\bsl{q}}\ .
}
(See \cref{eq:D-G_relation,eq:D-Q_relation})
In this way, we obtain a momentum-conserved $\prescript{2}{}{D}$, $\prescript{2}{}{Q}$ and $\prescript{2}{}{G}$ that has the right normalization, fermion antisymmetry, Hermiticity, and mutual relations.
The positive semidefiniteness of $\prescript{2}{}{D}$ is also guaranteed, and the remaining (2,2) conditions that we have not yet included are the positive semidefiniteness of $\prescript{2}{}{Q}$ and $\prescript{2}{}{G}$.

The way we include the positive semidefiniteness of $\prescript{2}{}{Q}$ and $\prescript{2}{}{G}$ is through the loss function for the training/optimization of the NN.
Explicitly, we treat $\prescript{2}{}{\tilde{Q}}_{\bsl{k}\bsl{k}'\bsl{q}}$ and $\prescript{2}{}{\tilde{G}}_{\bsl{k}\bsl{k}'\bsl{q}}$ at each fixed $\bsl{q}$ as a matrix indexed by $\bsl{k}$ and $\bsl{k}'$, compute their eigenvalues for each $\bsl{q}$, and include the absolute value of each negative eigenvalue as part of the loss function, which we label as $\L_{\text{PSD}_{Q/G}}$ (see \cref{eq:lambda_Q_G}) and \cref{app:losses} for more details).
In this way, the violation of the positive semidefiniteness of $\prescript{2}{}{Q}$ and $\prescript{2}{}{G}$ is penalized when minimizing the loss.
Note that in our design, the same NN allows input from different momentum meshes, which means the loss $\L_{\text{PSD}_{Q/G}}$ can be evaluated on multiple auxiliary meshes simultaneously even if we are targeting predicting the 2-RDM on one specific mesh.
We refer to this as \textit{interpolated} representability condition, which is not achievable in the conventional semidefinite programming methods for RDMs, including BPSDP.
The loss function also involves the predicted many-body energy, which has the following form
\eq{
E_{\text{pred}} = \frac{1}{N -1}\sum_{\bsl{k},\bsl{q}}\epsilon_{\bsl{k}}\tilde{D}_{\bsl{k},\bsl{k},\bsl{q}}  + \sum_{\bsl{k},\bsl{k}',\bsl{q}}V(\bsl{k}, \bsl{k}', \bsl{q})\tilde{D}_{\bsl{k},\bsl{k}',\bsl{q}} 
}
as the one-band projected many-body Hamiltonian with Bloch momentum conservation has the form
\eqa{
    H &= \sum_{\bsl{k}} c_{\bsl{k}}^\dagger c_{\bsl{k}} \epsilon_{\bsl{k}}  + \sum_{\bsl{k}, \bsl{k}', \bsl{q} } V(\bsl{k}, \bsl{k}', \bsl{q}) c_{-\bsl{k}+\bsl{q}}^\dagger c_{\bsl{k}}^\dagger c_{\bsl{k}'} c_{-\bsl{k}'+\bsl{q}}\ . \label{eq:ED_hamiltonian} 
}
The way we use the predicted many-body energy and the other components in the loss function depends on the way we use the NN, which we discuss next.

We will use the NN pipeline in two ways, interpolation (\cref{fig:nn_diagram}(b)) and variational optimization (\cref{fig:nn_diagram}(c)), which differ in their loss functions. 
For interpolation, we train the NN on $^2\tilde D^{\rm true}$, $^2\tilde Q^{\rm true}$, $^2\tilde G^{\rm true}$ calculated for the ground state by ED on small sizes, and let the NN predict the 2-RDM at large sizes.
Therefore, we will have the mean squared error between the predicted $^2\tilde D$, $^2\tilde Q$ and $^2\tilde G$ and the true $^2\tilde D^{\rm true}$, $^2\tilde Q^{\rm true}$ and $^2\tilde G^{\rm true}$ on the small training meshes as part of the loss, labeled as $\L_{\text{recon}}$.
The NN is then trained to minimize the total loss function $\L(\theta)$ that is a weighted sum of $\L_{\text{recon}}$ and the interpolated $\L_{\text{PSD}_{Q/G}}$, as shown in \cref{fig:nn_diagram}(b).

This interpolation training is motivated by the previous work~\cite{Azam2025mlRDM}, which has shown that NNs are able to predict pair-pair correlation functions (special contraction of 2-RDMs) of the 2D Richardson model ~\cite{Richardson1963PairPairCorr, Richardson1964Eigenstates, Dukelsky2004Richardson} and 1-RDM of two other 2D lattice models on larger $\bsl{k}$ meshes when trained on smaller $\bsl{k}$ meshes.
In \refcite{Azam2025mlRDM}, the NN is performing mainly a pure interpolation task, as the large momentum mesh is a denser version of the small momentum mesh, though there is no clear incorporation of any complete $(2,p)$ conditions.
In this work, we also benchmark each of our six architectures on the same task of interpolating the pair-pair correlation function of the Richardson model (see \cref{app:Richardson}) in the similar way as \refcite{Azam2025mlRDM}. 
We train each NN on the pair-pair correlation function on the conventional $6\times 6$ mesh, and predict the correlation function evaluated on the conventional $18\times 18$ mesh, and find accuracy as high as $98.57\%$, showing improvement from $94\%$ in \refcite{Azam2025mlRDM}.
This improvement comes from the fact that we implement the scaled principal component structure in the NN which is missing in \refcite{Azam2025mlRDM}, as elaborated in \cref{app:Richardson}.
Nevertheless, the consistently high performance across architectures confirms that each is capable of performing interpolation in momentum space, giving us faith in using them for the more demanding task of predicting full $2$-RDMs with $N$-representability conditions. 

The second way in which we use the NNs is variational optimization, where we use the NN as a variational ansatz for the $2$-RDM.
In this way, we have no training data from ED calculations, and directly treat the predicted energy $E_{\text{pred}}$ as part of the loss function, labeled as $\L_{E_{min}} = E_{\text{pred}}$.
Then, the NN is optimized to minimize the total loss $\L(\theta)$ that is a weighted sum the interpolated $\L_{\text{PSD}_{Q/G}}$ and $\L_{E_{min}}$, as shown in \cref{fig:nn_diagram}(c).

The two uses are not independent from each other.
Next we will see that the interpolation training in \cref{fig:nn_diagram}(b) can prepare the NN architecture for the variational optimization in \cref{fig:nn_diagram}(c) for predictions on system sizes well beyond the reach of ED.

\begin{figure}[t]
  \centering
  \includegraphics[width=\columnwidth]{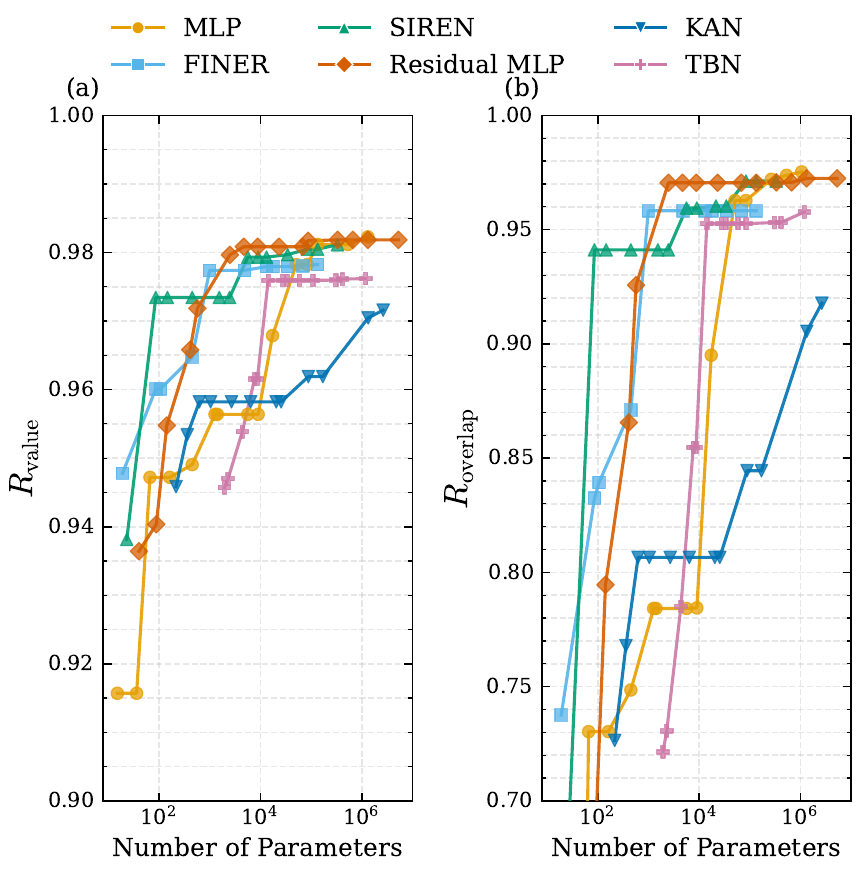}
  \caption{\textbf{NN Parameter Efficiency for Interpolation of the $2$-RDM of the one-band projected $t$MoTe$_2$.} 
  We plot the NN architectures' maximum achievable (a) $R_{\text{value}}$ and (b) $R_{\text{overlap}}$ at the indicated parameter count. 
  The prediction is tested against ED on the conventional $6\times 6$ mesh.
  } 
  \label{fig:param_efficiency}
\end{figure}

\begin{figure}[t]
  \centering
  \includegraphics[width=\columnwidth]{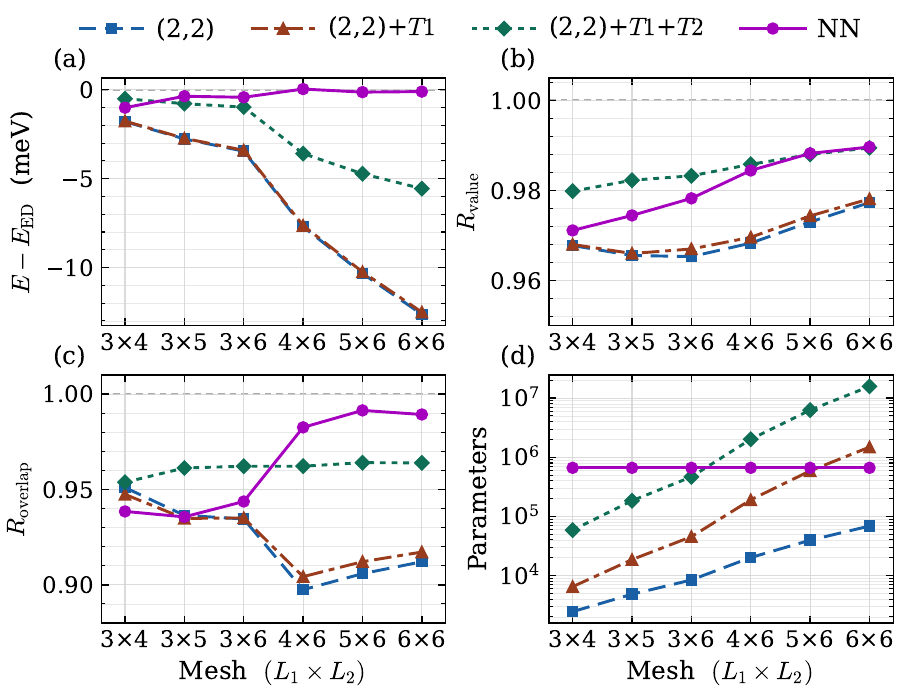}
  \caption{
  \textbf{Comparison of BPSDP and NN Variational Optimization for 2-RDM on several conventional meshes.} We provide BPSDP incorporating $(2,2)$ (blue), BPSDP incorporating $(2,2)$ and $T1$ (brown), and BPSDP incorporating $(2,2)$, $T1$, and $T2$ (green). We also provide the prediction of the NN variational optimization with all the listed meshes as target meshes (purple). 
  We plot (a) $E-E_{ED}$ with $E$ the converged/predicted energy and $E_{ED}$ the ED ground-state energy, (b) $R_{\text{value}}$ against the ED results, and (c) $R_{\text{overlap}}$ against the ED results over various conventional $L_1\times L_2$ meshes. 
  In (d), we plot the required number of real parameters for the each calculation, in log-scale. 
  }
  \label{fig:ansatz_comparison}
\end{figure}

\section{Application to FCI in $t\text{MoTe}_2$}

We apply the NN established in \cref{sec:NN_with_rep_conditions} to FCI in $t$MoTe$_2$. 
The model we consider is the $K$-valley moiré model with parameter values from \refcite{Zhang2024UniversalMoireModel} for twist angle $3.89^\circ$, together with the double-gate screened Coulomb interaction with gate distance $20\,\mathrm{nm}$ and relative dielectric constant $10$.
We convert the many-body model to the hole basis, and project it to the lowest hole band, which leads to a one-band many-body model of the form in \cref{eq:ED_hamiltonian}.
We would like to note that we choose the eigenvectors of the lowest hole band in the same parallel-transport gauge for all momentum meshes, to enable interpolation training.
We focus on the $2/3$ hole filling, which leads to a FCI state.
In the following, we will first apply the interpolation-based approach and show the performance of different NN architectures. Following this, we examine using the NN as a variational ansatz for predicting the 2-RDM. 
For details, we refer the readers to \cref{app:FCI}.

For interpolation, we train on either a $3\times 6$ mesh, or a combined set of $2\times 6$, $6\times 2$, and  $3\times 4$ meshes.
We include $6\times 6$ as an auxiliary mesh for the $\L_{\text{PSD}_{Q/G}}$.
All training and auxiliary meshes are conventional except the $6\times 2$, which is tilted (see \cref{fig:Momentum_space} for a visualization of the meshes).  
The training data are generated by the ED calculation, and we also benchmark energy prediction on the conventional $6\times 6$ to the ED calculation.
For the benchmark, we use two quantities to evaluate the accuracy.
One is the R-value, which is defined as
\begin{equation}
    R\utext{value} = \max(0,R).
    \label{eqn:r_value}
\end{equation}
with
\begin{equation}
    R = 1-\frac{ \sqrt{\mathrm{MSE}(T,T_{\text{true}})}}{\max(\Re T_{\text{true}},\Im T_{\text{true}})-\min(\Re T_{\text{true}},\Im T_{\text{true}})}\ ,
\end{equation}
where $\mathrm{MSE}(T,T_{\text{true}})$ is the MSE between the predicted tensor $T$ (\eg, $\prescript{2}{}{\tilde{D}}$ in this work) and the true tensor $T_{\text{true}}$, and $\Re$ and $\Im$ refer to the real and imaginary components of the tensor. 
The other is the normalized overlap
\begin{equation}
R_{\text{overlap}} = \frac{|\langle T_{\text{true}}, T\rangle|}{\max(\lVert T_{\text{true}}\rVert^2,\lVert T\rVert^2)}
\label{eqn:norm_overlap}
\end{equation}
where $\langle T_1 , T_2 \rangle = \sum_{\text{all indices}} (T_1)^*_{ij\dots} (T_2)_{ij\dots}$ is the Frobenius inner product and $\|T\|= \sqrt{\langle T, T \rangle }$ is the Frobenius norm.
\cref{fig:param_efficiency} presents the parameter efficiency of each model, that is, the maximum achievable $R\utext{value}$ and corresponding $R\utext{overlap}$ at each number of parameters.For the parameter efficiency, we include results from both the NN trained on the $3\times 6$ mesh and the NN trained on the three $12-\bsl{k}$ meshes. All models except the KAN reach similar levels of performance, with diminishing returns after $10^5$ parameters. Notably, the SIREN achieves strong performance at very few parameters (see \cref{app:Architectures} for architecture details). 

Combining both $2$-RDM prediction accuracy and the energy prediction accuracy, we select residual MLP as the model with best performance.
For that model, training on a single $3\times 6$ mesh versus a set of three 12-$\bsl{k}$ meshes yields comparable results: the single-mesh training leads to $R_{\text{value}}=98.18\%$ and $R_{\text{overlap}}=97.07\%$ on $6\times 6$, while the three-mesh training leads to NN achieves $R_{\text{value}}=98.17\%$ and $R_{\text{overlap}}=96.63\%$ on $6\times 6$. 
Nevertheless, training on multiple 12-$\boldsymbol{k}$ meshes is computationally much cheaper than training on the $3\times 6$ mesh, because ED on several 12-$\boldsymbol{k}$ meshes is much faster than ED on the $3\times 6$ mesh, owing to the exponential growth of the many-body Hilbert-space dimension.
However, its predicted energy on $6\times6$ is $2097.771$ meV, which is $77.353$ meV above the ED ground state of $2020.418$ meV, though it is already the best among all NNs (see \cref{tab:MoTe2_mesh_comparison} for a full comparison). 
The deviation between the predicted many-body energy and the ED energy is high, as the neutral excitation gap is about $2$ meV, which motivates us to turn to variational optimization.

For variational optimization, we choose the residual MLP with 663,298 parameters, and initialize the NN from interpolation-trained weights (\ie, model parameter values), and use our NN pipeline to minimize energy together with the interpolated $\L_{\text{PSD}_{Q/G}}$.
Specifically, we choose the conventional $3\times 4$, $3\times 5$, $3\times 6$, $4\times 6$, $5\times 6$ and $6\times 6$ meshes as the target meshes, and choose the conventional $2\times 6$ and tilted $6\times 2$ meshes as the auxiliary meshes.
The $\L_{E_{min}}$ involves all the target meshes, and the interpolated $\L_{\text{PSD}_{Q/G}}$ involves all the target and auxiliary meshes.
We also use the predicted energy on the conventional $2\times 6$ as validation loss, which tells us when to stop the training.
The results are in \cref{fig:ansatz_comparison}.
In particular, the variational NN predicts an energy of $2020.314$ meV on $6\times 6$, just $0.104$ meV below the ED energy with $R_{\text{value}}=98.96\%$ and $R_{\text{overlap}}=98.94\%$.

To compare this results to classical methods, we perform the BPSDP calculations on various conventional meshes up to $6\times 6$ with three scenarios, (i) only $(2,2)$ condition, (ii) $(2,2)+T1$ and  (iii) $(2,2)+T1 +T2$.
As shown in \cref{fig:ansatz_comparison}(a-c), the best accuracy is achieved by the BPSDP with $(2,2)+T1 +T2$, which gives an energy  $5.560$ meV below the ED ground-state energy, $R_{\text{value}}=98.94\%$ and $R_{\text{overlap}}=96.40\%$ on $6\times 6$.
Although the $R_{\text{value}}$ and $R_{\text{overlap}}$ of the BPSDP with $(2,2)+T1 +T2$ is comparable with that of the prediction of the NN, the energy of the former is much worse than that of the latter.
Since both energies fall below the ED ground state, they signal incomplete $N$-representability: the optimized 2-RDM does not correspond to any physical $N$-particle state. The closer the energy is to the ED value, the more physical the result. In this regard, the NN ($0.104$) meV below ED is more physical than the BPSDP ($5.560$ meV below ED).
The NN also outperform the BPSDP with $(2,2)+T1 +T2$ on $4\times 6 $ and $5\times 6$ meshes, as shown in \cref{fig:ansatz_comparison}(a-c).
Nevertheless, our NN only uses less than $1/20$ as many parameters as the BPSDP with $(2,2)+T1 +T2$ on $6\times 6$, as shown in \cref{fig:ansatz_comparison}(d), suggesting strong efficiency of the NN.
The reason is that the NN only needs to parameterized ${}^2\tilde{D}$, while BPSDP needs to parameterize ${}^2\tilde{D}$, ${}^2\tilde{Q}$, ${}^2\tilde{G}$ and two additional five-index tensors for $T1$ and $T2$.
In fact, if we include more $N$-representability conditions in BPSDP, the parameter number will keep increasing, but the parameters in NN can stay the same as the addition constraints are just in the loss function, as long as the NN is not in the underfitting regime.

\begin{figure}[t]
  \centering
  \includegraphics[width=\columnwidth]{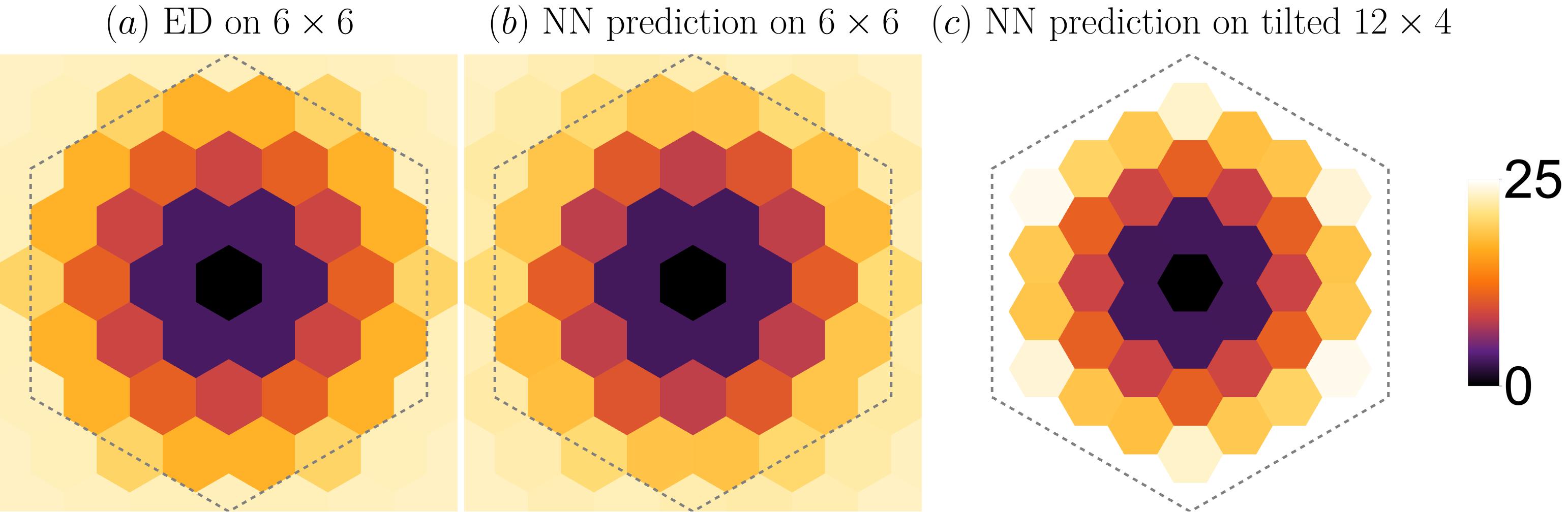}
  \caption{
  \textbf{The plot of $S(\bsl{q})$ for the one-band projected $t$MoTe$_2$.} 
  We plot $S(\bsl{q})$  from (a) the ED calculation on the conventional $6\times 6$,  (b) the prediction from NN variational optimization for the conventional $6\times 6$ mesh, and (c) the prediction from NN variational optimization for the tilted $12\times 4$ mesh.
  Here we plot all $|\bsl{q}|$ points that are smaller than half the length of the primitive reciprocal moir\'e lattice vector, the dahsed line labels the edge of the \BZ, and all plots use the same color bar.
  Note that, we cap the color range at 25  to better visualize the behavior near the $\Gamma$ point, and because of this color range, the outer part of (c) is out of range.
  }
  \label{fig:S_q}
\end{figure}

To visualize the comparison, we plot the static structure factor $S(\boldsymbol{q})= \langle \rho_{\boldsymbol{q}}\rho_{-\boldsymbol{q}}\rangle
-
\langle \rho_{\boldsymbol{q}}\rangle
\langle \rho_{-\boldsymbol{q}}\rangle$ obtained from ED and from NN variational optimization on the conventional $6\times6$ mesh in \cref{fig:S_q}(a-b).
Here $\rho_{\boldsymbol{q}}$ is the Fourier transform of the real-space density operator,
$
\rho_{\boldsymbol{q}}
=
\int d^2\boldsymbol{r}\,
e^{-i\boldsymbol{q}\cdot\boldsymbol{r}}
\rho(\boldsymbol{r})
$.
(See \cref{app:mb_QG} for explicit expression.)
In principle, $\boldsymbol{q}$ is not restricted to the \BZ, but the small $\bsl{q}$ is particularly interesting as it allows us to estimate the integrated many-body quantum metric $\G^{\rm mb}$ \cite{Provost1980FSMetric,SWM2000,Resta2006Polarization,Regnault2013FCI,OnishiFu2025QuantumWeight,Yu2024_QG_Charge_Fluctuation,Wu2024QGCornerChargeFluctuationsManyBody} (which characterizes the quantum geometry of the many-body state as discussed in \cref{app:mb_QG}) from the NN predicted $S(\bsl{q})$, since 
\eq{
\Tr[\G^{\rm mb}] = \left. \frac{\pi}{\V} \nabla_{\bsl{q}}^2 S(\bsl{q}) \right|_{\bsl{q}=0}
}
holds in the thermodynamic limit.
As shown in \cref{fig:S_q}(a-b), the match is very good for all $\bsl{q}$ inside and around $\BZ$.
The good match would suggest similar estimations of the many-body quantum metric on the conventional $6\times 6$ mesh from ED and the NN.
Indeed, we have $\Tr[\G^{\rm mb}] = 0.8695$ from the ED calculation and $\Tr[\G^{\rm mb}] =0.8161$ from the NN variational optimization, which shows a relative error of about $6.14\%$.
This value is larger but not too far away from the fractional Chern number $2/3$ of a single ground state, which is reasonable as the fractional Chern number is a lower bound of $\Tr[\G^{\rm mb}]$.

To go beyond the ED calculation, we perform another variational optimization with a titled $12\times 4$ mesh added to the target mesh, while keeping the auxiliary mesh and the validation as the previous optimization without the $12\times 4$ mesh.
The calculation is certainly beyond the reach of ED, as the Hilbert space dimension for ED is about $10^{10}$.
Our NN optimization gives a many-body energy of 2703.495 meV, and the resulting $S(\bsl{q})$ is shown in \cref{fig:S_q}(c), which gives a  $\Tr[\G^{\rm mb}] = 0.7974$.

\section{Discussion}
In summary, we develop a representability-aware NN pipeline for incorporating the $N$-representability conditions, and demonstrate its effectiveness in predicting the $2$-RDM of the FCI in tMoTe$_2$.  
Our method can be applied to systems both in the continuum limit or in the tight-binding model.
Unlike methods based on variational quantum Monte Carlo, the effectiveness of our method does not depend on whether the interaction has a density-density form.
Our work provides an effective tool for the study of strongly correlated systems.

\section{Acknowledgments}
We would like to thank Zhaoyu Han for helpful discussion.
J.Y.'s work is supported by startup funds at University of Florida.
J.B.H's and A.A.A's work is supported by the University Scholars Program at the University of Florida. A.A.A's work is also supported by the UF CCMS Undergraduate Fellowship.
H.P. is supported by US-ONR grant No.~N00014-23-1-2357.
The authors acknowledge UFIT Research Computing for
providing computational resources and support.

\bibliographystyle{apsrev4-2}
\bibliography{bibrefs,bibfile_references.bib}

\onecolumngrid
\tableofcontents

\appendix
\crefalias{section}{appendix}
\crefalias{subsection}{appendix}
\crefalias{subsubsection}{appendix}
\section{Basics}
\label{app:basics}
\begin{figure}[t]
    \centering
    \includegraphics[width=0.9\linewidth]{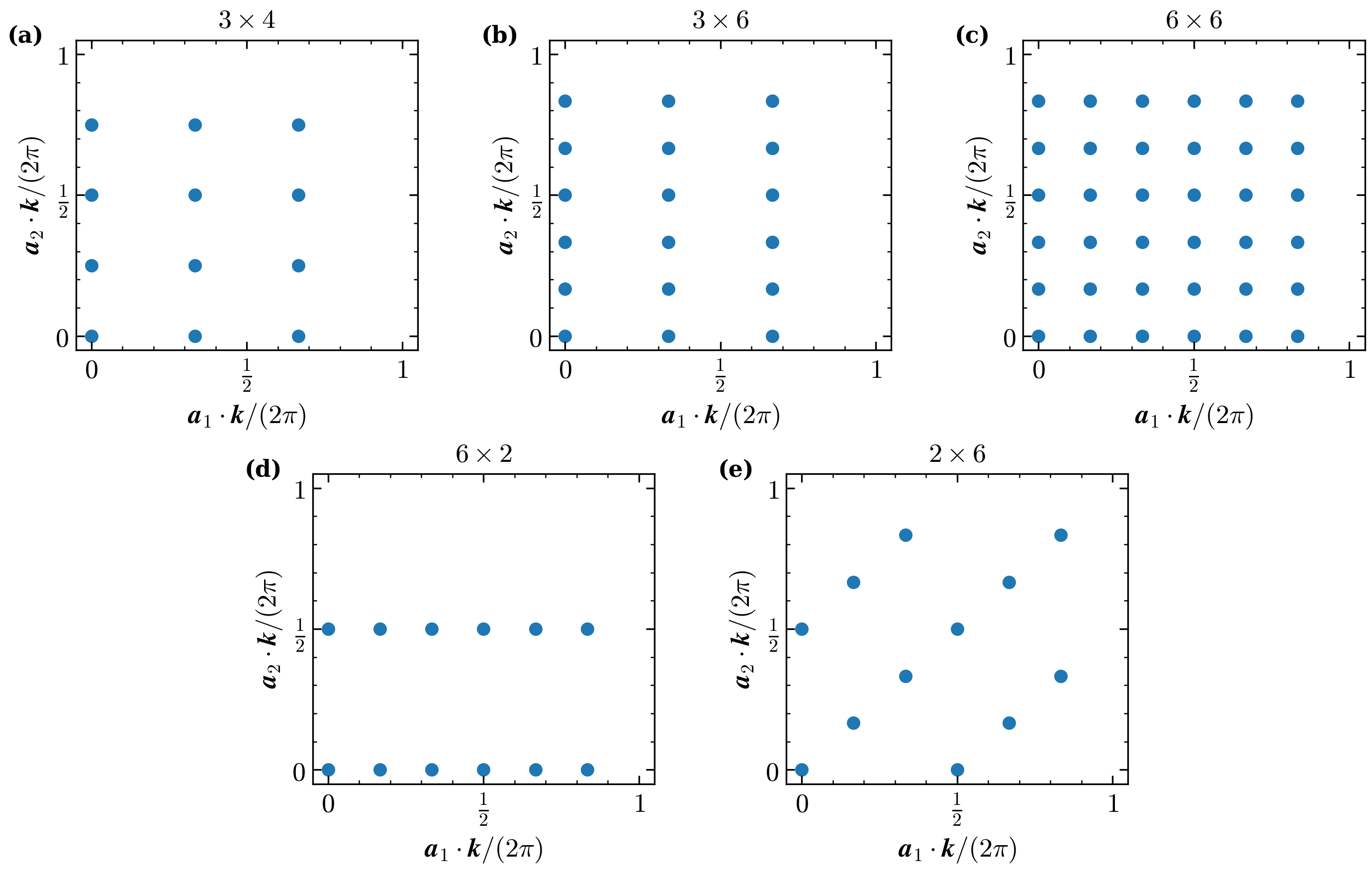}
    \caption{Visualization of the sampling of the \BZ in the conventional meshes (a-d)  and the tilted mesh (e) with $M=\begin{pmatrix}
        1 & 1 \\0 & 1 \end{pmatrix}$. 
        $\bsl{a}_i$ is the primitive lattice vector with $\bsl{a}_i\cdot\bsl{b}_j = 2\pi \delta_{ij}$.  }
    \label{fig:Momentum_space}
\end{figure}

In this work, we will only focus on the 2D system.
The Bloch momentum mesh of such systems have the general form:
\begin{equation}
    \bsl{k} = \frac{k_1}{L_1} (M_{11}\bsl{b}_1+M_{12}\bsl{b}_2) + \frac{k_2}{L_2} (M_{21}\bsl{b}_1+M_{22}\bsl{b}_2)\ ,
    \label{eqn:tilted_convention}
\end{equation}
where $\bsl{b}_1$ and $\bsl{b}_2$ are the primitive lattice vectors and $k_1 \in \{0,1,2, \dots, L_1-1\}$ and  $k_2 \in \{0,1,2, \dots, L_2-1\} $, and $M_{ij}\in\mathbb{Z}$ represent indices of a matrix $M=\begin{pmatrix}
    M_{11} & M_{12} \\
    M_{21} & M_{22}
\end{pmatrix}$.
We force $M$ to have $\det(M)=1$ to avoid duplicated $\bsl{k}$ points in the mesh---avoiding the case where two $\bsl{k}$ points in the mesh are related by a reciprocal lattice vector.
When $M=\diag(1,1)$, it is called a conventional $L_1\times L_2$ mesh:
\begin{equation}
    \bsl{k} = \frac{k_1}{L_1}\bsl{b}_1 + \frac{k_2}{L_2}\bsl{b}_2
    \ . \label{eqn:linear_convention}
\end{equation}
When $M$ is not identity, it is called a tilted $L_1\times L_2$ mesh.
Visualizations of both conventions are shown in \cref{fig:Momentum_space}.
We use $N_k= L_1  L_2$ to label the number of lattice sites.

Given a generic mesh (tilted or not), we can define the linearized momentum index for the convenience of plotting:
\begin{equation}
    k = L_2 k_1 + k_2.
    \label{eqn:linearlize}
\end{equation} so that $k \in \{0, \dots, N_k-1\}$. 
We use $\bsl{a}_1$ and $\bsl{a}_2$ to label the primitive lattice vectors that satisfies $\bsl{a}_i\cdot \bsl{b}_j = 2\pi \delta_{ij}$

\section{$n$-Particle Reduced Density Matrices and $N$-representability Condition}
\label{app:n_part_RDM_n_rep}

The $n$-particle reduced density matrices ($n$-RDMs) and  their $N$-representability conditions have been discussed in \refcite{Mazziotti_2012_N_rep}.
In this section, we will review the concepts that we will use in this work.

In this work, we will mainly study the 2-RDM, which in general has the expression
\eq{
\prescript{2}{}{D}_{ijkl}^{\rm true} = \left\langle c_i^\dagger c_j^\dagger c_k c_l \right\rangle \ ,\label{eqn:D} 
}
where $c^\dagger_i$ and $c_{i}$ are the fermion creation and annihilation operators, respectively, and $i,j,k,l$ label the single particle indices.
Physically, the average is taken with respect to a mixed (or pure) state of $N$ particles.
 $\prescript{2}{}{D}_{ijkl}^{\rm true}$ is antisymmetric $\prescript{2}{}{D}_{ijkl}^{\rm true} = - \prescript{2}{}{D}_{jikl}^{\rm true}= - \prescript{2}{}{D}_{ijlk}^{\rm true}$ and hermitian $\prescript{2}{}{D}_{ijkl}^{\rm true} = (\prescript{2}{}{D}_{lkji}^{\rm true})^*$.
It also satisfies the particle number constraint $\sum_{ij} \prescript{2}{}{D}_{ijji}^{\rm true} =  N (N-1)$.
However, given a generic tensor $\prescript{2}{}{D}_{ijkl}$ that is antisymmetric $\prescript{2}{}{D}_{ijkl} = - \prescript{2}{}{D}_{jikl}= - \prescript{2}{}{D}_{ijlk}$, is hermitian $\prescript{2}{}{D}_{ijkl} = \prescript{2}{}{D}_{lkji}^*$ and preserves $\sum_{ij} \prescript{2}{}{D}_{ijji}^{\rm true} = N (N-1)$, it is not guaranteed that $\prescript{2}{}{D}_{ijkl}$ is given by any $N$-particle state.
The complete set of $N$-representability conditions answer this question: $\prescript{2}{}{D}_{ijkl}$ is given by a certain $N$-particle state if and only if it satisfies the complete set of $N$-representability conditions.

Nevertheless, the optimization with the complete $N$-representability conditions is  a NP-complete problem ~\cite{Liu_2007_NRep_NP_Complete,mazziottiTwoelectronReducedDensity2012,Mazziotti_2012_N_rep}, and in practice, we only incorporate part of them.
\refcite{Mazziotti_2012_N_rep} provide a way to classify the conditions.
$(2,p)$ condition for $\prescript{2}{}{D}_{ijkl}^{\rm true}$ reads
\eq{
\label{eq:n-rep}
\left\langle \hat{C} \right\rangle \geq 0 
}
with $\hat{C}$ satisfying the following conditions: (i) $\hat{C}=\sum_{i} T_i T^\dagger_i$, (ii) $T_i$ is a polynomial of $c$ and $c^\dagger$ of degree $p$, and (iii) $\hat{C}$ only contains 2-body operators, 1-body operators, and constants, \ie, $\hat{C}$ must have the form
\eq{
\hat{C} = \sum_{ijkl} c^\dagger_{i} c^{\dagger}_j c_k c_l X_{ijkl} + \sum_{ij} c^\dagger_{i} c_j Y_{ij} + Z
}
with $Z$ a scalar constant.
Then, \cref{eq:n-rep} reads
\eq{
\sum_{ijkl} \prescript{2}{}{D}_{ijkl}^{\rm true} X_{ijkl} + \sum_{ij} Y_{ij} \frac{1}{N - 1}\sum_{l}\prescript{2}{}{D}_{illj}^{\rm true} + Z \geq 0\ .
}
Replacing $\prescript{2}{}{D}_{ijkl}^{\rm true}$ by the predicted $\prescript{2}{}{D}_{ijkl}$ provides the actual $(2,p)$ condition for $D$:
\eq{
\sum_{ijkl} \prescript{2}{}{D}_{ijkl} X_{ijkl} + \sum_{ij} Y_{ij} \frac{1}{N - 1}\sum_{l}\prescript{2}{}{D}_{illj} + Z \geq 0\ .
\label{eq:n-rep-explicit}
}

Explicitly, the $(2,2)$ conditions becomes making the following three quantities positive semidefinite (PSD): (i) $\prescript{2}{}{D}_{ijkl}$, (ii) $G_{ijkl}$ that satisfies 
\eq{
\label{eq:D-G_relation}
\prescript{2}{}{G}_{ijkl}
=
\frac{\delta_{jk}}{N-1}
\sum_m \prescript{2}{}{D}_{i m m l}
+
\prescript{2}{}{D}_{k i j l}\ ,
}
and (iii) $\prescript{2}{}{Q}_{ijkl}$ that satisfies
\eq{
\prescript{2}{}{Q}_{ijkl}
=
\delta_{jk}\delta_{il}
-
\delta_{ik}\delta_{jl}
-
\frac{\delta_{jk}}{N-1}
\sum_m \prescript{2}{}{D}_{l m m i}
+
\frac{\delta_{ik}}{N-1}
\sum_m \prescript{2}{}{D}_{l m m j}
+
\frac{\delta_{jl}}{N-1}
\sum_m \prescript{2}{}{D}_{k m m i}
-
\frac{\delta_{il}}{N-1}
\sum_m \prescript{2}{}{D}_{k m m j}
+
\prescript{2}{}{D}_{k l i j}\ .
\label{eq:D-Q_relation}
}
Explicitly, the PSD means that $\prescript{2}{}{D}_{ijkl} $ is a PSD matrix by treating $(i,j)$ as one index and $(l,k)$ as the other; similarly for  $\prescript{2}{}{Q}$ and $\prescript{2}{}{G}$.

The underlying physical reason is that $\prescript{2}{}{D}^{\rm true}_{ijkl}$ is PSD, and so are the physical version of $\prescript{2}{}{Q}$ and $\prescript{2}{}{G}$ 
\begin{align}
    \prescript{2}{}{Q}_{ijkl}^{\rm true} &= \left\langle c_i c_j c_k^\dagger c_l^\dagger \right\rangle \label{eqn:Q} \\
    \prescript{2}{}{G}_{ijkl}^{\rm true} &= \left\langle c_i^\dagger c_j c_k^\dagger c_l \right\rangle \ .\label{eqn:G}
\end{align}

\refcite{Mazziotti_2012_N_rep} also provides two $(2,3)$ conditions, called the $T1$ and generalized $T2$ conditions.
The $T1$ condition corresponds to having $\hat{C}$ in \cref{eq:n-rep} as
\eq{
\label{eq:C_T1}
\hat{C}_{T1} = \frac{1}{2}\left(\hat{C}_{T1,1}\hat{C}_{T1,1}^\dagger + \hat{C}_{T1,2}\hat{C}_{T1,2}^\dagger \right)  \ ,
}
with 
\eqa{
\hat{C}_{T1,1} &= \sum_{jkl} a_{jkl} \hat{c}_j^\dagger \hat{c}_k^\dagger \hat{c}_l^\dagger, \\
\hat{C}_{T1,2} &= \sum_{jkl} a_{jkl}^* \hat{c}_j \hat{c}_k \hat{c}_l, \ 
}
and the generalized $T2$ condition has $\hat{C}$ as
\eq{
\label{eq:C_T2}
\hat{C}_{T2}  = \frac{1}{2}\left(\hat{C}_{T2,1}\hat{C}_{T2,1}^\dagger + \hat{C}_{T2,2}\hat{C}_{T2,2}^\dagger \right)
}
with 
\eqa{
\hat{C}_{T2,1} &= \sum_{jkl} b_{jkl} \hat{c}_j^\dagger \hat{c}_k^\dagger \hat{c}_l + \sum_j d_j \hat{c}_j^\dagger,  \\
\hat{C}_{T2,2} &= \sum_{jkl} b_{jkl}^* \hat{c}_j \hat{c}_k \hat{c}_l^\dagger + \sum_j e_j \hat{c}_j. 
}
Explicitly, the $T1$ and generalized $T2$ conditions become ensuring \cref{eq:n-rep-explicit} is satisfied for (i) $\hat C_{T1}$ given by
\begin{align}
\begin{aligned}
\hat C_{T1} &= \frac{1}{2}\left(\hat{C}_{T1,1}\hat{C}_{T1,1}^\dagger + \hat{C}_{T1,2}\hat{C}_{T1,2}^\dagger \right)\\
    &= \frac{1}{2} \bigg[ 
\sum_{rsmn} \bigg( \sum_i ( -a_{irs}^* a_{imn} + a_{irs}^* a_{min} - a_{irs}^* a_{mni} ) + \sum_j ( a_{rjs}^* a_{jmn} - a_{rjs}^* a_{mjn} + a_{rjs}^* a_{mnj} ) \\&\qquad+ \sum_k ( -a_{rsk}^* a_{kmn} + a_{rsk}^* a_{mkn} - a_{rsk}^* a_{mnk} ) \Bigg) c_m^\dagger c_n^\dagger c_r c_s
\\&\qquad+\sum_{pq} \bigg( \sum_{ij} \big( - a_{ijq}^* a_{ijp} + a_{ijq}^* a_{ipj} + a_{ijq}^* a_{jip} - a_{ijq}^* a_{pij} - a_{ijq}^* a_{jpi} + a_{ijq}^* a_{pji} \Big) \\&\qquad+ \sum_{ik} \Big( a_{iqk}^* a_{ikp} - a_{iqk}^* a_{ipk} - a_{iqk}^* a_{kip} + a_{iqk}^* a_{pik} + a_{iqk}^* a_{kpi} - a_{iqk}^* a_{pki} \Big) \\&\qquad+ \sum_{jk} \Big( - a_{qjk}^* a_{jkp} + a_{qjk}^* a_{jpk} + a_{qjk}^* a_{kjp} - a_{qjk}^* a_{pjk} - a_{qjk}^* a_{kpj} + a_{qjk}^* a_{pkj} \Big) \bigg) c_p^\dagger c_q
\\&\qquad+ \sum_{ikj} \left( a_{ijk}^*a_{ijk}- a_{ijk}^*a_{ikj} - a_{ijk}^*a_{jik} + a_{ijk}^*a_{kij}+ a_{ijk}^*a_{jki}-a_{ijk}^*a_{kji} \right) \bigg]
\\ &= \sum_{ijkl} X^{T1}_{ijkl} c_i^\dagger c_j^\dagger c_k c_l + \sum_{il} Y^{T1}_{il} c_i ^\dagger c_l + Z^{T1}
\end{aligned}
\end{align}
and (ii) $\hat C_{T2}$ given by
\begin{equation}
\begin{aligned}
    \hat{C}_{T2} &= \frac{1}{2}\left(\hat{C}_{T2,1}\hat{C}_{T2,1}^\dagger + \hat{C}_{T2,2}\hat{C}_{T2,2}^\dagger \right)\\
    &= \frac{1}{2}\bigg[\sum_{ijkl}\Big(d_i b_{lkj}^* + d_l^*b_{ijk} + b_{jil}e_k + b_{kli}^* e_j^* + \sum_{m}\big(b_{mkj}^* b_{mil} - b_{mkj}^* b_{iml}+ b_{mkm}^* b_{ijl} \\&\qquad- b_{kmj}^* b_{mil} + b_{klj}^* b_{mim} + b_{kmj}^* b_{iml}- b_{kmm}^* b_{ijl} - b_{klj}^* b_{imm} + b_{ijm} b_{lkm}^*\big)\Big)c_i^{\dagger} c_j^{\dagger} c_k c_l\\
    &\qquad + \sum_{il} \Big(d_i d_l^* -e_l e_i^* + \sum_{j}\big(-b_{ijj}e_l + b_{ijl}e_j+b_{jij}e_l - b_{jil}e_j-b_{ljj}^* e_i^* + b_{lji}^* e_j^* + b_{jlj}^* e_i^* - b_{jli}^* e_j^* \\&\qquad+ \sum_{k}(-b_{jkk}^*b_{jil} - b_{jli}^* b_{jkk} - b_{jki}^* b_{kjl}+ b_{jli}^* b_{kjk} + b_{jkk}^* b_{ijl} + b_{jkj}^* b_{kil} - b_{jlj}^* b_{kik} - b_{jkj}^* b_{ikl} + b_{jlj}^* b_{ikk}\\
    &\qquad + b_{lji}^* b_{jkk} - b_{lki}^* b_{jkj} + b_{lkk}^* b_{jij} - b_{lkk}^* b_{ijj} + b_{jki}^* b_{jkl})\big)\Big)c_i^{\dagger} c_l\\
    &\qquad + \sum_{i}\Big(|e_i|^2 + \sum_{j}\big(b_{jii}e_j - b_{jij}e_i + b_{jii}^* e_j^* - b_{jij}^* e_i^* + \sum_{k}(b_{jkk}^*b_{jii} - b_{jkk}^* b_{iji} - b_{jkj}^* b_{kii} + b_{jkj}^* b_{iki})\big)\Big)\bigg]
    \\ &= \sum_{ijkl} X^{T2}_{ijkl} c_i^\dagger c_j^\dagger c_k c_l + \sum_{il} Y^{T2}_{il} c_i ^\dagger c_l + Z^{T2}\ .
\end{aligned}
\end{equation}
Given a physical $D_{ijkl}$, the $T1$ and generalized $T2$ condition should hold for any complex $a_{ijk}$, $b_{ijk}$, $d_i$ and $e_i$.

Alternatively speaking, the $(2,3)$ conditions comes from linear combination of 3RDMs:
\eqa{
\prescript{3}{}{D}_{ijk lmn}^{\rm true}
&=
\left\langle
c_i^\dagger c_j^\dagger c_k^\dagger
c_l c_m c_n
\right\rangle,
\\
\prescript{3}{}{Q}_{ijk lmn}^{\rm true}
&=
\left\langle
c_i c_j c_k
c_l^\dagger c_m^\dagger c_n^\dagger
\right\rangle,
\\
\prescript{3}{}{E}_{ijk lmn}^{\rm true}
&=
\left\langle
c_i^\dagger c_j^\dagger c_k
c_l^\dagger c_m c_n
\right\rangle,
\\
\prescript{3}{}{F}_{ijk lmn}^{\rm true}
&=
\left\langle
c_i c_j c_k^\dagger
c_l c_m^\dagger c_n^\dagger
\right\rangle .
}
The $T1$ condition can come from 
\eq{
\left[T1^{\rm true}\right]_{ijk,lmn} = \prescript{3}{}{D}_{ijk lmn}^{\rm true} + \prescript{3}{}{Q}_{ lmn ijk}^{\rm true} \text{ being PSD}\ ,
}
or more explicitly, the following is PSD
\eqa{
\label{eq:T1_components}
\left[T1^{\rm true}\right]_{ijk,lmn} =&
-\delta_{il}\delta_{jm}\delta_{kn}
-\delta_{im}\delta_{jn}\delta_{kl}
-\delta_{in}\delta_{jl}\delta_{km}
+\delta_{il}\delta_{jn}\delta_{km}
+\delta_{im}\delta_{jl}\delta_{kn}
+\delta_{in}\delta_{jm}\delta_{kl}
\\[3pt]
&+
\gamma_{il}
\left(
\delta_{jm}\delta_{kn}
-
\delta_{jn}\delta_{km}
\right)
-
\gamma_{im}
\left(
\delta_{jl}\delta_{kn}
-
\delta_{jn}\delta_{kl}
\right)
+
\gamma_{in}
\left(
\delta_{jl}\delta_{km}
-
\delta_{jm}\delta_{kl}
\right)
\\
&-
\gamma_{jl}
\left(
\delta_{im}\delta_{kn}
-
\delta_{in}\delta_{km}
\right)
+
\gamma_{jm}
\left(
\delta_{il}\delta_{kn}
-
\delta_{in}\delta_{kl}
\right)
-
\gamma_{jn}
\left(
\delta_{il}\delta_{km}
-
\delta_{im}\delta_{kl}
\right)
\\
&+
\gamma_{kl}
\left(
\delta_{im}\delta_{jn}
-
\delta_{in}\delta_{jm}
\right)
-
\gamma_{km}
\left(
\delta_{il}\delta_{jn}
-
\delta_{in}\delta_{jl}
\right)
+
\gamma_{kn}
\left(
\delta_{il}\delta_{jm}
-
\delta_{im}\delta_{jl}
\right)
\\[3pt]
&+
\delta_{il}\,\prescript{2}{}{D}_{jkmn}^{\rm true}
-
\delta_{im}\,\prescript{2}{}{D}_{jkln}^{\rm true}
+
\delta_{in}\,\prescript{2}{}{D}_{jklm}^{\rm true}
\\
&-
\delta_{jl}\,\prescript{2}{}{D}_{ikmn}^{\rm true}
+
\delta_{jm}\,\prescript{2}{}{D}_{ikln}^{\rm true}
-
\delta_{jn}\,\prescript{2}{}{D}_{iklm}^{\rm true}
\\
&+
\delta_{kl}\,\prescript{2}{}{D}_{ijmn}^{\rm true}
-
\delta_{km}\,\prescript{2}{}{D}_{ijln}^{\rm true}
+
\delta_{kn}\,\prescript{2}{}{D}_{ijlm}^{\rm true}\ ,
}
where $\gamma_{ij}
=
\sum_{l} \prescript{2}{}{D}_{illj}^{\rm true}/(N-1)$.

In this work, we will not completely adopt the generalized $T2$ condition.
Instead, we just adopt the ordinary $T2$, which corresponds to $d_i=e_i=0$ in the generalized $T2$ condition.
It comes from $E+F$:
\eq{
\left[T2^{\rm true}\right]_{ijk,lmn} 
=
\prescript{3}{}{E}_{ijk lmn}^{\rm true} + \prescript{3}{}{F}_{nml kji }^{\rm true} \text{ being PSD.}
}
The tensor element is
\eqa{
\label{eq:T2_components}
\left[T2^{\rm true}\right]_{ijk,lmn}
&=\left\langle
c_i^\dagger c_j^\dagger c_k c_l^\dagger c_m c_n
+
c_n c_m c_l^\dagger c_k c_j^\dagger c_i^\dagger
\right\rangle \\
&=
\delta_{kl}\prescript{2}{}{D}_{ijmn}^{\rm true}
+\delta_{ml}\prescript{2}{}{D}_{ijkn}^{\rm true}
-\delta_{nl}\prescript{2}{}{D}_{ijkm}^{\rm true}
+\delta_{kj}\prescript{2}{}{D}_{ilmn}^{\rm true}
-\delta_{mj}\prescript{2}{}{D}_{ilkn}^{\rm true}
+\delta_{nj}\prescript{2}{}{D}_{ilkm}^{\rm true}
\\
& \qquad -\delta_{ki}\prescript{2}{}{D}_{jlmn}^{\rm true}
+\delta_{mi}\prescript{2}{}{D}_{jlkn}^{\rm true}
-\delta_{ni}\prescript{2}{}{D}_{jlkm}^{\rm true}
\\
& \qquad -\delta_{kj}\delta_{ml}\gamma_{in}
+
\left(
-\delta_{mj}\delta_{nl}
+
\delta_{ml}\delta_{nj}
\right)\gamma_{ik}
+
\delta_{ki}\delta_{ml}\gamma_{jn}
+
\left(
\delta_{mi}\delta_{nl}
-
\delta_{ml}\delta_{ni}
\right)\gamma_{jk}
+
\delta_{kj}\delta_{nl}\gamma_{im}
-
\delta_{ki}\delta_{nl}\gamma_{jm}
\\
& \qquad +
\left(
-\delta_{ki}\delta_{mj}
+
\delta_{kj}\delta_{mi}
\right)\gamma_{ln}
+
\left(
\delta_{ki}\delta_{nj}
-
\delta_{kj}\delta_{ni}
\right)\gamma_{lm}
+
\left(
-\delta_{mi}\delta_{nj}
+
\delta_{mj}\delta_{ni}
\right)\gamma_{lk}
-\delta_{ml}
\left(
\delta_{ki}\delta_{nj}
-
\delta_{kj}\delta_{ni}
\right) \\
& \qquad +
\delta_{nl}
\left(
\delta_{ki}\delta_{mj}
-
\delta_{kj}\delta_{mi}
\right)\ ,
}
which is PSD.
The T1 and T2 conditions of $\prescript{2}{}{D}$ are given by replacing $\prescript{2}{}{D}^{\rm true}$ in \cref{eq:T1_components,eq:T2_components} by $\prescript{2}{}{D}$.

\section{Neural Network (NN) Architectures}
\label{app:Architectures}
For this paper, we adopt six different architectures.
The architecture is independent of the loss function---we can train all architectures on the same loss function and data.
The architectures are (i) a multi-layer perceptron (MLP) \cite{Hornik_1989_MLP}, (ii) a Residual MLP \cite{He2015Residual, Lee-Thorp2022FNet}, (iii) a  Kolmogorov–Arnold Network (KAN) \cite{Liu_2025_KAN}, (iv) an implicit neural representation (INR) in the style of SIREN \cite{Sitzmann2020SIREN}, (v) an INR in the style of FINER \cite{Liu2024FINER, Essakine2025ImplictNeural}, and (vi) a Transformer-based Network (TBN) \cite{vaswani2023attentionneed}.

Each architecture represents a mapping
\begin{equation}
\Phi:\mathbb{R}^m\to\mathbb{R}^2 \label{eqn:NN_eqn}
\end{equation}
where $m$ and $n$ are the input and output dimensions, respectively. 
As we study $n$-RDM in the momentum space, the inputs are the fermion indices, including momenta  $(\bsl{k}_1,\bsl{k}_2,...,)$ and all other degrees of freedom per momentum.
The outputs are the values of the RDM evaluated at these indices. 

In the following, we will review the six architectures.

\subsection{MLP}

Our first architecture is the MLP. The MLP is the standard NN architecture, owing to its ability to universally approximate any vector valued function with arbitrary precision \cite{Hornik_1989_MLP}. Based on this property, given an  arbitrary input $\bsl{x} \in \mathbb{R}^m$ , the MLP is able to predict an output $f(\bsl{x})\in \mathbb{R}^n$. The function described in \cref{eqn:NN_eqn} which represents the NN is thus an approximation of this function and, in our work an approximation of the true function of the $n$-RDM. For the MLP, we can choose the number of hidden layers, with each hidden layer  defined as
\begin{equation}
    \bsl{h}^{(l+1)} = \sigma\left(\bsl{W}^{(l)}\bsl{h}^{(l)} + \bsl{f}^{(l)}\right) 
\end{equation}
where $\bsl{W}^{(l)} \in \mathbb{R}^{h \times h}$ and  $\bsl{f}^{(l)}\in\mathbb{R}^{h}$ are the weights and biases at layer $l$. Here $\bsl{W}^{(l)}\bsl{h}^{(l)}$ represents standard matrix multiplication. For these layers, $h$ is the hidden dimension and $\sigma$ is the Rectified Linear Unit (ReLU) activation function that acts component-wise, 
\begin{equation}
\sigma(x) =    \max (0,x).
\label{eqn:RELU}
\end{equation}
where the ReLU function adds the non-linearity to the network. This gives us the full MLP 
architecture
\begin{equation}
    \begin{aligned}
            \bsl{h}^{(1)} &= \sigma\left(\bsl{W}^{(0)}\bsl{x} + \bsl{f}^{(0)}\right) \in \mathbb{R}^h\\
            \bsl{h}^{(2)} &= \sigma\left(\bsl{W}^{(1)}\bsl{h}^{(1)} + \bsl{f}^{(1)}\right)\in \mathbb{R}^h \\
            &\vdots \\
            \bsl{h}^{(d-1)}&=  \sigma\left(\bsl{W}^{(d-2)}\bsl{h}^{(d-2)} + \bsl{f}^{(d-2)}\right)\in \mathbb{R}^h \\
            \bsl{T} &=  \bsl{W}^{(d-1)}\bsl{h}^{(d-1)} + \bsl{f}^{(d-1)}\in \mathbb{R}^n
        \label{eqn:MLP}
    \end{aligned}
\end{equation}
where $\bsl{x\in\mathbb{R}}^m$ is our input vector, $\bsl{T}$ is our output, and $d$ is the depth, the total number of layers. We note that $\bsl{W}^{(d-1)}\in\mathbb{R}^{n\times h}$ and $\bsl{f}^{(d-1)}\in \mathbb{R}^n$ instead of $\mathbb{R}^h$; for future equations we assume the dimension of the final weights and biases allows an evaluation to $\mathbb{R}^n$. $\bsl{T}\in \mathbb{R}^n$ is the same output as the $\Phi$ mapping defined above. The weights and biases are initialized from a uniform distribution $\mathcal{U}\left(-d_{in}^{-\frac{1}{2}},d_{in}^{-\frac{1}{2}}\right)$ where $d_{in}$ is the number of input features of each respective layer. Our full analysis of this architecture performs a sweep over the depth and hidden dimension, using values in \cref{tab:MLP_hyper}.

\begin{table}[htb]
    \centering
    \setlength{\tabcolsep}{8pt} 
    \renewcommand{\arraystretch}{1.20}  
    \begin{tabular}{c|c}
         \toprule
         \textbf{Hyperparameter} & \textbf{Value} \\
         \midrule
         Depth & $[1,2,3,4,5, 6]$ \\
         Hidden Dimension & $[1, 2, 4,8,16,32,64,128,256]$ \\
         \bottomrule
    \end{tabular}
    \caption{List of hyperparameters for training an MLP. The depth, $d$, refers to the number of layers in the model, while the hidden dimension, $h$, corresponds to the number of nodes in each hidden layer. }
    \label{tab:MLP_hyper}
\end{table}

\subsection{Residual MLP}
A Residual MLP modifies the MLP by replacing each hidden layer with a residual block. Each block now computes
\begin{equation}
\label{eq:residual_MLP}
    \bsl{h}^{(l+1)} = \text{LayerNorm}\!\left(\bsl{h}^{(l)} + \mathcal{F}(\bsl{h}^{(l)}, W^{(l)}_1, W^{(l)}_2, \bsl{f}^{(l)}_1, \bsl{f}^{(l)}_2)\right)
\end{equation}
where
\begin{equation}
    \mathcal{F}(\bsl{h}^{(l)}, W^{(l)}_1, W^{(l)}_2, \bsl{f}^{(l)}_1, \bsl{f}^{(l)}_2) = W_2^{(l)}\,\sigma\!\left(W_1^{(l)}\bsl{h}^{(l)} + \bsl{f}_1^{(l)}\right) + \bsl{f}_2^{(l)},
\end{equation}
with $W_1^{(l)}, W_2^{(l)} \in \mathbb{R}^{h \times h}$ and $\bsl{f}_1^{(l)}, \bsl{f}_2^{(l)} \in \mathbb{R}^h$. Here $\sigma$ is the Gaussian Error Linear Unit (GeLU) activation function 
\begin{equation}
    \sigma(x) = x \cdot \frac{1}{2}\left(1+\operatorname{erf}\left(\frac{x}{\sqrt{2}}\right) \right)
    \label{eqn:GELU}
\end{equation}
and LayerNorm \cite{Ba2016LayerNorm} normalizes the activations across features to stabilize training. 
The addition of $\bsl{h}^{(l)}$ to the output of $\mathcal{F}$ forms a residual connection, allowing gradients to flow directly through the network and counteracting the vanishing gradient problem that limits the trainable depth of plain MLPs \cite{He2015Residual}. Our initialization is the same as the MLP.
The final architecture is just given by replacing the layer expression in \cref{eqn:MLP} by \cref{eq:residual_MLP}.
Our analysis performs a sweep over the depth and hidden dimension, as defined in \cref{tab:MLP_residual_hyper}.

\begin{table}[htb]
    \centering
    \setlength{\tabcolsep}{8pt} 
    \renewcommand{\arraystretch}{1.20}  
    \begin{tabular}{c|c}
         \toprule
         \textbf{Hyperparameter} & \textbf{Value} \\
         \midrule
         Depth & $[1,2,3,4,5, 6, 10, 15, 20, 40]$ \\
         Hidden Dimension & $[1, 2, 4,8,16,32,64,128,256]$ \\
         \bottomrule
    \end{tabular}
    \caption{List of hyperparameters for training a Residual MLP. The depth, $d$, refers to the number of residual blocks in the model, while the hidden dimension, $h$, corresponds to the number of nodes in each hidden layer. }
    \label{tab:MLP_residual_hyper}
\end{table}

\subsection{KAN}
While an MLP is based on the universal approximation theorem, KANs are based on the Kolmogorov-Arnold representation theorem, which presents a way to represent multivariate continuous functions as a sum of functions of a single variable \cite{Liu_2025_KAN}.
For this representation, a KAN replaces the fixed activation functions of an MLP with a learnable activation function defined on each edge of the network. Each KAN layer $\bsl{h}^{(l+1)}$ is a matrix of $1$D functions 
\begin{equation}
    {h}^{(l+1)}_q = \sum_{p=1}^{n_{in}}\phi_{q,p}\left(h_p^{(l)}\right)
\end{equation}
where $q=1,2,\dots,n_{out}$, $p=1,2,\dots,n_{in}$, $n_{in}$ and $n_{out}$  are the input and output dimensions of the layers respectively. For our implementation, $n_{in}=n_{out}=h$ for hidden layers, while $n_{in}$ of the first layer is $m$ and $n_{out}$ of the last layer is $n$, using the values defined for $\Phi$ in \cref{eqn:NN_eqn}. For these activation functions, we have
\begin{align}
\phi_{q,p}(x) &= w_b b(x)+w_s \text{spline}(x), \\
    b(x) &= \frac{x}{1+e^{-x}}\\
    \text{spline}(x) &= \sum_i^{G+k} c_i B_i(x).
\end{align}
where $c_i$ are trainable, and $B_i$ are B-splines, piecewise polynomial functions of order $k$ defined over a uniform grid. $G$ represents the grid size, and we set $k=3$.  A larger grid size increases the number of basis functions per edge, allowing for more complex activations and potentially better representation of the data. Note that  $w_b$ and $w_s$ are trainable parameters initialized via Kaiming uniform initialization \cite{Kaiming2015Initialization}, where weights are drawn from a uniform distribution $\mathcal{U}\left(-\sqrt{\frac{6}{(1+a^2)n_{in}}}, \sqrt{\frac{6}{(1+a^2)n_{in}}}\right)$. The base scale $s_b$ and spline scale $s_s$ are fixed hyperparameters that control the initialization variance of those weights by setting $a=\sqrt{5}s_b$ or $a=\sqrt{5}s_s$ for the scale and spline weights respectively.
The spline coefficients, in contrast, are initialized by fitting to small noise near zero, encouraging a simpler model at the start of training. The full set of hyperparameters is listed in  \cref{tab:KAN_hyper}.

\begin{table}[htb]
    \centering
    \setlength{\tabcolsep}{8pt} 
    \renewcommand{\arraystretch}{1.20}  
    \begin{tabular}{c|c}
         \toprule
         \textbf{Hyperparameter} & \textbf{Value} \\
         \midrule
         Depth & $[1,3,5]$ \\
         Hidden Dimension & $[ 2, 4, 8,  64, 256]$ \\
         Grid Size & $[1,  3, 5]$ \\
         Base Scale & $[1, 5]$ \\
         Spline Scale & $[1, 5]$ \\
         \bottomrule
    \end{tabular}
    \caption{List of hyperparameters for training a KAN. Depth, $d$, is the number of layers and hidden dimension $h = n_{in} =n_{out}$  is the width of each hidden layer. Grid size, $G$, controls the number of B-spline  basis functions per edge, so a larger $G$ allows the activation functions to represent more complex shapes. The base scale $s_b$ and spline scale $s_s$ both proportionally modify the slope parameter $a$ of the Kaiming uniform initialization. A larger value shrinks the initialization interval.}
    
    \label{tab:KAN_hyper}
\end{table}

\subsection{SIREN}
A SIREN replaces the standard activations of the MLP with a sinusoidal
representation such that each layer has the component-wise activation function
\begin{equation}
    \sigma(x) = \sin(\omega x)
\end{equation}
where $\omega$ is either $\omega_0$ or $\omega_1$ defined below.
This method has been shown to be effective in representing continuous signals and complex function mappings in the context of implicit neural representations \cite{Essakine2025ImplictNeural}. 

We perform a sweep over the model depth, $d$ and hidden dimension, $h$ as with the MLP. For the SIREN, we additionally sweep over two frequency parameters: First $\omega_0$ which controls the frequency of the first layer, and Hidden $\omega_1$ which controls the frequency of all subsequent hidden layers. These values scale the argument of the activation, and their different initialization has been shown to benefit training \cite{Sitzmann2020SIREN}. A final linear layer is appended after the last sinusoidal layer without an activation function, allowing 
predictions outside the range $[-1, 1]$ imposed by the sinusoidal 
activation. The entire NN architecture, then, takes the form 
\begin{equation}
\begin{aligned}
    \bsl{h}^{(1)} &= \sin\left(\omega_0 \middle(W^{(0)}\bsl{x} + \bsl{f}^{(0)}\middle) \right) \in \mathbb{R}^h\\
    \bsl{h}^{(2)} &= \sin\left(\omega_1 \middle(W^{(1)}\bsl{h}^{(1)} + \bsl{f}^{(1)}\middle) \right) \in \mathbb{R}^h\\ 
    & \vdots \\
    \bsl{h} ^{(d-1)} &=  \sin\left(\omega_1 \middle(W^{(d-2)}\bsl{h}^{(d-2)} + \bsl{f}^{(d-2)}\middle) \right) \in \mathbb{R}^h\\ 
    \bsl{T} & = W^{(d-1)} \bsl{h}^{(d-1)} + \bsl{f}^{(d-1)} \in \mathbb{R}^n
\end{aligned}
    \label{eqn:SIREN}
\end{equation}
where $\bsl{x}\in\mathbb{R}^m$ is our input coordinate tensor and $\bsl{T}$ is our output. 
The SIREN uses the initialization scheme of \refcite{Sitzmann2020SIREN} where we take the first layer weights $W^{(0)}$ as drawn from a uniform distribution $\mathcal{U}\left(\frac{-1}{m},\frac{1}{m}\right)$ and the subsequent layer weights $W^{(l)}$ are taken from the distribution $\mathcal{U}\left(-\sqrt{\frac{6}{h}}\cdot \omega_1^{-1},\sqrt{\frac{6}{h}}\cdot \omega_1^{-1} \right)$.

 The full set of hyperparameters is listed in \cref{tab:SIREN_hyper}

\begin{table}[htb]
    \centering
    \setlength{\tabcolsep}{8pt} 
    \renewcommand{\arraystretch}{1.20}  
    \begin{tabular}{c|c}
         \toprule
         \textbf{Hyperparameter} & \textbf{Value} \\
         \midrule
         Depth & $[1,2,3,4,5]$ \\
         Hidden Dimension & $[1, 2, 4,  16, 32, 64, 128, 256]$ \\
         First $\omega_0$ & $[3, 6, 12]$ \\
         Hidden $\omega_1$ & $[5, 10]$ \\
         Linear Layer & [True] \\
         \bottomrule
    \end{tabular}
    \caption{List of hyperparameters for training a SIREN. Depth, $d$ is the number of layers  and hidden dimension, $h$, is the number of nodes in a hidden layer. The $\omega_0$ parameters control the frequency of the activation function on the first and hidden layers respectively, and the Linear Layer parameter exists to demonstrate the design of the output layer. }
    \label{tab:SIREN_hyper}
\end{table}

\subsection{FINER}
A FINER is a modification of the SIREN in order to better represent a broad spectrum of frequencies. While a SIREN has a fixed scaling value, a FINER uses the activation function
\begin{equation}
    \sigma(x) = \sin(\omega_i(|x|+1)x).
\end{equation}
where the factor $(|x|+1)$ modifies the frequency based on the value of the neuron before the activation function is applied \cite{Essakine2025ImplictNeural, Liu2024FINER}. This allows different neurons to operate at a broader range of frequencies. The full NN architecture is then, the same as \cref{eqn:SIREN} with the modified activation function.

As with SIREN, we sweep over depth and hidden dimension, as well as First $\omega_0$ and Hidden $\omega_1$, which initialize the frequency scale for the first and further layers. A final linear layer is appended, providing the same role as in the SIREN.  The FINER uses the same weight initialization as SIREN. The full set of hyperparameters is listed in \cref{tab:FINER_hyper}.

\begin{table}[htb]
    \centering
    \setlength{\tabcolsep}{8pt} 
    \renewcommand{\arraystretch}{1.20}  
    \begin{tabular}{c|c}
         \toprule
         \textbf{Hyperparameter} & \textbf{Value} \\
         \midrule
         Depth & $[1,2,3]$ \\
         Hidden Dimension & $ [1, 4, 8, 16,  64, 128, 256]$ \\
         First $\omega_0$ & $[0.1, 1, 3, 6]$ \\
         Hidden $\omega_1$ & $[1, 5, 10]$ \\
 Linear Layer & [True] \\
         \bottomrule
    \end{tabular}
    \caption{List of hyperparameters for training a FINER. Depth, $d$ is the number of layers  and hidden dimension, $h$, is the number of nodes in a hidden layer. The $\omega_0$ parameters control the frequency of the activation function on the first and hidden layers respectively, and the Linear Layer parameter exists to demonstrate the design of the output layer.  }
    \label{tab:FINER_hyper}
\end{table}

\subsection{TBN}

The TBN is a Transformer-Based Neural Network that uses a global self-attention mechanism \cite{vaswani2023attentionneed} to map input coordinates to n-RDM values. Our implementation is largely based on the linear attention variant from \refcite{Katharopoulos2020LinearAttention}. Unlike the MLP and residual MLP architectures, the TBN processes all input coordinates simultaneously, allowing each point to attend to all other points in the input sequence and thereby capture long-range structure in momentum space.

To address the low-frequency spectral bias of standard networks, the TBN begins with a Fourier feature (FF) encoder \cite{Rahimi2007RandomFeatures,Mildenhall2020NeRF,Tancik2020FourierFeatures} that maps each input coordinate $\widetilde{\bsl{x}} \in [-1,1]^m$ (where $m\in\mathbb{N}$ is the input dimensionality) to a higher-dimensional sinusoidal representation. The coordinates are first shifted to the interval $[0, 1]$ via $\bsl{x} = \left(\widetilde{\bsl{x}}+\bsl{1}_{m}\right)/2$ (where $\bsl{1}_{m}$ denotes an $m$-length vector of all ones) to produce a new vector $\bsl{x} \in [0,1]^m$, and then for $N_f$ log-spaced frequencies $f_i = 2^{i-1}$ (where $i = 1, \ldots, N_f$), each coordinate component $x_j$ is mapped as
\begin{equation}
    \phi(\bsl{x}) = \left[\sin(2\pi f_i x_j),\; \cos(2\pi f_i x_j)\right]_{i,j},
\end{equation}
producing a feature vector $\phi(\bsl{x})\in \mathbb{R}^{1\times 2mN_f}$. This provides the network with multi-scale sinusoidal basis functions, which allows the TBN to represent fine-scale variation in the n-RDM without relying only on learned weights to recover high-frequency content. Next, define the matrix $X_{\phi}\in\mathbb{R}^{N_k\times 2mN_f}$ (where $N_k$ is the number of $\bsl{k}$ points, $\ie$ the number of input coordinates we have) by:
\begin{equation}
    X_{\phi} = \begin{bmatrix}
      \phi(\bsl{x}_1) \\
      \vdots\\
      \phi(\bsl{x}_{N_k}) \\
    \end{bmatrix}\in\mathbb{R}^{N_k\times 2mN_f},
\end{equation}
where the subscript $i\in\{1, \dots N_k\}$ indexes the $N_k$ vectors $\phi(\bsl{x}_i)$. The encoded features are then projected into a hidden space of dimension $h$ via a learned linear map, $\ie$, we have:
\begin{equation}
     H^{(0)} = X_{\phi}W^{(0)} + \bsl{1}_{N_k}\bsl{f}^{(0)},
\end{equation}
where $\bsl{1}_{N_k}\in \mathbb{R}^{N_k\times 1}$ is a column vector of all ones, while $W^{(0)}\in\mathbb{R}^{2 mN_f\times h}$ and $\bsl{f}^{(0)}\in\mathbb{R}^{1\times h}$ are the input projection weights and biases, and $H^{(0)}\in\mathbb{R}^{N_k\times h}$. After this transformation, we apply $d$ transformer layers. For each $\ell\in\{0, \dots, d-1\}$, the transformer layer then applies
\begin{equation}
\begin{aligned}
    A^{(\ell)}
    &=
    H^{(\ell)}
    +
    \operatorname{Attn}\left(
    \operatorname{LayerNorm}(H^{(\ell)})
    \right), \\
    \hat{A}^{(\ell)} &= \operatorname{LayerNorm}\left(A^{(\ell)}\right), \\
    H^{(\ell+1)}
    &=
    A^{(\ell)}
    +\operatorname{Dropout}\left(
    \operatorname{Dropout}\left(
    \sigma\left(
    \hat{A}^{(\ell)}W_1^{(\ell)}
    +
    \bsl{1}_{N_k}\bsl{f}_1^{(\ell)}
    \right)\right)
    W_2^{(\ell)}
    +
    \bsl{1}_{N_k}\bsl{f}^{(\ell)}_2\right),
\end{aligned}
\label{eqn:TBN_Layer}
\end{equation}
where $\sigma$ is the GELU activation \cref{eqn:GELU}, $\operatorname{LayerNorm}$ \cite{Ba2016LayerNorm} normalizes the activations across features, and the dropout operation \cite{Srivastava2014Dropout} (denoted here as ``$\operatorname{Dropout}$'') deactivates a certain percentage of the NN nodes each training epoch, forcing the NN to create redundancy and learn a finer structure. Here, $W_1^{(\ell)} \in \mathbb{R}^{h\times h_{\text{ffn}}}$, $W_2^{(\ell)} \in \mathbb{R}^{h_{\text{ffn}}\times h}$, and $\bsl{f}_1^{(\ell)} \in \mathbb{R}^{1\times h_{\text{ffn}}}$, $\bsl{f}_2^{(\ell)} \in \mathbb{R}^{1\times h}$ are the weights and biases of the feed-forward network, with $h_{\text{ffn}}=4h$ the intermediate dimension. Notably, $\operatorname{Attn}(\cdot)$ in \cref{eqn:TBN_Layer} is the attention mechanism, which we now define as follows. Let $\widetilde{H}^{(\ell)} = \operatorname{LayerNorm}(H^{(\ell)})$, $N_h\in\mathbb{N}$ be the number of attention heads, and define $D_{\mathrm{Attn}}\in\mathbb{N}$ as $D_{\mathrm{Attn}} = h/N_h$ (we choose $N_h$ so that $h\bmod N_h = 0$). Define a function $\psi_{\mathrm{Attn}}: \mathbb{R}\rightarrow\mathbb{R}$ via
\begin{align}
    \psi_{\mathrm{Attn}}(u) &= \operatorname{ELU}(u) +1, \qquad\forall u\in \mathbb{R} \\
    \operatorname{ELU}(u) &= \begin{cases}
        u &\text{ if } u > 0\\
        \exp(u)-1 &\text{ if } u\leq 0
    \end{cases}
\end{align}
Construct learnable weight matrices $W_Q^{(\ell)}, W_K^{(\ell)}, W_V^{(\ell)}, W_O^{(\ell)}\in\mathbb{R}^{h\times h}$ and learnable bias vectors $\bsl{f}_Q^{(\ell)}, \bsl{f}_K^{(\ell)}, \bsl{f}_V^{(\ell)}, \bsl{f}_O^{(\ell)}\in\mathbb{R}^{1\times h}$, and compute
\begin{align}
    \widetilde{Q}^{(\ell)} &= \psi_{\mathrm{Attn}}\left(\widetilde{H}^{(\ell)}W_Q^{(\ell)}+\bsl{1}_{N_k}\bsl{f}_Q^{(\ell)}\right), \\
    \widetilde{K}^{(\ell)} &= \psi_{\mathrm{Attn}}\left(\widetilde{H}^{(\ell)}W_K^{(\ell)}+\bsl{1}_{N_k}\bsl{f}_K^{(\ell)}\right), \\
    \widetilde{V}^{(\ell)} &= \widetilde{H}^{(\ell)}W_V^{(\ell)}+\bsl{1}_{N_k}\bsl{f}_V^{(\ell)}, 
    \qquad
    \widetilde{Q}^{(\ell)},\widetilde{K}^{(\ell)},\widetilde{V}^{(\ell)}\in\mathbb{R}^{N_k\times h}
\end{align}
where the action of $\psi_{\mathrm{Attn}}$ on a matrix-valued input is understood to be applied elementwise. Next, for all $i_h\in\{1, \dots N_h\}$, define the index sets $I_x=\{1, \dots N_k\}$ and $I_y(i_h) = \{(i_h-1)D_{\mathrm{Attn}}+1, \dots i_h D_{\mathrm{Attn}}\}$, and define
\begin{equation}
    Q_{i_h}^{(\ell)} = \widetilde{Q}^{(\ell)}_{I_x,\, I_y(i_h)},
    \quad
    K_{i_h}^{(\ell)} = \widetilde{K}^{(\ell)}_{I_x,\, I_y(i_h)},
    \quad
    V_{i_h}^{(\ell)} = \widetilde{V}^{(\ell)}_{I_x,\, I_y(i_h)},
\end{equation}
where we have used the notation $\widetilde{Q}^{(\ell)}_{I_x,\, I_y(i_h)}$ to represent the submatrix $\left(\widetilde{Q}^{(\ell)}_{ij}\right)_{i\in I_x,\,j\in I_y(i_h)}$ (and similarly for $\widetilde{K}^{(\ell)}_{I_x,\, I_y(i_h)}$ and $\widetilde{V}^{(\ell)}_{I_x,\, I_y(i_h)}$). Let $Y_{i_h}^{(\ell)}\in \mathbb{R}^{N_k \times D_{\mathrm{Attn}}}$ be defined by
\begin{equation}
    \left(Y_{i_h}^{(\ell)}\right)_{j} = \frac{\sum_{k'=1}^{N_k} \left(\left(Q_{i_h}^{(\ell)}\right)_j \left(K_{i_h}^{(\ell)}\right)_{k'}^T\right)\left(V_{i_h}^{(\ell)}\right)_{k'}}{\max\left(\sum_{k'=1}^{N_k}\left(Q_{i_h}^{(\ell)}\right)_j \left(K_{i_h}^{(\ell)}\right)_{k'}^T,\, 1\times10^{-6}\right)},
\end{equation}
where we use the notation $\left(Y_{i_h}^{(\ell)}\right)_{j}$ to indicate the $j$th row of $Y_{i_h}^{(\ell)}$ (and similarly for the matrices $Q_{i_h}^{(\ell)}, K_{i_h}^{(\ell)}, V_{i_h}^{(\ell)}$). We compute the attention output $\operatorname{Attn}\left(\widetilde{H}^{(\ell)}\right)\in\mathbb{R}^{N_k\times h}$ by:
concatenating the matrices $Y_{i_h}^{(\ell)}$ for each $i_h\in\{1, \dots, N_h\}$, applying dropout, and then projecting to the output dimension via a matrix multiplication with $W_O^{(\ell)}$:
\begin{equation}\label{eqn:attn}
    \operatorname{Attn}\left(\widetilde{H}^{(\ell)}\right) = \operatorname{Dropout}\left(\operatorname{Concat}\left(Y_{1}^{(\ell)}, \dots, Y_{N_h}^{(\ell)}\right)\right)W_O^{{(\ell)}}+\bsl{1}_{N_k}\bsl{f}_O^{(\ell)}\in\mathbb{R}^{N_k\times h},
\end{equation}
where ``Concat'' refers to the concatenation operation. Next, let
\[
    \theta_{\rm ffn}^{(\ell)}
    =
    \left(
    W_1^{(\ell)},
    W_2^{(\ell)},
    \bsl{f}_1^{(\ell)},
    \bsl{f}_2^{(\ell)}
    \right)
    \qquad
    \text{and}
    \qquad
    \theta_{\rm Attn}^{(\ell)}
    =
    \left(
    W_Q^{(\ell)},
    W_K^{(\ell)},
    W_V^{(\ell)},
    W_O^{(\ell)},
    \bsl{f}_Q^{(\ell)}, \bsl{f}_K^{(\ell)}, \bsl{f}_V^{(\ell)}, \bsl{f}_O^{(\ell)}
    \right)
\]
denote the learnable parameters for the feed-forward network and the attention mechanism in the $\ell$th layer, respectively (note that every $\operatorname{LayerNorm}$ also carries learnable affine parameters, which we suppress from the notation for simplicity). Define $\theta_{\rm TBN}^{(\ell)} = \left(\theta_{\rm ffn}^{(\ell)},\,\theta_{\rm Attn}^{(\ell)}\right)$. For ease of notation,
we write
\begin{equation}
    H^{(\ell+1)}
    =
    \mathcal{F}^{(\ell)}
    \left(
    H^{(\ell)};\theta_{\rm TBN}^{(\ell)}
    \right),
\end{equation}
where
$\mathcal{F}^{(\ell)}:\mathbb{R}^{N_k\times h}\rightarrow
\mathbb{R}^{N_k\times h}$ denotes the update in \cref{eqn:TBN_Layer}. This gives us a full architecture
\begin{equation}
    \begin{aligned}
            H^{(0)} &= X_{\phi}W^{(0)} + \bsl{1}_{N_k}\bsl{f}^{(0)} \in\mathbb{R}^{N_k\times h}\\
            H^{(1)} &= \mathcal{F}^{(0)}\left(H^{(0)};\theta_{\rm TBN}^{(0)}\right) \in\mathbb{R}^{N_k\times h}\\
            &\vdots \\
            H^{(d)}&= \mathcal{F}^{(d-1)}\left(H^{(d-1)};\theta_{\rm TBN}^{(d-1)}\right) \in\mathbb{R}^{N_k\times h}\\
            T &= \operatorname{LayerNorm}\!\left(H^{(d)}\right)W^{(d)} + \bsl{1}_{N_k}\bsl{f}^{(d)} \in\mathbb{R}^{N_k\times n}
        \label{eqn:TBN}
    \end{aligned}
\end{equation}
where $W^{(d)}\in\mathbb{R}^{h\times n}$ and $\bsl{f}^{(d)}\in\mathbb{R}^{1\times n}$ are the output weights and biases. Hidden dimension weights and biases are described above (see \cref{eqn:TBN_Layer}).
\begin{table}[htb]
    \centering
    \setlength{\tabcolsep}{8pt}
    \renewcommand{\arraystretch}{1.20}

    \begin{tabular}{c|c}
        \toprule
        \textbf{Hyperparameter} & \textbf{Value} \\
        \midrule
        Depth & $[1, 2, 3, 4, 6]$ \\
        Hidden Dimension & $[8, 16, 32, 64, 128]$ \\
        Number of Heads & $[1, 2, 8]$ \\
        Number of Frequencies & $[4, 8]$ \\
        Dropout & $[0.0, 0.1]$ \\
        \bottomrule
    \end{tabular}
    \caption{List of hyperparameters for training a TBN. Depth, $d$, and hidden dimension $h$ control the size of the model. Number of heads, $N_h$, gives the number of parallel heads in the Attention layer, described in \cref{eqn:attn}. Number of frequencies, $N_f$, determines the initial number of features in the FF encoding. Dropout percentage impacts training, and fights against overfitting.}
    \label{tab:ITRFM_hypr}
\end{table}

\section{ Interpolation Training, Variational Optimization and Testing}

In this work, we will use the NNs in two ways.
The first way is called the interpolation.
We train NNs on $n$-RDMs computed on small $L_1 \times L_2$ lattices to predict $n$-RDMs on larger systems. 
This task is an interpolation, as the larger-size $\bsl{k}$ mesh is a denser version of the smaller-size mesh.
We note that in our training and predictions, the smaller meshes do not need to be subsets of the larger meshes. 
The second way is to use NN as variational ansatze, which will be optimized according to the chosen loss functions.
The ansatze is usually directly optimized for large sizes.

In both interpolation training and variational optimization, we need to define the loss functions, and we will need to test them based on certain criteria.
In this section, we will discuss the loss functions and test criteria that we may adopt.
We will also provide a way to benchmark the efficiency of the model.

\subsection{Loss Functions for Training and Optimization}
\label{app:losses}
 
In this part, we will discuss all the loss functions.
For each loss function, we will specify the form of NN it is used for.
Even if we stick to one way of using NN, we may not adopt all loss functions in all scenarios and the specific ways of using the same loss function may vary, which will be specified in \cref{app:Richardson} and \cref{app:FCI}.

We begin with 4 loss functions that we only used for the interpolation training.
First, a fundamental loss for interpolation training is the mean square error defined on our true tensor $T_{\text{true}}$ (RDMs on small sizes) and our predicted tensor $T$ as
\begin{equation}
    \mathcal{L}_{mse} = {\frac{1}{\operatorname{Card}(T_{\text{true}})}\sum_{i=1}^{\operatorname{Card}(T_{\text{true}})}\left|T_i - T_{\text{true},i}\right|^2}
    \label{eqn:MSE_Loss}
\end{equation}
where $\operatorname{Card}(T_{\text{true}})$ is the number of complex elements of $T_{\text{true}}$, $i$ labels the $i$th element, and $|\cdot|$ is the modulus. This loss forces the NN output to be similar to the true data on the small meshes, providing information to interpolate from.  When training on multiple meshes, the total reconstruction loss is the sum of \cref{eqn:MSE_Loss} evaluated on each training tensor prediction pair ($T_{\text{true}}^s$, $T^s$), normalized by the mean squared magnitude of that tensor,
\begin{equation}
    \mathcal{L}_{\text{recon}} = \sum_s \frac{\mathcal{L}_{\text{mse}}(T_{\text{true}}^s, T_s)}{\frac{1}{\operatorname{Card}(T_s)}\sum_i |T_{\text{true},i}^s|^2},
    \label{eqn:recon_loss}
\end{equation}
so that tensors of different magnitudes contribute equally to the total loss.
To be consistent, we will also use \cref{eqn:recon_loss} even if there is only one training mesh.

Second, we define a downsampling loss for the interpolation training. For this loss, we use a bilinear interpolation to transform the predicted tensor $T$ on a larger $\bsl{k}$ mesh to the smaller $\bsl{k}$ mesh on which we have the ground truth. We call this new tensor $T_{downsampled}$. We then use \cref{eqn:MSE_Loss} to find
\begin{equation}
    \mathcal{L}_{downsample}= \mathcal{L}_{mse}(T_{\text{true}}, T_{downsampled}).
    \label{eqn:downsample_loss}
\end{equation}
This is a loss to test the interpolation power.

Third, we define a symmetry loss to encourage systems to satisfy underlying symmetries (\ie, rotational symmetry) in the interpolation training.
For this loss, we apply a transformation to our predicted tensor $T$  and calculate the MSE between the transformed tensor $T'$ and its non-transformed counterpart, giving
\begin{equation}
    \mathcal{L}_{sym} = \mathcal{L}_{mse} (T, T'). 
    \label{eqn:sym_loss}
\end{equation}
The loss will be zero if the two tensors are identical, that is when they satisfy the symmetry. This is a dimension-agnostic loss---it can be applied to any sizes mesh without knowing the ground truth.

Fourth for the interpolation training, we define a normalization loss when we wish to encourage an output to have a specific magnitude. For a scalar $s$, this loss is defined as 
\begin{equation}
    \mathcal{L}_{norm} = (||{T}||-s)^2
    \label{eqn:norm_loss}
\end{equation}
where $||\cdot ||$ represents the Frobenius norm of the tensor $T$, \ie, 
\eq{
 \left\|T\right\| = \sqrt{\sum_{i} \left|T_i\right|^2}\ .
}
 This is a dimension-agnostic loss.

We note that for the Richardson model in \cref{app:Richardson}, we only use the above 4 losses as we will only study the pair-pair correlation function with interpolation.
For the study of FCI \cref{app:FCI}, we will not use point-group symmetry loss, downsampling loss and norm losses, but we will explicitly incorporate fermionic antisymmetry and Hermiticity.
We will use the following additional losses, as we will deal with the full 2-RDM.

Regarding the NN for the full 2-RDM, we will let the NN to predict $\prescript{2}{}{D}$, and then predict $\prescript{2}{}{Q}$ and $\prescript{2}{}{G}$ using the relation in \cref{eq:D-G_relation} and \cref{eq:D-Q_relation}.
In particular, we will enforce the Hermiticity, PSD and antisymmetry of $D$, as detailed in \cref{app:FCI}.
Then, \cref{eq:D-G_relation} and \cref{eq:D-Q_relation} guarantee the Hermiticity of $\prescript{2}{}{Q}$ and $\prescript{2}{}{G}$ and antisymmetry $\prescript{2}{}{Q}$, but they cannot guarantee the PSD of $\prescript{2}{}{Q}$ and  $\prescript{2}{}{G}$.
Then, we introduce the violation of PSD $\prescript{2}{}{Q}$ and  $\prescript{2}{}{G}$ as a loss: 
\begin{equation}
    \mathcal{L}_{PSD_{Q/G}} = \frac{1}{A} \sum_{\alpha=1}^A\operatorname{ReLU}\left(-\lambda_\alpha^{Q/G}\right)^2
    \label{eq:lambda_Q_G}
\end{equation}
where $\lambda^{Q/G}_\alpha$ is the $\alpha$th eigenvalue of $\prescript{2}{}{Q}$ or $\prescript{2}{}{G}$. As these tensors must be PSD, this pushes the prediction to be more physical, and this is a dimension-agnostic loss. 
This loss is used in both the interpolation training and the variational optimization.

When predicting the full 2-RDM, one can immediately predict the many-body energy $E_{\text{pred}}$, as we only consider two-body interaction in this work.
Then, we may include the energy in the loss function.
For the interpolation training, we know the ground-state energy $E_{ED}$ at small sizes, and thus we can define a loss as the distance between this energy and our predicted energy $E_{\text{pred}}$ as
\begin{equation}
    \mathcal{L}_{E_{dist}} = (E_{ED}-E_{\text{pred}})^2.
    \label{eqn:energy_dist}
\end{equation}
For the variational optimization on large sizes,  we do not know the ground-state energy in general, and thus we directly treat the predicted $E_{\text{pred}}$ as the loss to minimize:
\begin{equation}
    \mathcal{L}_{E_{min}} = E_{\text{pred}} \ .
    \label{eqn:energy_min}
\end{equation}

\subsection{Testing Criteria}

To evaluate the effectiveness of the architectures, two different quantities were used to benchmark the accuracy prediction. We use (i) R-value accuracy and (ii) normalized overlap, as scores defining the NN accuracy at predicting the large-size $n$-RDM.
For the NNs that predict the complete 2-RDMs, the PSD of predicted $\prescript{2}{}{Q}$ and $\prescript{2}{}{G}$ in \cref{eq:lambda_Q_G} and the many-body energy are also quantities that can benchmark the accuracy, which we will not repeat here.
In the following, we discuss the R-value accuracy and normalized overlap.

\subsubsection{R-value Accuracy}
The R-value accuracy is previously defined   \cite{Azam2025mlRDM}. We define a matrix $T_{\text{true}}$ as our true data, and $T$ as our predicted data. 
We then have a mean square error (MSE)  as defined in \cref{eqn:MSE_Loss}.
With the mean square error, we then define an accuracy 
\begin{equation}
    R = 1-\frac{ \sqrt{MSE}}{\max(\Re T_{\text{true}},\Im T_{\text{true}})-\min(\Re T_{\text{true}},\Im T_{\text{true}})}
\end{equation}
where $\Re$ and $\Im$ refer to extracting the real and imaginary components of the tensor. 
This gives us 
\begin{equation}
    R\utext{value} = \max(0,R).
    \label{eqn:r_value}
\end{equation}
This metric equals $1$ for identical matrices and approaches $0$ as the reconstruction error reaches or exceeds the span of the data. Because of the normalization, the error is dimensionless and directly interpretable as a score. Additionally, unlike pure MSE this normalization allows for comparison across datasets of different magnitude.

While generally useful, this metric can be misleading for sparse data. Suppose a $100\times100$ zero matrix, with 4 entries of value $1$ and all other entries $0$. If we then use the zero matrix as the prediction $T$, we will have
\begin{equation}
\text{R} = 4\cdot10^{-4}
\end{equation}
and
\begin{equation}
     R=0.98
\end{equation}
giving an accuracy $R_{\text{value}}=0.98$. Despite this accuracy, the model has not predicted any nonzero structure. 

We  note that, as the normalization depends on $\max(\Re T_{\text{true}},\Im T_{\text{true}})$ and $\min(\Re T_{\text{true}}, \Im T_{\text{true}})$, the accuracy is sensitive to outliers. Large outliers can cause the accuracy to appear far better than reality, inflating the score.
Owing to these caveats, it is beneficial to develop other tools to cross check with the R-value.

\subsubsection{Normalized Overlap}
To cross check with the R-value, we here adopt the normalized overlap.
This overlap focuses on the relative structure of the prediction and true matrix, based on the cosine of the angle between the two matrices.  Using $T$ as our true matrix and $T$ as our prediction, the normalized overlap is defined as
\begin{equation}R_{\text{overlap}} = \frac{|\langle T_{\text{true}}, T\rangle|}{\max(\lVert T_{\text{true}}\rVert^2,\lVert T\rVert^2)}
\label{eqn:norm_overlap}
\end{equation}
where $\langle\cdot ,\cdot \rangle$ is the complex dot product and $||T||= \sqrt{\langle T, T \rangle }$ . The value will be between $0$ and $1$, with a value of $1$ when the two matrices are identical and $0$ occurring when they are perpendicular.

As opposed to the R-value, the normalized overlap is $0$ for any $0$ vector prediction. Additionally, it also has an interpretable score that remains consistent across datasets of different magnitudes, and provides an understanding of how well the NN is predicting the overall structure. $R_{\text{overlap}}$ therefore provides a secondary score to compare with the R-value.

\subsection{Parameter Efficiency}

To benchmark the NN, we use the number of parameters needed to achieve a certain accuracy as a benchmark of the efficiency, leading to the parameter-efficiency plots.
We benchmark the parameter efficiency only for the interpolation tasks, evaluating both the R-value and normalized overlap as a function of trainable parameters. To generate these plots, we conduct a
comprehensive sweep over the architectures and hyperparameter configurations listed in \cref{app:Architectures}. For each combination, we train the model for a fixed number of epochs, recording training loss throughout the whole training and creating model checkpoints. Upon completion of training, we predict the RDM using the checkpointed model with the lowest training loss throughout, and compute the accuracy metric.

From the set of all trained configurations, we produce a scatterplot of parameter count versus accuracy. To extract a performance frontier, we partition the parameter axis into 15 bins of equal width in $\log p$ space, where $p$ is the number of parameters. Within each bin $i\in 0,1,\dots,15$, where a higher $i$ denotes more parameters, we select the trial achieving the highest test R-value, giving us a set of frontier points $(p_i, s_i)$ where $p_i$ is the parameters of this trial and $s_i$ is the best test score, both within the bin. We then enforce monotonicity along the parameter axis by setting  $s_j'=\max(s_j, \{s_i\}_{i<j})$ so that $s_j'$ is the best achievable performance using at most $p_j$ parameters. 

These parameter-efficiency curves allow us to determine the model capacity required to accurately reconstruct large-size mesh $n$-RDMs from small-size mesh inputs. We are also able to quickly determine which NN classes are best at this form of machine learning, and compare different losses' effect on training. 

\section{Richardson Model Training}
\label{app:Richardson} 

In this section, we train NNs to predict the pair-pair correlation function of the ground states of the Richardson model. 
The models are trained on small sizes and thus this is the interpolation way of using NNs.
This is an exactly solvable model, allowing us to verify that our different NNs are able to interpolate in this simple case before attempting on a more complex material.

\subsection{Model Setup}
\label{sec:Richardson Model Setup}

The Richardson model of superconductivity~\cite{Richardson1963PairPairCorr,Richardson1964Eigenstates,Dukelsky2004Richardson} has the Hamiltonian
\begin{equation}
H = \sum_{\bsl{k}, s} \eps_{\bsl{k}}c^\dagger_{\bsl{k},s}c_{\bsl{k},s}+\frac{u}{L^2}\left(\sum_{\bsl{k}}A^\dagger_{\bsl{k}}\right)\left(\sum_{\bsl{k'}}A_{\bsl{k'}}\right).
\label{eqn:Richardson}
\end{equation}
where  $c^\dagger_{\bsl{k}s}$ creates an electron at Bloch momnetum $\bsl{k}$ and spin $s$, $A^\dagger_{\bsl{k}} = c^\dagger_{\bsl{k},\uparrow}c^\dagger_{-\bsl{k},\downarrow}$ is the Cooper-pair operator, $u <0$ denotes attractive interaction, $\eps_{\bsl{k}} = t(\cos k_x + \cos k_y) $, and $\bsl{k}$ takes value from an $L_1\times L_2$ mesh. 
We particularly focus on the pair-pair correlation function
\begin{equation}
    C_{true,\bsl{kk}'} \equiv \langle A^\dagger _{\bsl{k}} A_{\bsl{k'}} \rangle-\langle A^\dagger _{\bsl{k}}\rangle\langle A_{\bsl{k'} }\rangle .
    \label{eqn:pairpair}
\end{equation}
where $\langle \dots \rangle$ is the average with respect to the ground state in the total-spin-zero sector.
We will use $C_{\bsl{k}\bsl{k}'}$ as the predicted one.
Due to the spinless time-reversal and inversion symmetry, we have the pair-pair correlation function of the Richardson Model $C\in \mathbb{R}^{N_k \times N_k}$. 
The pair-pair correlation function is a special set of components of the 2-RDM, as $\langle A^\dagger _{\bsl{k}}\rangle=0$ for the particle-number-conserved states considered here.
The Hermiticity and PSD condition of $\prescript{2}{}{D}$ then are reduced to  $C_{\bsl{kk}'} $ being a symmetric PSD real matrix, and we will not consider other $N$-representability constraints.
Same as \refcite{Azam2025mlRDM}, we use a fixed electron filling of $\nu = N/N_k=1/6$.

\subsection{Training Data and Preprocessing}
\label{sec:Richardson_Loss}
\begin{figure}[t]
    \centering
    \includegraphics[width=0.6\linewidth]{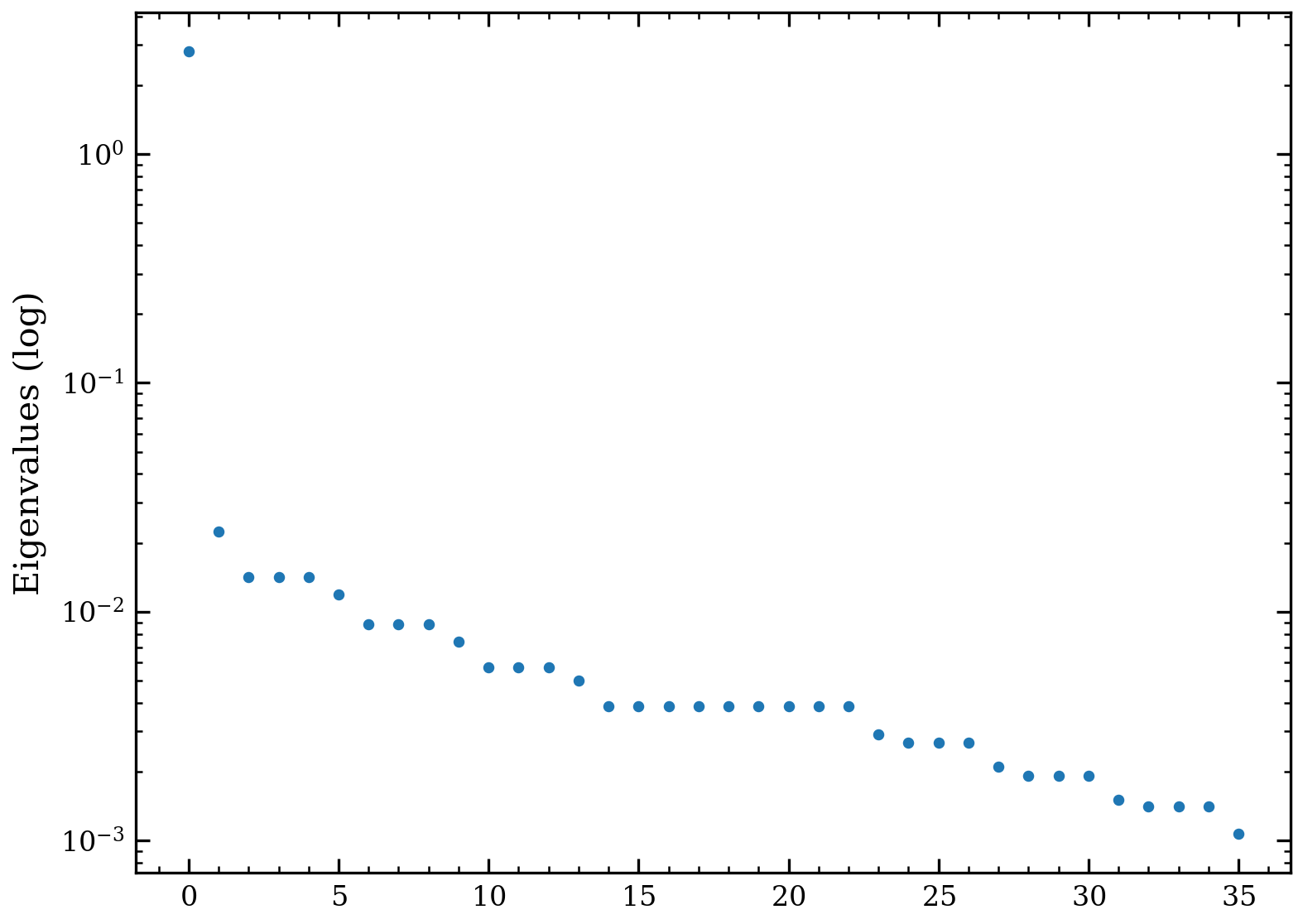}
    
    \caption{Eigenvalue of the pair-pair correlation function vs. index for the training data on the conventional $6\times 6$ mesh. The first eigenvalue is two orders of magnitude larger than the others, and all other eigenvalues have similar relative magnitude.
    }
    \label{fig:Richardson_eigenvalues}
\end{figure}

For the Richardson Model, we train the NN on both a conventional $6\times 6$ mesh and 4 tilted meshes each of which has 12 $\bsl{k}$ points.
We then predict on an conventional $18\times 18$ mesh.
We will {provide a general discuss on both ways}, and then describe the differences. 
Both methods train on the dominant component. 
Specifically, in \cref{fig:Richardson_eigenvalues} we see the ordered list of eigenvalues of the pair-pair correlation function for the training data and note that only one eigenvalue dominates. 
Because of this, we may express an approximation of the full pair-pair correlation function as
\begin{equation}
    C = \lambda  \tilde v \tilde v ^\dagger\ , 
    \label{eqn:eigenvalue_eqn}
\end{equation}
where $\lambda$ is the dominant eigenvalue and $\tilde v$ is the dominant normalized eigenvector.
To demonstrate the effectiveness of this calculation, we consider primary eigenvalue-eigenvector pair of the true pair-pair correlation function on the $6 \times 6$ $\bsl{k}$ point mesh, that is the pair with the highest eigenvalue. When using this 1-eigenvalue construction as a prediction on the true data could then have a normalized overlap of $0.99997$ and an R-value of $0.996$.

In order to find an appropriate eigenvalue, we note that the trace of the  physical pair-pair correlation function is always 
\begin{equation}
    \Tr(C) = \frac{N_k}{12}\ ,
\end{equation}
where the 1/12 comes from half the electron filling (which represents the Cooper pair filling).
In the training, we fix the value of $\lambda$ to be the largest eigenvalue from the training data, and in the prediction, we will scale it by
\begin{equation}
    \lambda_{pred} = \lambda_{train}\frac{N_k^{prediction}}{N_k^{training}}\ .
    \label{eqn:lambda_pred}
\end{equation}
Note that $N_k^{prediction}$ is the input for the prediction, as it is just the number of $\bsl{k}$ points in the mesh on which the prediction is performed.
So we are not using any large-size test data to train the model.
On the other hand, $\tilde{v}$ in \cref{eqn:eigenvalue_eqn} is what the NN will learn to predict.

To train the NN for $\tilde{v}$, we first compute the eigen decomposition of the training data, which gives the dominating eigenvector $v$, which we take as $\| v\|=1$.
After that, we preprocess the dominant eigenvector as a NN input. We first normalize the coordinates used for predicting the NN output, mapping each $\bsl{k}$ to a point in $[-1,1]^2$ and applying periodic boundary conditions; this transforms the eigenvector's shape to $(L_1+1)\times (L_2+1)$, and we refer to the periodic eigenvector as $v_w$. We then 
apply the element-wise standardization
\begin{equation}
    v_w' = \frac{v_w-\mu}{\sigma}
\end{equation}
where $v'_w$ is the modified vector and $\mu$ and $\sigma$ are the mean and standard deviation of $v_w$.
This process is reversible, with a de-scaling given by
\begin{equation}
    v_w = v_w' \sigma + \mu.
    \label{eqn:descaling}
\end{equation} 

When using the NN to predict our pair-pair correlation function we apply this descaling \cref{eqn:descaling} to the NN output $\tilde v_w'$, which has the same standardization as $v_w'$,  using the same $\sigma$ and $\mu$ of the dominant training eigenvector. We then enforce the four-fold rotational symmetry and Hermiticity on this output by computing and averaging each symmetry transformation, cropping the periodic boundary to recover  $\tilde v$, and enforcing  $\| \tilde v \| =1$ before reconstructing the pair-pair correlation function with \cref{eqn:eigenvalue_eqn}.

During training, we incorporate four different losses: a small data loss, a downsampling loss, a symmetry loss, and a unit-norm loss as defined in \cref{app:losses}.
For these losses, we consider the vector $ \tilde v_s$  predicted by our NN over the small $\bsl{k}$ and $\tilde v_l$ predicted over the large $\bsl{k}$ mesh. For the small data loss, we use \cref{eqn:MSE_Loss} to calculate the MSE between $\tilde v_s$ and $v$. For the downsampling loss, we use \cref{eqn:downsample_loss}, predicting the large-size tensor and downsampling to compare to the small-size tensor.  For the symmetry loss, \cref{eqn:sym_loss}, we use the four-fold rotational symmetry and Hermiticity (which is simplifed to transpose due to $C$ being real). This loss is applied after $20$ epochs, ensuring they don't overpower the loss of the NN.  For the Norm loss defined in \cref{eqn:norm_loss} we take $s=1$, ensuring the vector is normalized.

Hence, the total loss is 
\begin{equation}
    \mathcal{L} = \alpha\utext{small}\mathcal{L}\utext{small}  + \alpha\utext{interp}\mathcal{L}\utext{downsample} +\alpha_{\text{sym}}\mathcal{L_{\text{sym}}}  + \alpha\utext{norm}\mathcal{L}\utext{norm}
    \label{eqn:Richardson_loss}
\end{equation}
where the weightings of these losses can be seen in  \cref{tab:richardson_lambda}.

\begin{table}[htb]
    \centering
    \setlength{\tabcolsep}{8pt} 
    \renewcommand{\arraystretch}{1.20}  
    \begin{tabular}{c|c}
         \toprule
         \textbf{Weight} & \textbf{Value} \\
         \midrule
            $\alpha{\utext{small}}$ & $1$\\ 
            $\alpha_{\text{interp}}$ & $1$\\ 
            $\alpha_{\text{sym}}$ & $10^{-2}$ \\ 
            $\alpha_{\text{norm}}$ & $10^{-3}$ \\ 
         \bottomrule
    \end{tabular}
    \caption{List of loss weights for the training on the pair-pair correlation function of the Richardson Model as used in \cref{eqn:Richardson_loss}.}
    \label{tab:richardson_lambda}
\end{table}

We use the Adam optimizer \cite{Kingma2017Adam}  and train all models for 1000 epochs and use a learning rate of $10^{-4}$.

The above method is what we use for the training on the conventional $6\times 6$ mesh.
For the training on the 4 $12$-$\bsl{k}$ meshes, we note that the fact that there is only one Cooper pair for each mesh,  we always choose $\lambda_{training} = 1$ for all meshes.

\begin{figure}
    \centering
    \includegraphics[width=0.7\linewidth]{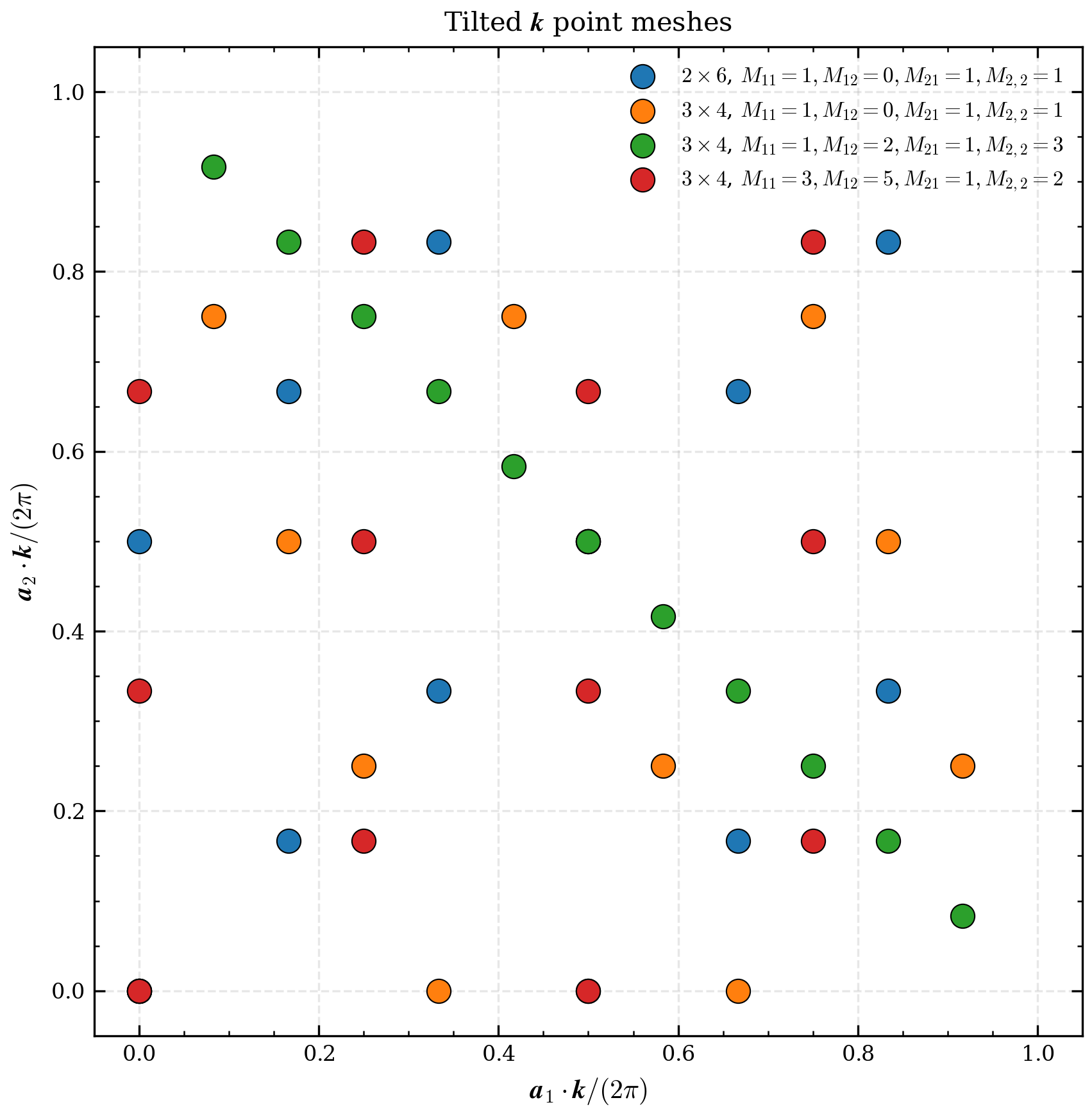}
    \caption{Visualization of the sampling of the \BZ for the tilted mesh training of the pair-pair correlation function. The convention used here is the same as that described by \cref{eqn:tilted_convention}.
    $\bsl{a}_i$ is the primitive lattice vector with $\bsl{a}_i\cdot\bsl{b}_j = 2\pi \delta_{ij}$.
    }
    \label{fig:k_points_tilted_mesh}
\end{figure}

For the dominant eigenvectors of all meshes, we combine them by merging the 4 momentum meshes into a single set of $\bsl{k}$ points, with each point in each mesh carrying the corresponding component of the eigenvector in its original mesh.
A visualization of the chosen $\bsl{k}$ points can be seen in \cref{fig:k_points_tilted_mesh}.
Because the tilted mesh training consists of scattered $\bsl{k}$ rather than a regular grid, the  interpolation loss $\mathcal{L}_{interp}$ and the unit-norm loss $\mathcal{L}_{norm}$ are not applicable.
Everything else about the training is in the same way as that on the conventional $6\times 6$ mesh.

\subsection{ML Results}

In this section, we evaluate the performance of our NN architectures for training on both a $6\times 6$ $\bsl{k}$ point mesh and on 4 tilted $12-\bsl{k}$ meshes of the Richardson model's pair-pair correlation function. The prediction is then evaluated on the $18\times 18$ mesh.

\subsubsection{Trained on $6\times 6$ Mesh}
\begin{figure}[t]
    \centering
    \includegraphics[width=0.8\linewidth]{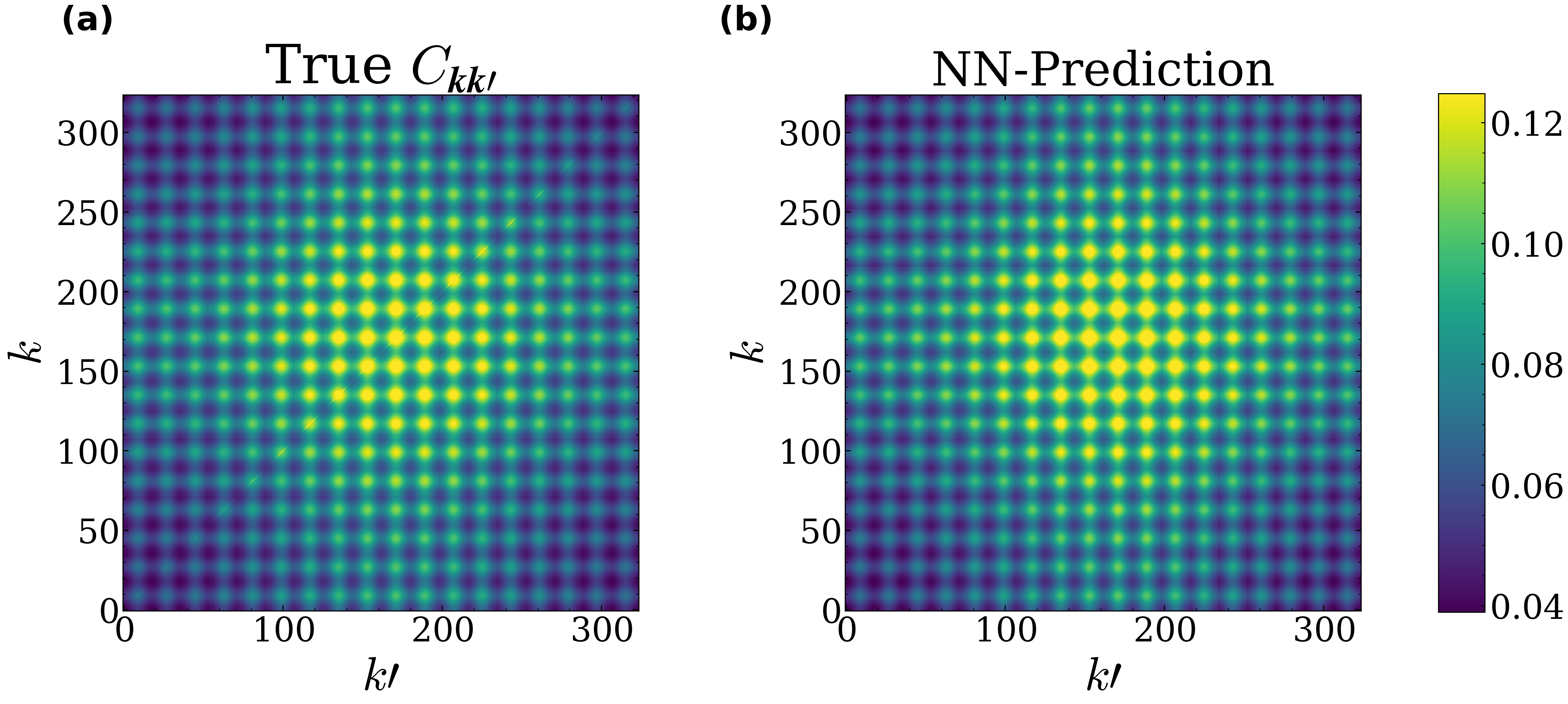}
    \vspace{0.5em}

    \caption{True pair-pair correlation function (left) and predicted pair-pair correlation function (right) by a KAN for the Richardson model, \cref{eqn:Richardson},  on $18\times 18$ mesh . Here $k$ and $k'$ refer to the linearized momentum indices of the pair-pair correlation function, \cref{eqn:pairpair}, linearized with \cref{eqn:linearlize}.
    The KAN is trained on $6\times 6$ mesh and the prediction has an R-value of $0.981607$ and a normalized overlap of $0.999928$.  
    }
    \label{fig:Richardson_general_best_pred}
\end{figure}

When trained on the $6\times 6$ mesh, each architecture is able to reach a similar level of performance, though the number of needed parameters is different, as shown in  \cref{tab:Richardson_performance}. This table provides the best R-value of each architecture over our parameter sweep. We find that KAN achieves the best R-value of $0.98572$ and a normalized overlap of $0.97028$, while the TBN is the worst NN with an R-value of $0.98371$ and a normalized overlap of $0.97013$. Nevertheless, the values are close, and do not provide much information about the true differences between these models. We also examine the best prediction, in \cref{fig:Richardson_general_best_pred}.  This provides the true data and the prediction. The NN does a very good job of predicting the original data, only missing a perturbation along the diagonal of the original data.

\begin{table}[htb]
    \centering
    \setlength{\tabcolsep}{8pt}
    \renewcommand{\arraystretch}{1.20}
    \begin{tabular}{c|c|c|c}
         \toprule
         \textbf{Architecture} & $\bsl{R_{\text{value}}}$ & $\bsl{R_{\text{overlap}}}$  & \textbf{Number of Parameters} \\
         \midrule
            MLP & 0.98555 & 0.97027 & 66,817 \\
            Residual MLP & 0.98491 & 0.97022 & 11,585 \\
            KAN & 0.98572 & 0.97028 & 791,040 \\
            SIREN & 0.98571 & 0.97028 & 1,185 \\
            FINER & 0.98553 & 0.97027 & 4,417 \\
            TBN & 0.98371 & 0.97013 & 1,588,737 \\
         \bottomrule
    \end{tabular}
    \caption{List of best performance for each model trained on the pair-pair correlation function of the Richardson model, \cref{eqn:Richardson}, on the $6\times 6$ mesh. The best performance is picked based on the R-value, and we provide the corresponding normalized overlap.}
    \label{tab:Richardson_performance}
\end{table}

\begin{figure}[t]
    \centering
    \includegraphics[width=0.8\linewidth]{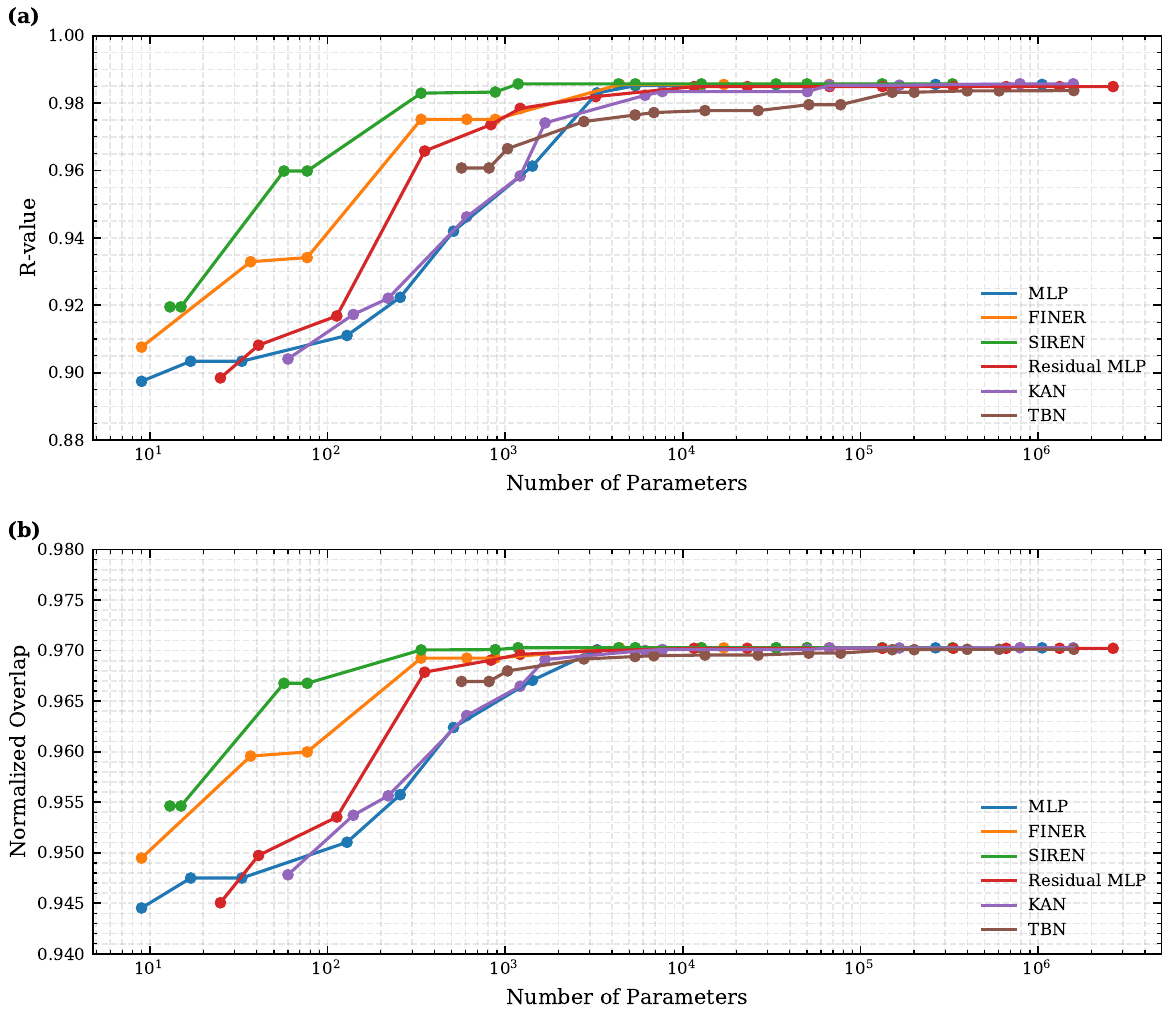}

    \caption{R-value (a) and normalized overlap (b) versus number of parameters when training on the $6\times 6$  point mesh and predicting on the $18\times 18$ $\bsl{k}$ point mesh.
    }
    \label{fig:richardson_parameter_efficiency}
\end{figure}

We can also examine the parameter efficiency, as seen in \cref{fig:richardson_parameter_efficiency}. These plots show us how the NNs adapt in accuracy with the number of parameters. Both the SIREN and the FINER, which are designed for INR, outperform all models at lower parameters, with SIREN performing the best. The performance continues until around $1000$ total parameters, when the accuracy saturates and all models reach a similar level of performance. 

For the KAN, we found the optimal model with $791,040$ parameters, a depth of $3$, a hidden dimension of $256$ a grid size of 1, a base scale of 1, and a spline scale of 5. For the MLP, we found the optimal model with $33,537$ parameters, a depth of $3$ and a hidden dimension of $128$. For the Residual MLP, we found the optimal model with $1,982,465$ parameters, a depth of $40$, and a hidden dimension of $256$. For the SIREN, we found the optimal model with $3,297$ parameters, a depth of $3$, a hidden dimension of $32$, a $\omega_0$ of 6, and an $\omega_1$ of $5$.  For the FINER, we found the optimal model with $12,737$ parameters, a depth of $3$, a hidden dimension of $64$, and both $\omega_0$ and $\omega_1$ of $1$. Finally, for the TBN, we found an optimal model with 339,105 parameters, a depth of 2, a hidden dimension of $128$, no dropout, 2 heads, and 4 frequencies. 

Finally, as our current data is dominated by a single eigenvalue, we evaluate the performance of just a simple Fourier representation of the data. When computing the fast Fourier Transform (FFT) on the small-size $6\times 6$ mesh, applying a zero-padding to the frequency domain to provide the same dimension of the large-size data, and using an inverse FFT to reconstruct the data, we find an R-value of $0.98379$ and a normalized overlap of $0.970056$. Hence, most of our results can be closely approximated by simple techniques. However, we were able to achieve scores slightly higher than this result, showing the promise of our method.

\subsubsection{Trained on Tilted Meshes}
\label{sec:richard_tilted_mesh_results}

\begin{figure}[t]
    \centering
    \includegraphics[width=0.8\linewidth]{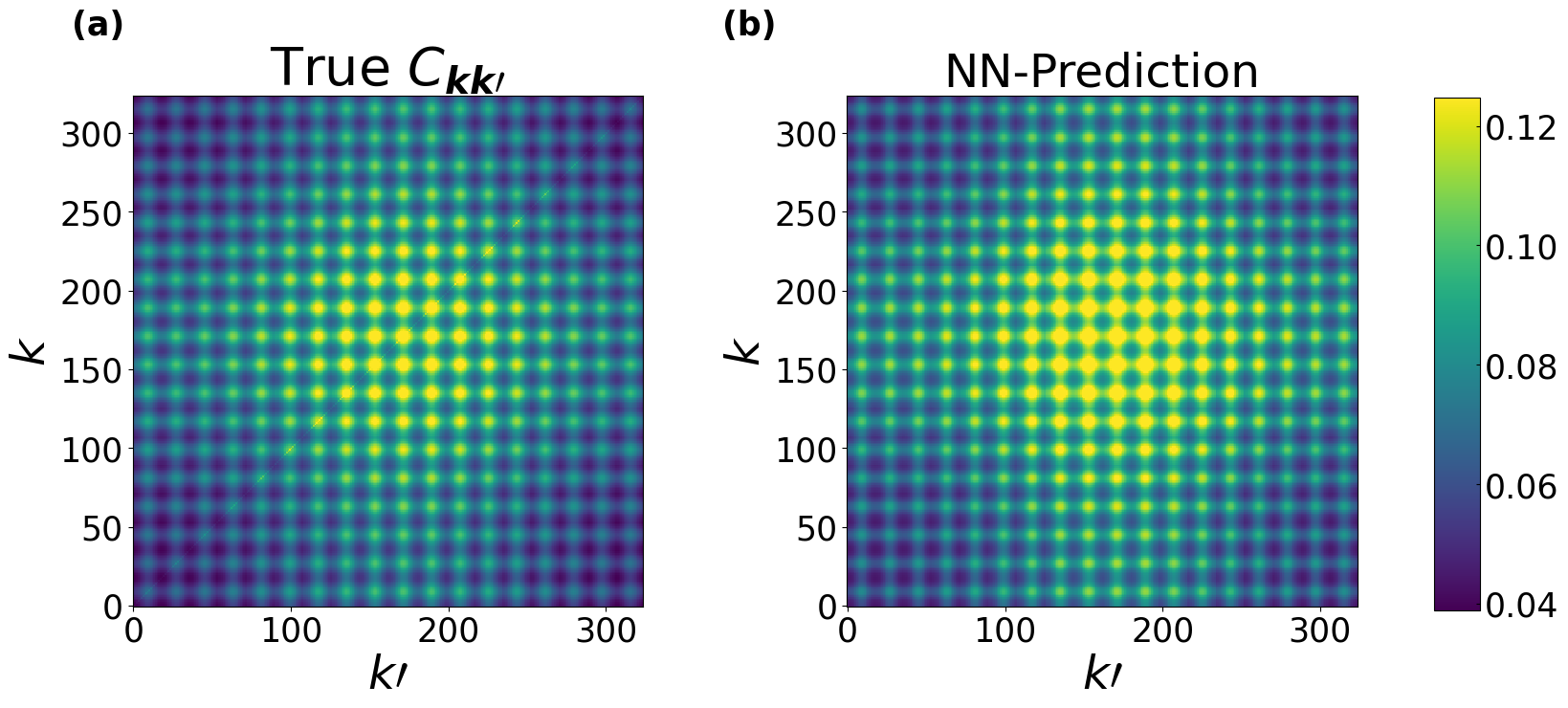}    
    
    \caption{True  pair-pair correlation function (left) and predicted pair-pair correlation function (right) by a KAN for the Richardson model, \cref{eqn:Richardson} trained on 4 tilted meshes with 12 $\bsl{k}$ points each. Here $k$ and $k'$ refer to the linearized momentum indices of $\bsl{k}$ and $\bsl{k}'$ of the pair-pair correlation function, \cref{eqn:pairpair}, in the same way as \cref{fig:Richardson_general_best_pred}. The KAN is trained on the tilted meshes and the prediction has an R-value of $0.942494$ and a normalized overlap of $0.907553$. The poorer scores are  related to an improperly chosen eigenvalue.  }
    \label{fig:Richardson_tilted_best_pred}
\end{figure}

When trained on tilted meshes (each of which has 12 momenta), each architecture is again able to reach a similar level of performance, as shown in \cref{tab:Richardson_performance_tilted}. This table provides the best R-value of each architecture over our parameter sweep. We find that KAN achieves the best R-value of $0.95427$ and a normalized overlap of $0.907805$, while the FINER  is the worst NN with an R-value of $0.95408$ and a normalized overlap of $0.907553$. Again, the values are extremely close, and do not provide much information about the true differences between these models. We may also  examine a plot of the  best prediction in \cref{fig:Richardson_tilted_best_pred}. Here, we see a slightly less accurate prediction, as $\lambda_{training}=1$ is fixed for the tilted meshes. With a properly chosen eigenvalue, that is the actual eigenvalue found during eigendecomposition of the large-size data, we have a potential R-value as high as $0.987520$. We note that, when doing any future interpolation based on training the eigendecomposition, special care must be taken in finding eigenvalues for reconstruction. 

\begin{table}[htb]
    \centering
    \setlength{\tabcolsep}{8pt} 
    \renewcommand{\arraystretch}{1.20}  
    \begin{tabular}{c|c|c|c}
         \toprule
         \textbf{Architecture} & {$\bsl{R_{\text{value}}}$} &  $\bsl{R_{\text{overlap}}}$ & \textbf{Number of Parameters} \\ 
         \midrule
            MLP & 0.954651 & 0.907665 & 17025 \\
            Residual MLP & 0.954272 & 0.907589 &3233 \\
            KAN  & 0.955362 &0.907805  &  7680\\
            SIREN  & 0.954633 & 0.907662 & 337 \\
            FINER & 0.954085 & 0.907553 & 337 \\
            TBN  & 0.954785 & 0.907692 & 301185 \\
         \bottomrule
    \end{tabular}
    \caption{List of best performance for each model trained on the pair-pair correlation function of the Richardson model, \cref{eqn:Richardson}, on 4 tilted mesh with 12 $\bsl{k}$ points each. The best performance is picked based on the R-value, and we provide the corresponding normalized overlap and number of parameters.}
    \label{tab:Richardson_performance_tilted}
\end{table}

\begin{figure}[t]
    \centering
    \includegraphics[width=0.8\linewidth]{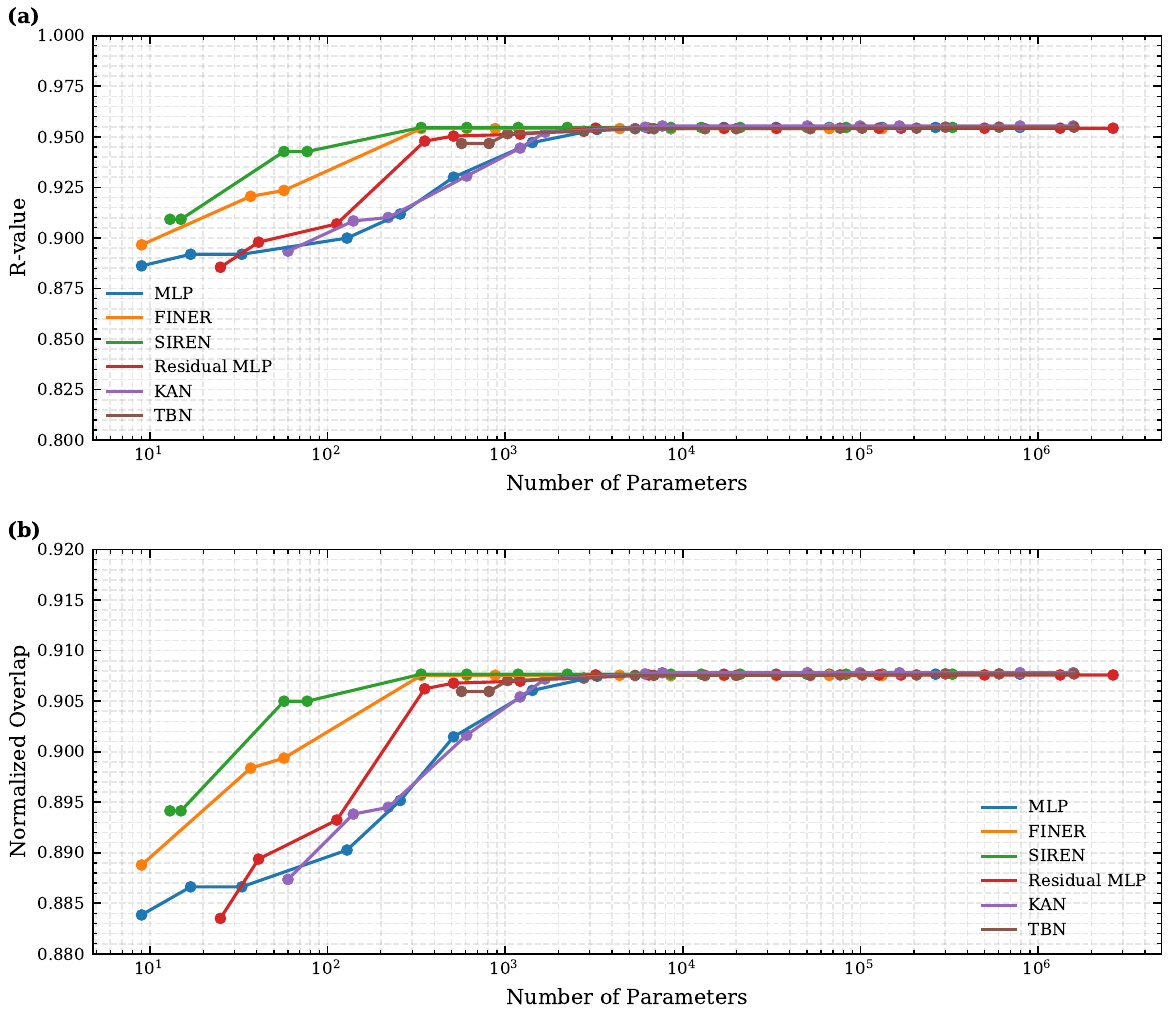}
    
    \caption{R-value (a) and {normalized overlap} (b) versus number of parameters for the training on 4 tilted $\bsl{k}$ point meshes of the pair-pair correlation function of the Richardson model, \cref{eqn:Richardson}, where each of the tilted meshes has $12$ $\bsl{k}$ points.
    }
    \label{fig:richardson_parameter_efficiency_tilted}
\end{figure}

We also examine the parameter efficiency, as seen in \cref{fig:richardson_parameter_efficiency_tilted}. These plots show us how the architectures
change with number of parameters,  and we can see a leveling off around $1000$ parameters, as before. The performance of each model is analogous to those trained on the $6\times 6$ mesh. 

For the KAN, we found the optimal model with 7,680 parameters, a depth of 1, a hidden dimension of 256, a grid size of 5, a base scale of 1, and a spline scale of 5. For the MLP, we found the optimal model with 17,025 parameters, a depth of 2, and a hidden dimension of 128. For the Residual MLP, we found the optimal model with $23,105$ parameters, a depth of $20$, and a hidden dimension of 8. For the SIREN, we found the optimal model with 337 parameters, a depth of 1, a hidden dimension of 16, a first $\omega_0$ of $12$, and a hidden $\omega_1$ of 5. For the FINER we found the optimal model with 337 parameters, a depth of 1, a hidden dimension of 16, a first $\omega_0$ of 1.0, and a hidden $\omega_1$ of $0.1$. Finally, for the TBN, we found the optimal model with 301,185 parameters, a depth of 6, a hidden dimension of 64, 1 head, 4 frequencies, and no dropout.

Finally, as our current data is dominated by a single eigenvalue, we evaluate the performance of just a simple Fourier representation of the data. Unlike the previous method, we must first modify the data in order for FFT upscaling to work. First, the dominant eigenvector is computed exactly from each training file's pair-pair correlation function, then its values at the $\bsl{k}$ points are interpolated onto a uniform $12\times 12$ $\bsl{k}$ point grid using cubic interpolation. Cubic interpolation fits piecewise cubic polynomials between data points and matching first derivatives, creating a smooth transition between the points. We then do the same upscaling as before, going to our $18\times 18$ $\bsl{k}$ point mesh with the FFT zero-padding. The pair-pair correlation function is then reconstructed using $\frac{18^2}{27} {v}{v}^T$. With this method, we find a normalized overlap of $0.907113$ and an R-value of $0.51947$ providing a non-learned baseline. We note that the additional cubic interpolation step required by irregular sampling introduces an extra approximation not present in the $6\times 6$ mesh baseline. The neural networks are able to exceed this baseline, demonstrating they provide some useful structure beyond simple Fourier methods.

\section{Twisted Bilayer MoTe$_2$}
\label{app:FCI} 

In this section, we examine how different NNs are able to interpolate the 2-RDM of twisted bilayer MoTe$_2$, and further examine their ability to be variationally minimized and predict the ground-state 2-RDMs and energy of the system. 

\subsection{Model Formulation}

We first start with a brief review of the $\K$-valley model for $t$MoTe$_2$~\cite{wu_topological_2019} that we use in this work.
The non-interacting part of the model has a general form 
\eq{
H_{\K,0}^h = - \sum_{M_x, M_y \in \dsN} \sum_{l,l'=t,b}\int d^2 r \left(\ii^{M_x+M_y} \partial_x^{M_x} \partial_y^{M_y} \widetilde{c}^\dagger_{\K,l,\bsl{r}} \right) t^{M_x M_y}_{ll'}(\bsl{r})\widetilde{c}_{\K,l',\bsl{r}} 
}
where $\widetilde{c}_{\K, l,\bsl{r}}^\dagger$ is the creation operator for a hole in the $\K$ valley  in the $l$th layer at position $\bsl{r}$, and $\dsN$ denotes the set of non-negative integers.
The detailed form of $t^{M_x M_y}_{ll'}$ is \refcite{Zhang2024UniversalMoireModel}, and we use the parameter values for twist angle $3.89^\circ$.
Those parameters are derived from density functional theory (DFT) calculations without any continuous parameter fitting.

For the interaction, we adopt
\eq{
H_{\K,int}^h = \frac{1}{2}\sum_{l l' } \int d^2 r d^2 r' V(\bsl{r}-\bsl{r}')  \widetilde{c}^\dagger_{\K, l, \bsl{r}} \widetilde{c}^\dagger_{\K, l', \bsl{r}'} \widetilde{c}_{\K, l', \bsl{r}'} \widetilde{c}_{\K, l, \bsl{r}}  \ ,
}
which is the double-gated screened Coulomb potential $V(\bsl{r})$ with gate distance $\xi$:
\eq{
V(\bsl{r}) =\int_{\dsR^2} \frac{d^2 p }{ (2\pi)^2} V(\bsl{p}) e^{\ii \bsl{p}\cdot\bsl{r}} \text{ with } V(\bsl{p}) =   \frac{e^2 }{4 \epsilon_r \epsilon_0 } \frac{\tanh(\xi |\bsl{p}|/2)}{|\bsl{p}|/2}\ ,
}
where $\epsilon_0$ is the vacuum permittivity, and $\epsilon_r$ is the relative dielectric constant.

We then diagonalize the noninteracting part of the Hamiltonian in the momentum space, project continuum many-body Hamiltonian to the subspace of the lowest hole band, and obtain 
\eqa{
    H &= \sum_{\bsl{k}} c_{\bsl{k}}^\dagger c_{\bsl{k}} \epsilon_{\bsl{k}}  + \sum_{\bsl{k}, \bsl{k}', \bsl{q} } V(\bsl{k}, \bsl{k}', \bsl{q}) c_{-\bsl{k}+\bsl{q}}^\dagger c_{\bsl{k}}^\dagger c_{\bsl{k}'} c_{-\bsl{k}'+\bsl{q}}\ , \label{eq:ED_hamiltonian_app}
}
where
\eq{
c_{\bsl{k}}^{\dagger} = \sum_{\bsl{Q}} \widetilde{c}_{\K, \bsl{k},\bsl{Q}}^{\dagger} U_{\bsl{Q}}(\bsl{k})
}
with $U(\bsl{k})$ the eigenvector of the lowest hole band with eigenvalue $\epsilon_{\bsl{k}}$ at the Bloch momentum $\bsl{k}$ and $\widetilde{c}_{\K, \bsl{k},\bsl{Q}}^\dagger$ the Fourier transformation of $\widetilde{c}^\dagger_{\K,l,\bsl{r}}$. 
We choose the shortest 60 $\bsl{Q}$ vectors for the projection, $\xi = 20\,\mathrm{nm}$ and $\epsilon_r = 10$.
All the $U(\bsl{k})$ on all the meshes are selected from the $U(\bsl{k})$ on a conventional $60\times 60$ mesh in the parallel transport gauge to make sure the continuity inside the \BZ, though the continuity cannot be across the \BZ boundary due to the obstruction of the Chern number.
(See details in \refcite{Zhang2024UniversalMoireModel}.)

\subsection{Parametrization of RDMs}
In the case of the projected $t$MoTe$2$, the physical 2-RDMs reads
\eqa{
& \prescript{2}{}{D}_{\bsl{k}_1\bsl{k}_2\bsl{k}_3\bsl{k}_4}^{\rm true} =  \left\langle c_{\bsl{k}_1}^\dagger c_{\bsl{k}_2}^\dagger c_{\bsl{k}_3} c_{\bsl{k}_4} \right\rangle \\
& \prescript{2}{}{Q}_{\bsl{k}_1\bsl{k}_2\bsl{k}_3\bsl{k}_4}^{\rm true}  = \left\langle c_{\bsl{k}_1} c_{\bsl{k}_2} c_{\bsl{k}_3}^\dagger c_{\bsl{k}_4}^\dagger \right\rangle  \\
& \prescript{2}{}{G}_{\bsl{k}_1\bsl{k}_2\bsl{k}_3\bsl{k}_4}^{\rm true}  =  \left\langle c_{\bsl{k}_1}^\dagger c_{\bsl{k}_2} c_{\bsl{k}_3}^\dagger c_{\bsl{k}_4} \right\rangle\ .}

We always consider the 2-RDMs that are generated by one ground state with a definite many-body momentum.
Therefore, the physical and the predicted 2-RDMs must respect the momentum conservation: 
\eqa{
& \prescript{2}{}{D}_{\bsl{k}_1\bsl{k}_2\bsl{k}_3\bsl{k}_4} = 0 \text{ if } \bsl{k}_1+\bsl{k}_2 \neq \bsl{k}_3+\bsl{k}_4\ \mod\ \bsl{G}_M \\
& \prescript{2}{}{Q}_{\bsl{k}_1\bsl{k}_2\bsl{k}_3\bsl{k}_4} = 0 \text{ if } 
\bsl{k}_1+\bsl{k}_2 \neq \bsl{k}_3+\bsl{k}_4\ \mod\ \bsl{G}_M\\
& \prescript{2}{}{G}_{\bsl{k}_1\bsl{k}_2\bsl{k}_3\bsl{k}_4} = 0 \text{ if } 
\bsl{k}_1+\bsl{k}_3=\bsl{k}_2+\bsl{k}_4\ \mod\ \bsl{G}_M\ ,
}
where $\bsl{G}_M$ is the moiré reciprocal lattice vector.
With this knowledge, we are able to include momentum conservation to reduce the number of parameters per model from $\sim N_k^4$ to $\sim N_k^3$, improving the computational feasibility of the training. 
Explicitly, the elements that are allowed to be nonzero are simply
\eqa{
     & \prescript{2}{}{\tilde{D}}_{\bsl{k},\bsl{k}',\bsl{q}} = \prescript{2}{}{D}^{}_{-\bsl{k}+\bsl{q},\, \bsl{k},\, \bsl{k}',\, -\bsl{k}'+\bsl{q}} \\
     & \prescript{2}{}{\tilde{Q}}_{\bsl{k},\bsl{k}',\bsl{q}}  = \prescript{2}{}{Q}^{}_{-\bsl{k}+\bsl{q},\, \bsl{k},\, \bsl{k}',\, -\bsl{k}'+\bsl{q}} \\
     & \prescript{2}{}{\tilde{G}}_{\bsl{k},\bsl{k}',\bsl{q}} = \prescript{2}{}{G}^{}_{\bsl{k}+\bsl{q},\, \bsl{k},\, \bsl{k}',\, \bsl{k}'+\bsl{q}} .
     \label{eqn:D_Q_G_tilde}
}
Given a predicted $\prescript{2}{}{\tilde{D}}_{\bsl{k},\bsl{k}',\bsl{q}}$, the corresponding many-body energy reads
\eq{
E_{\text{pred}} = \frac{1}{N -1}\sum_{\bsl{k},\bsl{q}}\epsilon_{\bsl{k}}\prescript{2}{}{\tilde{D}}_{\bsl{k},\bsl{k},\bsl{q}}(x)  + \sum_{\bsl{k},\bsl{k}',\bsl{q}}V_{\bsl{k},\bsl{k}',\bsl{q}}\prescript{2}{}{\tilde{D}}_{\bsl{k},\bsl{k}',\bsl{q}} \ .
}

The translational invariance reduces the Hermiticity to
\eqa{
     & \prescript{2}{}{\tilde{D}}_{\bsl{k},\bsl{k}',\bsl{q}}^* = \prescript{2}{}{\tilde{D}}_{\bsl{k}',\bsl{k},\bsl{q}}\\
     & \prescript{2}{}{\tilde{Q}}_{\bsl{k},\bsl{k}',\bsl{q}}^*  = \prescript{2}{}{\tilde{Q}}_{\bsl{k}',\bsl{k},\bsl{q}} \\
     & \prescript{2}{}{\tilde{G}}_{\bsl{k},\bsl{k}',\bsl{q}}^* = \prescript{2}{}{\tilde{G}}_{\bsl{k}',\bsl{k},\bsl{q}}
     \label{eqn:D_Q_G_tilde_Hermiticity}
}
equivalent to $\prescript{2}{}{\tilde{D}}_{\bsl{k},\bsl{k}',\bsl{q}}$ being a hermitian matrix with index $\bsl{k}\bsl{k}'$ for each $\bsl{q}$ (similarly for $\prescript{2}{}{\tilde{Q}}$ and $\prescript{2}{}{\tilde{G}}$), and the antisymmetry of $\prescript{2}{}{D}$ and $\prescript{2}{}{Q}$ to
\eqa{
     & \prescript{2}{}{\tilde{D}}_{-\bsl{k}+\bsl{q},\bsl{k}',\bsl{q}} = \prescript{2}{}{\tilde{D}}_{\bsl{k},-\bsl{k}'+\bsl{q},\bsl{q}} = -\prescript{2}{}{\tilde{D}}_{\bsl{k},\bsl{k}',\bsl{q}}\\
     & \prescript{2}{}{\tilde{Q}}_{-\bsl{k}+\bsl{q},\bsl{k}',\bsl{q}} = \prescript{2}{}{\tilde{Q}}_{\bsl{k},-\bsl{k}'+\bsl{q},\bsl{q}} = -\prescript{2}{}{\tilde{Q}}_{\bsl{k},\bsl{k}',\bsl{q}} \ .
     \label{eqn:D_Q_tilde_AS}
}
The particle number constraints are
\eqa{
\label{eq:particle_number_constraints}
& \sum_{\bsl{k},\bsl{q}} \prescript{2}{}{\tilde{D}}_{\bsl{k},\bsl{k},\bsl{q}} = N (N-1) \\
& \sum_{\bsl{k},\bsl{q}} \prescript{2}{}{\tilde{Q}}_{\bsl{k},\bsl{k},\bsl{q}} = (N_k - N) (N_k - N-1) \\
& \sum_{\bsl{k},\bsl{k}'} \prescript{2}{}{\tilde{G}}_{\bsl{k},\bsl{k}',\bsl{0}} = N^2 \ .
}
The PSD of $\prescript{2}{}{D}$ now can be simplified as
\eqa{
& \sum_{\bsl{k}_1\bsl{k}_2\bsl{k}_3\bsl{k}_4} Y_{\bsl{k_1}\bsl{k_2}} \prescript{2}{}{D}_{\bsl{k_1}\bsl{k_2}\bsl{k_3}\bsl{k_4}} Y^\star_{\bsl{k_4}\bsl{k_3}}\geq 0 \\
& \Leftrightarrow \sum_{\bsl{k}_1\bsl{k}_2\bsl{k}_3\bsl{k}_4} Y_{\bsl{k_1}\bsl{k_2}} \sum_{\bsl{k},\bsl{k}',\bsl{q}}\delta_{\bsl{k}_1,-\bsl{k}+\bsl{q}\mod\bsl{G}_M}\delta_{\bsl{k}_2,\bsl{k}}\prescript{2}{}{D}_{-\bsl{k}+\bsl{q},\bsl{k},\bsl{k}',-\bsl{k}'+\bsl{q}} \delta_{\bsl{k}_4,-\bsl{k}'+\bsl{q}\mod\bsl{G}_M}\delta_{\bsl{k}_3,\bsl{k}'}Y^\star_{\bsl{k_4}\bsl{k_3}}\geq 0 \\
& \Leftrightarrow  \sum_{\bsl{k},\bsl{k}',\bsl{q}} Y_{-\bsl{k}+\bsl{q},\bsl{k}} \prescript{2}{}{D}_{-\bsl{k}+\bsl{q},\bsl{k},\bsl{k}',-\bsl{k}'+\bsl{q}}Y^\star_{-\bsl{k}'+\bsl{q},\bsl{k}'}\geq 0 \\
& \Leftrightarrow  \sum_{\bsl{q}}\sum_{\bsl{k},\bsl{k}'} y_{\bsl{q},\bsl{k}} \prescript{2}{}{D}_{-\bsl{k}+\bsl{q},\bsl{k},\bsl{k}',-\bsl{k}'+\bsl{q}}y_{\bsl{q},\bsl{k}}^\star \geq 0 \\
& \Leftrightarrow  \sum_{\bsl{q}}\sum_{\bsl{k},\bsl{k}'} y_{\bsl{q},\bsl{k}} \prescript{2}{}{\tilde{D}}_{\bsl{k},\bsl{k},\bsl{q}}y_{\bsl{q},\bsl{k}}^\star \geq 0 
}
for any complex $y_{\bsl{q},\bsl{k}} = Y_{-\bsl{k}+\bsl{q},\bsl{k}}$.
It means that the PSD of $\prescript{2}{}{D}$ is equivalent to that $\prescript{2}{}{\tilde{D}}_{\bsl{k},\bsl{k}',\bsl{q}}$ being a PSD matrix of index $\bsl{k}\bsl{k}'$ for each $\bsl{q}$.
Similarly, the PSD of $\prescript{2}{}{Q}$ equivalent to that $\prescript{2}{}{\tilde{Q}}_{\bsl{k},\bsl{k}',\bsl{q}}$ being a PSD matrix of index $\bsl{k}\bsl{k}'$ for each $\bsl{q}$, and the PSD of $\prescript{2}{}{G}$ equivalent to that $\prescript{2}{}{\tilde{G}}_{\bsl{k},\bsl{k}',\bsl{q}}$ being a PSD matrix of index $\bsl{k}\bsl{k}'$ for each $\bsl{q}$.

For our NNs, we parameterize $\prescript{2}{}{\tilde{D}}_{\bsl{k},\bsl{k}',\bsl{q}}$ as
\eq{
\label{eq:Dtilde_as_A}
\prescript{2}{}{\tilde{D}}_{\bsl{k},\bsl{k}',\bsl{q}} = (A_{\bsl{q}} A_{\bsl{q}}^\dagger)_{\bsl{k},\bsl{k}'}
}
with $A_{\bsl{q}}$ a complex matrix of size $N_k\times N_k$.
Our NN directly predicts $A_
{\bsl{q}}$, and antisymmetrize it via
\eq{
\label{eq:A_AS}
\left[A_{\bsl{q}}\right]_{\bsl{k},\bsl{p}} \rightarrow (\left[A_{\bsl{q}}\right]_{\bsl{k},\bsl{p}}  -\left[A_{\bsl{q}}\right]_{-\bsl{k}+\bsl{q},\bsl{p}})/2
}
for all $\bsl{k},\bsl{q},\bsl{p}$.
It then predicts $\prescript{2}{}{\tilde{D}}_{\bsl{k},\bsl{k}',\bsl{q}}$ via \cref{eq:Dtilde_as_A}, and then predicts $\prescript{2}{}{\tilde{Q}}$ and $\prescript{2}{}{\tilde{G}}$ via \cref{eq:D-Q_relation} and \cref{eq:D-G_relation}, respectively.
Due to the Hermiticity and antisymmetry of $\prescript{2}{}{\tilde{D}}_{\bsl{k},\bsl{k}',\bsl{q}}$, the AS and Hermiticity of $\prescript{2}{}{\tilde{Q}}$ and the Hermiticity of $\prescript{2}{}{\tilde{G}}$ are automatically guaranteed, owing to \cref{eq:D-Q_relation,eq:D-G_relation}.
In addition, we also guarantee the particle number constraint by normalizing the predicted $\prescript{2}{}{\tilde{D}}$ to $N(N-1)$, which particle number constraints of $\prescript{2}{}{\tilde{Q}}$ and the Hermiticity of $\prescript{2}{}{\tilde{G}}$, owing to \cref{eq:D-Q_relation,eq:D-G_relation}.
Then, the only remaining $(2,2)$ conditions are the PSD of $\prescript{2}{}{\tilde{Q}}$ and $\prescript{2}{}{\tilde{G}}$, which we treated as the PSD loss, as discussed in \cref{app:losses}.

In addition, the translational invariance can also simplify the parametrization of the 3-RDM such as
\eqa{
\prescript{3}{}{D}_{-\bsl{k}_1'-\bsl{k}_2' +\bsl{q},\bsl{k}_1,\bsl{k}_2,\bsl{k}_2',\bsl{k}_1',-\bsl{k}_1-\bsl{k}_2 +\bsl{q}}^{\rm true}  & = \left\langle c^\dagger_{-\bsl{k}_1-\bsl{k}_2 +\bsl{q}}c^\dagger_{\bsl{k}_1}c^\dagger_{\bsl{k}_2}   c_{\bsl{k}_2'} c_{\bsl{k}_1'} c_{-\bsl{k}_1'-\bsl{k}_2' +\bsl{q}} \right\rangle = \prescript{3}{}{\widetilde{D}}_{\bsl{k}_1,\bsl{k}_2,\bsl{k}_2',\bsl{k}_1',\bsl{q}}^{\rm true}  \\
\prescript{3}{}{Q}_{-\bsl{k}_1-\bsl{k}_2 +\bsl{q}, \bsl{k}_1,\bsl{k}_2,\bsl{k}_2',\bsl{k}_1',-\bsl{k}_1'-\bsl{k}_2' +\bsl{q}}^{\rm true} & =\left\langle c_{-\bsl{k}_1-\bsl{k}_2 +\bsl{q}} c_{\bsl{k}_1} c_{\bsl{k}_2}   c^\dagger_{\bsl{k}_2'} c^\dagger_{\bsl{k}_1'} c^\dagger_{-\bsl{k}_1'-\bsl{k}_2' +\bsl{q}}\right\rangle = \prescript{3}{}{\widetilde{Q}}_{\bsl{k}_1,\bsl{k}_2,\bsl{k}_2',\bsl{k}_1',\bsl{q}}^{\rm true}\ .
}

\begin{figure}[t]
    \centering
    \includegraphics[width=1.0\linewidth]{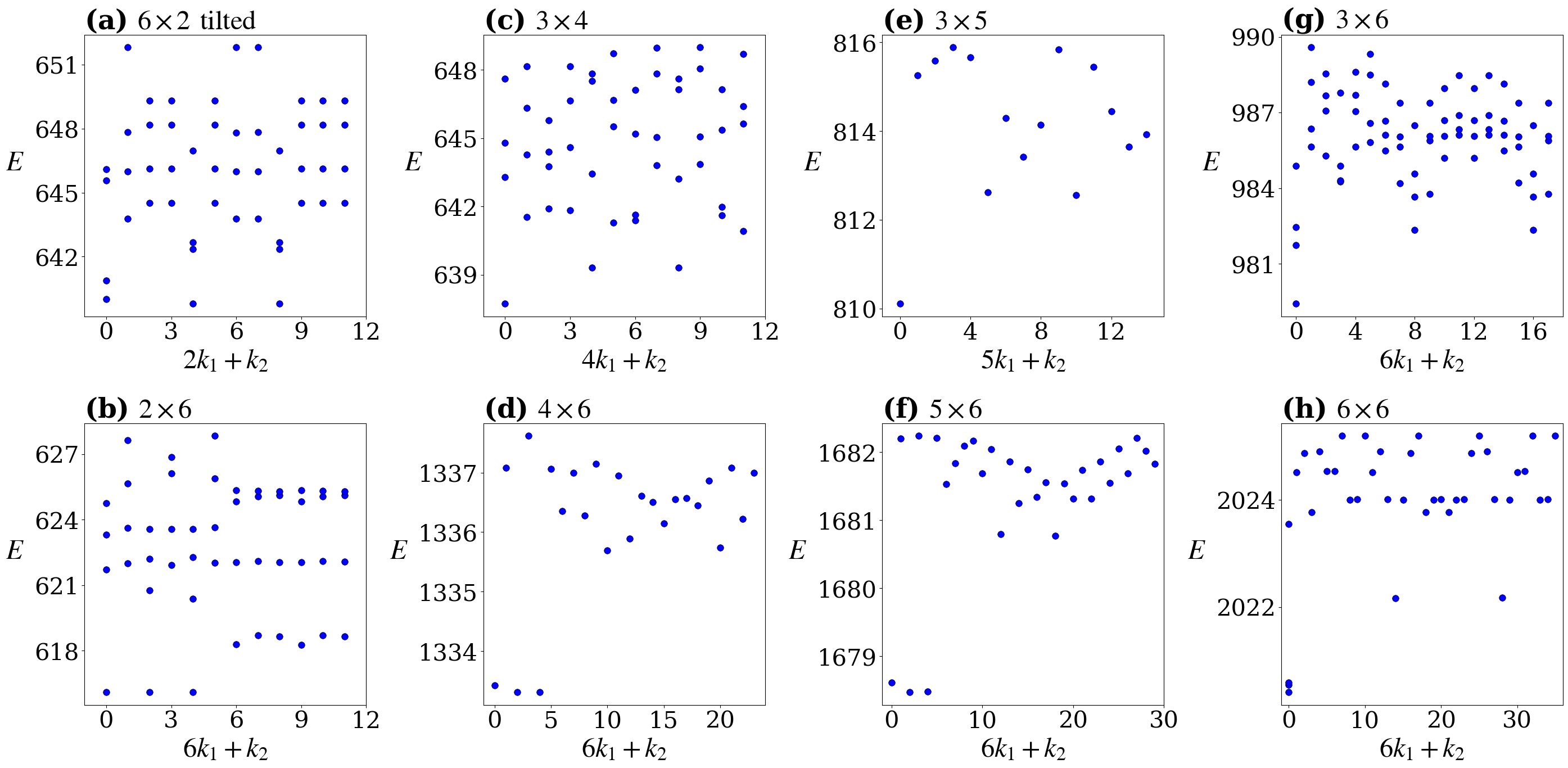}
    \caption{The ED spectrum for twisted bilayer MoTe$_2$ plotted for (a) $6\times 2$ (tilted mesh), (b) $2\times 6$, (c) $3\times 4$, (d) $4\times 6$, (e) $3\times 5$, (f) $5\times 6$, (g) $3\times 6$, and (h) $6\times 6$ lattice sizes. We plot the ground-state energy as a function of the total many-body momentum $\bsl{k}$, which is mapped to a linear index $L_2 k_1 + k_2$ for visualization via \cref{eqn:linearlize}. For (a), (b), (c), and (g), we calculate the lowest 4 eigenstates for each value of $\bsl{k}$. For (d), (e), and (f), we calculate only the ground state for each $\bsl{k}$. For (h), we calculate the lowest 4 eigenstates for $\bsl{k}=(0,0)$ and the ground state only for all other values of $\bsl{k}$. In each case the ED calculation is done for an electron filling of $\nu=\frac{2}{3}$, corresponding to the Fractional Chern Insulator (FCI) phase.}
    \label{fig:ED_spectrum}
\end{figure}

\subsection{Many-body quantum metric from 2-RDM}
\label{app:mb_QG}

The many-body quantum metric is defined by the tracking the evolution of the many-body state under twist boundary condition~\cite{Provost1980FSMetric,SWM2000,Resta2006Polarization,Regnault2013FCI,OnishiFu2025QuantumWeight,Yu2024_QG_Charge_Fluctuation,Wu2024QGCornerChargeFluctuationsManyBody}.
Specifically, given $\ket{\Psi(\bsl{q})}$ with twist-boundary condition $\bsl{q}$, we define
\begin{equation}
\ket{\Phi_{\bsl{q}}}
=
e^{-\ii \bsl{q} \cdot \bsl{X}}
\ket{\Psi(\bsl{q})}\ ,
\end{equation}
where $\bsl{X}$ is the many-body position operator
\eq{
\bsl{X} = \int d^2 r \sum_{l} \bsl{r} \widetilde{c}^\dagger_{\K, l, \bsl{r}} \widetilde{c}_{\K, l, \bsl{r}} \ .
}
Then, the many-body quantum metric reads
\eqa{
g_{ij}^{\mathrm{mb}}(\bsl{q}) & = \frac{1}{2}\bra{\partial_{q_i}\Phi_{\bsl{q}}}(1-\ket{\Phi_{\bsl{q}}}\bra{\Phi_{\bsl{q}}})\ket{\partial_{q_j}\Phi_{\bsl{q}}} + (i\leftrightarrow j) \\
& = \frac{1}{2}\bra{\Psi_{\bsl{q}}} X_i(1-\ket{\Psi_{\bsl{q}}}\bra{\Psi_{\bsl{q}}})X_j\ket{\Psi_{\bsl{q}}} + (i\leftrightarrow j)\ .
}
At $\bsl{q}=0$ which means the periodic boundary condition, we then have
\eq{
g_{ij}^{\mathrm{mb}}(\bsl{0}) = \frac{1}{2}\bra{\Psi_{\bsl{0}}} X_i X_j\ket{\Psi_{\bsl{0}}} - \frac{1}{2}\langle X_i\rangle \langle X_j\rangle  + (i\leftrightarrow j) = \frac{1}{2}\bra{\Psi_{\bsl{0}}} (X_i - \langle X_i\rangle) (X_j -  \langle X_j\rangle)\ket{\Psi_{\bsl{0}}}  + (i\leftrightarrow j)\ .
}
In the large size, we would have
\eq{
\G^{\rm mb} = \frac{1}{2\pi}\int d^2 q\  g_{ij}^{\mathrm{mb}}(\bsl{q}) = \frac{2\pi}{\V} g_{ij}^{\mathrm{mb}}(\bsl{0}) + O(1/\V)\ ,
}
which means we can use $g_{ij}^{\mathrm{mb}}(\bsl{0})$ to approximate $\G^{\rm mb}$: 
\eq{
\G^{\rm mb} \approx \frac{2\pi}{\V} g_{ij}^{\mathrm{mb}}(\bsl{0}) = \frac{\pi}{\V}\bra{\Psi_{\bsl{0}}} (X_i - \langle X_i\rangle) (X_j -  \langle X_j\rangle)\ket{\Psi_{\bsl{0}}}  + (i\leftrightarrow j)\ .
}

On the other hand, let's look at 
\eq{
S(\bsl{q}) = \bra{\Psi_{\bsl{0}}}\rho_{\bsl{q}}\rho_{-\bsl{q}} \ket{\Psi_{\bsl{0}}} - \bra{\Psi_{\bsl{0}}}\rho_{\bsl{q}}\ket{\Psi_{\bsl{0}}}\bra{\Psi_{\bsl{0}}}\rho_{-\bsl{q}} \ket{\Psi_{\bsl{0}}}
}
with
\eq{
\rho_{\bsl{q}} = \sum_{\bsl{k},\bsl{Q}} \tilde{c}^\dagger_{\K,\bsl{k}+\bsl{q},\bsl{Q}}\tilde{c}_{\K,\bsl{k},\bsl{Q}}\ .
}
Equivalently,
\begin{equation}
\rho_{\boldsymbol q}
=
\int d^2 r\,
e^{-\ii \boldsymbol q\cdot \boldsymbol r}
\hat n(\boldsymbol r),
\qquad
\hat n(\boldsymbol r)
=
\sum_l
\tilde c^\dagger_{\K,l,\boldsymbol r}
\tilde c_{\K,l,\boldsymbol r}.
\end{equation}
Expanding around $\boldsymbol q=\boldsymbol 0$ gives
\begin{equation}
\rho_{\boldsymbol q}
=
N
-
\ii \sum_i q_i X_i
-
\frac{1}{2}\sum_{ij}q_iq_j X_{ij}
+
O(q^3),
\qquad
X_i
=
\int d^2 r\, r_i\hat n(\boldsymbol r),
\qquad
X_{ij}
=
\int d^2 r\, r_i r_j\hat n(\boldsymbol r).
\end{equation}
Thus
\begin{equation}
\begin{aligned}
\langle \rho_{\boldsymbol q}\rho_{-\boldsymbol q}\rangle
&=
N^2
+
\sum_{ij} q_iq_j\langle X_iX_j\rangle
-
N \sum_{ij} q_iq_j\langle X_{ij}\rangle
+
O(q^3),
\\
\langle \rho_{\boldsymbol q}\rangle
\langle \rho_{-\boldsymbol q}\rangle
&=
N^2
+
\sum_{ij}q_iq_j\langle X_i\rangle\langle X_j\rangle
-
\sum_{ij}Nq_iq_j\langle X_{ij}\rangle
+
O(q^3).
\end{aligned}
\end{equation}
After subtracting the disconnected part, the $N^2$ and $X_{ij}$ terms cancel, giving
\begin{equation}
S_{\boldsymbol q}
=
\sum_{ij}q_iq_j
\left(
\langle X_iX_j\rangle
-
\langle X_i\rangle\langle X_j\rangle
\right)
+
O(q^3).
\end{equation}
Therefore,
\begin{equation}
g_{ij}^{\rm mb}(\bsl{0})
=
\frac{1}{2}
\left.
\frac{\partial^2 S_{\boldsymbol q}}
{\partial q_i\partial q_j}
\right|_{\boldsymbol q=\boldsymbol 0}\ ,
\end{equation}
which means
\eq{
\G^{\rm mb} \approx \frac{2\pi}{\V} g_{ij}^{\mathrm{mb}}(\bsl{0}) = 
\frac{\pi}{\V}
\left.
\frac{\partial^2 S_{\boldsymbol q}}
{\partial q_i\partial q_j}
\right|_{\boldsymbol q=\boldsymbol 0}\ .
}

Owing to the momentum definite ground state that we focus on, we have
\eqa{
S(\bsl{q}) & = \bra{\Psi_{\bsl{0}}}\rho_{\bsl{q}}\rho_{-\bsl{q}} \ket{\Psi_{\bsl{0}}} - \delta_{\bsl{q},0} N^2 = \sum_{\bsl{k},\bsl{k}'}\sum_{\bsl{Q}\bsl{Q}'} \langle\tilde{c}^\dagger_{\K,\bsl{k}+\bsl{q},\bsl{Q}}\tilde{c}_{\K,\bsl{k},\bsl{Q}} \tilde{c}^\dagger_{\K,\bsl{k}',\bsl{Q}'}\tilde{c}_{\K,\bsl{k}'+\bsl{q},\bsl{Q}'}\rangle   - \delta_{\bsl{q},0} N^2 \\
& = - \sum_{\bsl{k},\bsl{k}'}\sum_{\bsl{Q}\bsl{Q}'}  \langle\tilde{c}^\dagger_{\K,\bsl{k}+\bsl{q},\bsl{Q}}\tilde{c}^\dagger_{\K,\bsl{k}',\bsl{Q}'}\tilde{c}_{\K,\bsl{k},\bsl{Q}} \tilde{c}_{\K,\bsl{k}'+\bsl{q},\bsl{Q}'} \rangle + N  - \delta_{\bsl{q},0} N^2\\
& = - \sum_{\bsl{k},\bsl{k}'}\sum_{\bsl{Q}\bsl{Q}'}  \langle c^\dagger_{\bsl{k}+\bsl{q}}U_{\bsl{Q}}^*(\bsl{k}+\bsl{q})c^\dagger_{\bsl{k}'}U_{\bsl{Q}'}^*(\bsl{k}')c_{\bsl{k}}  U_{\bsl{Q}}(\bsl{k}) c_{\bsl{k}'+\bsl{q}} U_{\bsl{Q}'}(\bsl{k}'+\bsl{q})\rangle + N  - \delta_{\bsl{q},0} N^2\\
& = - \sum_{\bsl{k},\bsl{k}'} M_{\bsl{k},\bsl{q}} M_{\bsl{k}',\bsl{q}}^*  \langle c^\dagger_{\bsl{k}+\bsl{q}}c^\dagger_{\bsl{k}'}c_{\bsl{k}}  c_{\bsl{k}'+\bsl{q}}\rangle + N  - \delta_{\bsl{q},0} N^2\\
& = \sum_{\bsl{k},\bsl{k}'} M_{\bsl{k},\bsl{q}} M_{\bsl{k}',\bsl{q}}^*  \langle c^\dagger_{\bsl{k}+\bsl{q}} c_{\bsl{k}} c^\dagger_{\bsl{k}'}  c_{\bsl{k}'+\bsl{q}}\rangle - \sum_{\bsl{k}} M_{\bsl{k},\bsl{q}} M_{\bsl{k},\bsl{q}}^*  \langle c^\dagger_{\bsl{k}+\bsl{q}}  c_{\bsl{k}+\bsl{q}}\rangle + N  - \delta_{\bsl{q},0} N^2\\
& = \sum_{\bsl{k},\bsl{k}'} M_{\bsl{k},\bsl{q}} M_{\bsl{k}',\bsl{q}}^*  \langle c^\dagger_{\bsl{k}+\bsl{q}} c_{\bsl{k}} c^\dagger_{\bsl{k}'}  c_{\bsl{k}'+\bsl{q}}\rangle - \sum_{\bsl{k}} M_{\bsl{k},\bsl{q}} M_{\bsl{k},\bsl{q}}^*  \langle c^\dagger_{\bsl{k}+\bsl{q}}  c_{\bsl{k}+\bsl{q}}\rangle + N  - \delta_{\bsl{q},0} N^2\ .
}
Therefore, the predicted $S(\bsl{q})$ has the form
\eq{
S(\bsl{q}) = \sum_{\bsl{k},\bsl{k}'} M_{\bsl{k},\bsl{q}} M_{\bsl{k}',\bsl{q}}^*  \prescript{2}{}{\tilde{G}}_{\bsl{k},\bsl{k}',\bsl{q}}   - \sum_{\bsl{k}} ( |M_{\bsl{k},\bsl{q}}|^2 -1) \frac{1}{N-1} \sum_{\bsl{k}'} \prescript{2}{}{\tilde{D}}_{\bsl{k}',\bsl{k}',\bsl{q}+\bsl{k}'+\bsl{k}} - \delta_{\bsl{q},0} N^2    \ .
}

\subsection{Training Data and Testing Data}

\label{sec:MoTe2_preprocess}

In this section, we first discuss how we obtain our training and testing data, and explain the preprocessing necessary for the use of the training data in our NN. 

We obtain the training and testing data via ED on the projected Hamiltonian \cref{eq:ED_hamiltonian_app} for $2\times 6$, $3\times 4$, $3\times 5$, $3\times 6$,  $4\times 6$, $5\times 6$, $6\times 6$ conventional meshes and a $6\times 2$ tilted mesh, all  with $-2/3$ electron filling (\ie, $1/3$ hole filling).
The ED spectra are summarized in \cref{fig:ED_spectrum}, and the ground-state energies are listed in \cref{tab:Ground_state_energy}.
Note that the ground state does not always have zero many-body momentum.
From the ED ground state, we can then calculate the $\tilde{D}^{\rm true}$, $\tilde{Q}^{\rm true}$ and $\tilde{G}^{\rm true}$, with which we can perform the interpolation training and the testing of the predictions.
Specifically, for the interpolation tasks, we use $2\times 6$, $3\times 4$, $3\times 6$, and tilted $6\times 2$ for the training and $6\times 6$ for the testing. 
For the interpolation training, the training data is composed of 2-RDMs over 4 different $\bsl{k}$ point meshes: conventional $2\times 6$, tilted $6\times 2$, and conventional $3\times 4$ meshes, each containing 12 $\bsl{k}$-points, and a conventional $3\times 6$ mesh with 18 $\bsl{k}$-points. 
The tilted $6\times 2$ mesh has $M = \mat{1 & 1 \\ 0 & 1}$ in  \cref{eqn:tilted_convention}. 
All meshes are used for testing in the variational optimization.

\begin{figure}[t]
    \centering
    \includegraphics[width=0.6\linewidth]{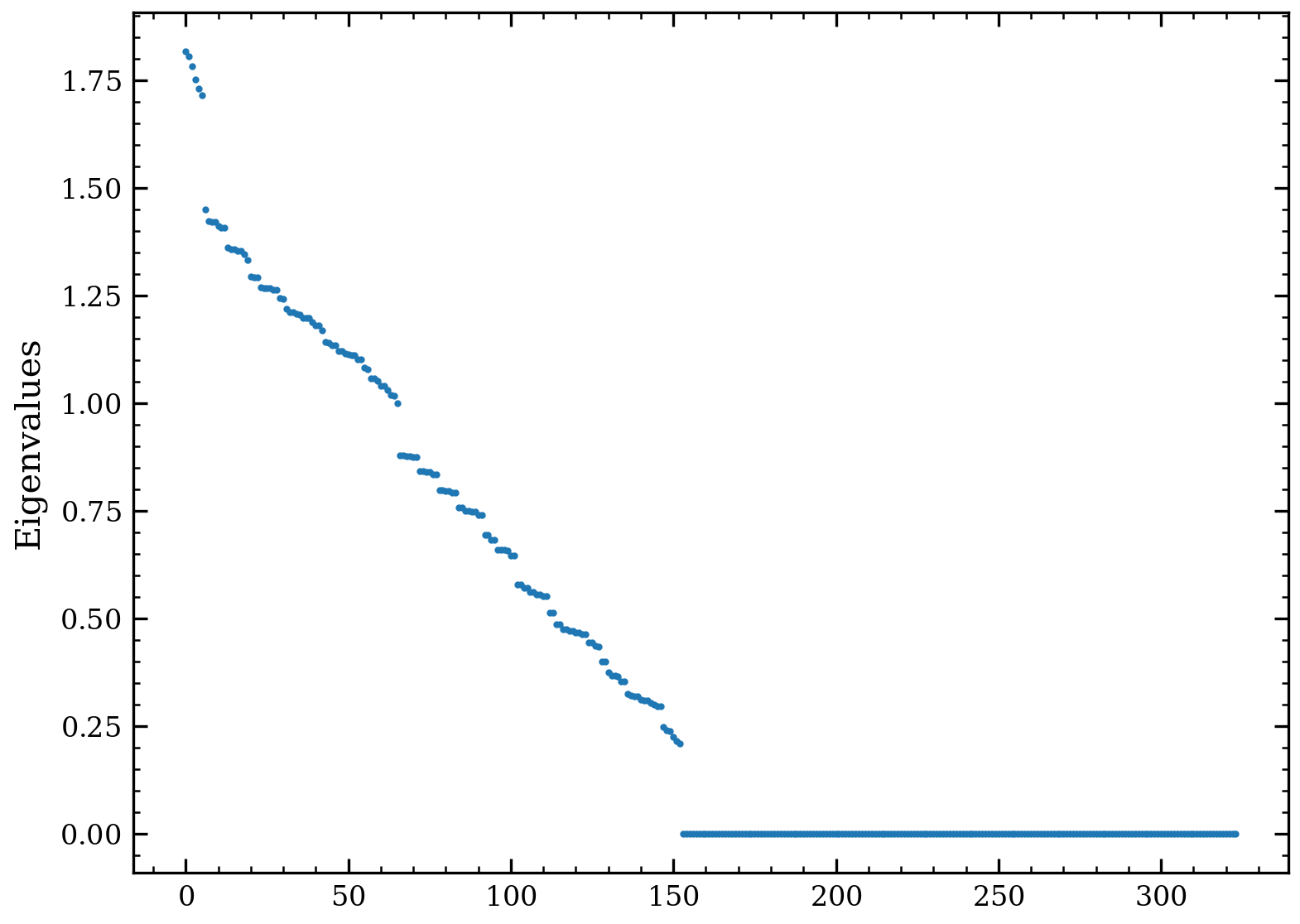}
    \caption{
    The eigenvalues of $\prescript{2}{}{\tilde{D}}_{\bsl{k},\bsl{k}',\bsl{q}}$ of the projected $t$MoTe$_2$ model on the conventional $3\times 6$ mesh for all fixed $\bsl{q}$ --- treating $\prescript{2}{}{\tilde{D}}_{\bsl{k},\bsl{k}',\bsl{q}}$ as a matrix with index $\bsl{k}\bsl{k}'$ for each fixed $\bsl{q}$.
    We note that, unlike in \cref{fig:Richardson_eigenvalues}, there is a large spectrum of significant eigenvalues leading to a more complicated interpolation. 
    }
    \label{fig:2D_MoTe2_Eigenvalues}
\end{figure}

For the interpolation training, we first examine the eigenvalues of $\prescript{2}{}{\tilde{D}}$ on the conventional $3\times 6$ mesh.
As shown in \cref{fig:2D_MoTe2_Eigenvalues}, there are no clear principal components, unlike the case for the Richardson model.
Therefore, in this case, we do not use eigendecomposition for the NNs.

\begin{table}[htb]
    \centering
    \setlength{\tabcolsep}{8pt} 
    \renewcommand{\arraystretch}{1.20}  
    \begin{tabular}{cccc}
        \toprule
        $L_1\times L_2$  &  $M$ & $E_{\mathrm{ED}}$ & $\bsl{k}_{\mathrm{GS}}$ \\
        \midrule
        $2\times 6$ & $\mat{ 1 & 0 \\ 0 & 1}$ & 616.108 & $(0,2)$ \\
        $3\times 4$ & $\mat{ 1 & 0 \\ 0 & 1}$ & 637.739 & $(0,0)$ \\
        $6\times 2$ & $\mat{ 1 & 1 \\ 0 & 1}$ & 639.800 & $(4,0)$ \\
        $3\times 5$ & $\mat{ 1 & 0 \\ 0 & 1}$ & 810.115 & $(0,0)$ \\
        $3 \times 6$ & $\mat{ 1 & 0 \\  0 & 1}$ & 979.428 & $(0,0)$ \\
        $4\times 6$ & $\mat{ 1 & 0 \\  0 & 1}$ & 1333.311 & $(0,4)$ \\
        $5\times 6$ & $\mat{ 1 & 0 \\  0 & 1}$ & 1678.474 & $(0,2)$ \\
        $6\times 6$ & $\mat{ 1 & 0 \\  0 & 1}$ & 2020.418 & $(0,0)$ \\
        \bottomrule
    \end{tabular}
    \caption{Ground-state energies $E_{\mathrm{ED}}$ (measured in meV) and the ground-state many-body momentum $\bsl{k}_{\mathrm{GS}}$ for twisted bilayer MoTe$_2$ calculated through ED. These data correspond to \cref{fig:ED_spectrum}.}
    \label{tab:Ground_state_energy}
\end{table}

\subsection{Framework of the NN}

In this section, we discuss the framework for the NN, which is shared by both the interpolation and variational optimization. An architecture plot of the  framework can be seen in \cref{fig:nn_diagram}, and we expand and generalize that description. Generally, the framework is given a momentum mesh. It then uses a NN to predict the value of an object (i.e. $A_{\bsl{q}}$) on that momentum mesh by evaluating the NN at each individual $\bsl{k}$ point and aggregating the results. With this predicted object, the framework may enforce constraints  (i.e. antisymmetry, PSD, particle number) by modifying the object before giving the final predicted 2-RDM object.  

To use this framework, we must only provide a few key pieces of information: $L_1$ and $L_2$, the momentum mesh Matrix $M=\begin{pmatrix}
    M_{11} & M_{12} \\
    M_{21} & M_{22}
\end{pmatrix}$) defined in  (\cref{app:basics}), and the electron filling $\nu$. From this, the framework is able to predict the 2-RDM over this mesh. 
The NN framework first generates the momentum mesh; in our case, the mesh is the set $(\bsl{k},\bsl{p},\bsl{q})$ where each element is over $L_1\times L_2$.  We map each $\bsl{k}$ in \cref{eqn:tilted_convention} to $[-1,1]^2$ via
\begin{align}
    v(\bsl{k}) \equiv  \left(2 \frac{(\bsl{a}_{M,1}\cdot\bsl{k}\ \mod 2\pi)}{2\pi}-1,2 \frac{(\bsl{a}_{M,2}\cdot\bsl{k}\ \mod 2\pi)}{2\pi}-1\right)  = 
2\left(\frac{k_1 M_{11}}{L_1} + \frac{k_2 M_{21}}{L_2} \mod\,1, \frac{k_1 M_{12}}{L_1} + \frac{k_2 M_{22}}{L_2} \mod\,1\right) -1 \  ,
\end{align} 
with $\bsl{a}_{M,1}$ and $\bsl{a}_{M,2}$ the primitive lattice vectors.
We then define the coordinate input as
\begin{equation}
    \bsl{x}_v = \bigl(v(\bsl{k}),\, v(\bsl{p}),\, v(\bsl{q})\bigr) \in [-1,1]^6.
\end{equation}
and the scale input as  $\bsl{s} = \left(\frac{1}{L_1}, \frac{1}{L_2}\right)$. This $\bsl{s}$ carries the information of the mesh, and can contain an arbitrary size amount of information about the mesh we predict on. The only requirement that the input is the same for the entire mesh, it only differs when we use the framework on different meshes.  
The full NN input is formed by appending $\bsl{s}$ to $\bsl{x}_v$, with
\begin{equation}
    \bsl{x} = \left(v(\bsl{k}),\, v(\bsl{p}),\, v(\bsl{q}), \frac{1}{L_1}, \frac{1}{L_2}, \frac{1}{N}\right) \in [-1,1]^6 \times (0,1]^3,
\end{equation}
so that the final NN is a mapping $\Phi: \mathbb{R}^9 \to \mathbb{R}^2$:
\eq{
\Phi(\bsl{x}) = ( \Re\left( [A_{\bsl{q}}]_{\bsl{k}\bsl{p}} \right), \Im\left( [A_{\bsl{q}}]_{\bsl{k}\bsl{p}} \right) )\ .
}

\subsection{Interpolation Training}

\label{sec:MoTe2_interpolation_training}

We first describe the interpolation training. This method is characterized by exact 2-RDM on small $\bsl{k}$ meshes, which we use to train the NN and  predict a large mesh which.  For this study, we use take two choices of sampling: a single-mesh interpolation with conventional $3\times 6$ training mesh, or  a three-mesh interpolation with three training  meshes (conventional $2\times 6$, conventional $3\times 4$, and tilted $6\times 2$) each of which has 12 $\bsl{k}$ points. Due to the nature of ED, these smaller meshes are computationally faster to calculate, and let us investigate if this form of training can lead to more desirable results.

For the interpolation, we build a loss function with  losses defined in  \cref{app:losses}. Specifically, we use the reconstruction loss between $\prescript{2}{}D$, $\prescript{2}{}Q$, and $\prescript{2}{}G$ with \cref{eqn:recon_loss},  the PSD of $\prescript{2}{}{Q}$ and $\prescript{2}{}{G}$ with \cref{eq:lambda_Q_G} and energy distance losses with \cref{eqn:energy_dist}. For each loss, we associate a weight $\alpha_i$ which determines its impact on training, giving us 
\begin{equation}
\begin{aligned}
    \mathcal{L}(\theta) &= \mathcal{L}_{{recon}}(\theta) 
    + \alpha_{{PSD}_Q}\mathcal{L}_{{PSD}_Q}(\theta) 
    + \alpha_{{PSD}_G}\mathcal{L}_{{PSD}_G}(\theta)   
    + \alpha_{E_{{dist}}} \mathcal{L}_{E_{{dist}}}(\theta)
\end{aligned}
\label{eqn:MoTe2_loss}
\end{equation}
where $\theta$ represents our NN's current parameters. 
Additionally, the PSD loss on $^2Q$ and $^2G$ is taken with respect to all meshes we train on (\ie, on which we have the ground truth), as well as the auxiliary mesh---conventional $6\times 6$.

We run the training for $N_{epochs}=1500$ and train the NN using the Adam optimizer \cite{Kingma2017Adam} with a cosine annealed \cite{Loshchilov2017CosineAnnealing} learning rate from $\eta$ to $\eta/100$. We set $N_w=800$ warmup epochs training the MSE loss alone (i.e. $\alpha_{PSD_{Q/G}}=\alpha_{T1}=\alpha_{E_{dist}}=0$) so that the model does not get stuck in a non-optimal local minimum based on the other losses. Following that, we spend the next $N_r=r \cdot (N_{epochs}-N_w)$ epochs gradually increasing the penalty weights to their full reported values, where $r=0.7$ is the ramping ratio. This  ramp-up phase allows the model to adapt to the new penalties. Specifically, we multiply each penalty loss except the MSE by a ramp factor \begin{equation}
    \rho = \min\left(1, \frac{N_{current}-N_w}{r(N_{epochs}-N_w)}\right).
    \label{eqn:ramp}
\end{equation}
 where $N_{current}$ refers to the current epoch we are training on. Once training has completed, we evaluate the NN which has the lowest loss (excluding those NNs evaluated in the warm-up and ramp phases) on the test set. The full set of hyperparameters can be seen in \cref{tab:MoTe_2Hyper}, where we provide the weights used in the loss.

As we have defined a method to interpolate from small-size $\bsl{k}$ meshes to larger ones, we wish to benchmark the method. Specifically, we examine three factors: the impact of physics guided constraints and losses, the impact of NN architecture, and the impact of BZ sampling. 
To perform this evaluation, we conduct a comprehensive grid-search over  every combination of NN architecture and their respective  hyperparameters defined in \cref{app:Architectures}, the training on a single $3\times 6$ mesh or three 12-$\bsl{k}$ meshes, and value of $\alpha_{PSD_{Q}}$, $\alpha_{PSD_{G}}$ and $\alpha_{E_{dist}}$.

We first, however, examine how a Fourier interpolation performs on this data. To do so, we perform the same method as in \cref{sec:richard_tilted_mesh_results}, calculating the FFT on the $3\times 6$ mesh and comparing to the $6\times 6$ mesh. With this interpolation, we find a normalized overlap of $0.64825$ and an R-value of $0.81049$, giving us a baseline for our examination.

\begin{table}[htb]
    \centering
    \setlength{\tabcolsep}{8pt} 
    \renewcommand{\arraystretch}{1.20}  
    \begin{tabular}{c|c}
         \toprule
         \textbf{Hyperparameter} & \textbf{Value} \\
         \midrule
         $N_{epochs}$ & 1600 \\
         $\eta$ & $10^{-3}$\\
         $N_w$ & 800\\
         $r$ & 0.7
         \\
         \midrule
         $\alpha_{PSD_Q}$ & 0 or 1 \\
         $\alpha_{PSD_G}$ & 0 or 1\\
         $\alpha_{E_{dist}}$ & 0 or 1\\
         \bottomrule
    \end{tabular}
    \caption{Hyperparameters and loss weights for the twisted MoTe$_2$ (see \cref{eq:ED_hamiltonian_app})  interpolation training.   $N_{epochs}$ is the number of training epochs, while $N_w$ is the number of warmup epochs where we only apply the reconstruction loss. $r$ is the ramping ration, which modifies \cref{eqn:ramp} and  $\eta$ is the learning rate. All $\alpha$ values are the loss function weights, used in \cref{eqn:MoTe2_loss}. 
    We note that these loss function weights may take different values for different NN.
    }
    \label{tab:MoTe_2Hyper}
\end{table}

\subsubsection{Effect of Constraints and Losses}
\label{app:ML_results_constraints_losses}
We  examine the contribution of each physical constraint and loss by providing the best $R\utext{value}$, and corresponding $R\utext{overlap}$ and predicted energy on the $6\times 6$ mesh, for predictions with the constraint or loss enabled or disabled. The training is comprehensive, and includes all 6 NNs and the training on one $3\times 6$ or three $12-\bsl{k}$ meshes.
These results are given in \cref{tab:MoTe2_loss_importance}.
We find that while the $\alpha_{PSD_Q}=1$ and $\alpha_{PSD_G}=1$ can be helpful, $\alpha_{E_{dist}}=1$ is detrimental.
Therefore, we will always keep $\alpha_{E_{dist}}=0$.
The values of $\alpha_{PSD_Q}$ and $\alpha_{PSD_G}$ depends on the architecture as shown in the following.

We first note the $\Delta R\utext{value}=+0.0200$ (the difference between on and off the constraint) for the ${}^2D$ PSD constraint is the same as the ${}^2D$ antisymmetry constraint and the $QG$ reconstruction loss. This indicates that these three features together enable the best possible performance, while all three disabled account for the worst performance. Additionally, we note this performance difference is significant, as $\Delta R\utext{overlap}=+0.1482$  and $\Delta E_{pred}=-102.918$. Hence, we use  these three constraint and losses for further tests.  For the other losses tested, we see that only PSD on ${}^2Q$ and ${}^2G$ both lead to small improvements with $\Delta R\utext{value}=+0.0016$ and $\Delta R\utext{overlap}=0.0001$ for ${}^2Q$ and $\Delta R\utext{value}=0.0021$ and $\Delta R\utext{overlap}=0.0032$  for ${}^2G$.  Finally, we see that the energy distance loss had a highly negative effect, with a $\Delta R\utext{value}=-0.0260$  and $\Delta R\utext{overlap}=-0.1833$.

\begin{table}[htb]
    \centering
    \setlength{\tabcolsep}{8pt}
    \renewcommand{\arraystretch}{1.20}
    \begin{tabular}{l|cc|c|cc|c|cc|c}
        \toprule
        & \multicolumn{3}{c|}{\textbf{$R_\text{value}$}} & \multicolumn{3}{c|}{\textbf{$R_\text{overlap}$}} & \multicolumn{3}{c}{\textbf{Energy (meV)}} \\
        \textbf{Loss Weights} & \textbf{1} & \textbf{0} & \textbf{$\Delta$} & \textbf{1} & \textbf{0} & \textbf{$\Delta$} & \textbf{1} & \textbf{0} & \textbf{$\Delta$} \\
        \midrule
        $ \alpha_{PSD_Q}$              & $0.9823$ & $0.9807$ & $+0.0016$ & $0.9737$ & $0.9736$ & $+0.0001$ & $2123.942$ & $2127.876$ & $-3.933$ \\
        $\alpha_{PSD_G}$              & $0.9823$ & $0.9798$ & $+0.0025$ & $0.9737$ & $0.9704$ & $+0.0032$ & $2123.942$ & $2128.510$ & $-4.567$ \\
        $\alpha_{E_{dist}}$         & $0.9563$ & $0.9823$ & $-0.0260$ & $0.7903$ & $0.9737$ & $-0.1833$ & $2125.221$ & $2123.942$ & $+1.279$ \\
        \bottomrule
    \end{tabular}
    \caption{Effect of each $\alpha$ in \cref{eqn:MoTe2_loss} on interpolation  performance for the one-band projected twisted bilayer MoTe$_2$ model (see \cref{eq:ED_hamiltonian_app}). 
    The values are shown for the best the best $R\utext{value}$ on $6\times 6$ among all results of all six NNs that are trained on either a single $3\times 6$ mesh or three $12-\bsl{k}$ meshes.
    When the value of one $\alpha$ is fixed, the other two $\alpha$'s are allowed to take any values of $0,1$ to give the best results.
    }
    \label{tab:MoTe2_loss_importance}
\end{table}

\subsubsection{Architecture Comparison}
For our grid search, we also examine each of our NN architectures and their hyperparameters, as defined in \cref{app:Architectures}. The result for the NN's  optimal performance are given  in \cref{tab:MoTe2_architectures}. From this, we see that the  MLP has the best performance, with an R-value of $0.9823$ and normalized overlap of $0.9737$. We note that this is for a relatively large NN with $1,318,914$ parameters, while the KAN and TBN  also benefit from having a larger number of parameters. The NNs that are designed explicitly for INR, SIREN and FINER, have their performance maximize at a smaller 331,778 and 134,402 parameters respectively, while the residual MLP has a similar peak at $334,210$ parameters.

We note that, while the MLP has the best performance, both the SIREN and Residual MLP can reach similar levels of success with  $R\utext{value}$ only $0.0012$ and $0.0005$ less than that of the Residual MLP. respectively.  The FINER and TBN perform marginally worse, while the KAN performs the most poorly, with a peak $R \utext{value}$ $0.0107$ less than the Residual MLP and a peak $R_{\text{overlap}}$ $0.0559$ less. 

\begin{figure}[t]
    \centering
    \includegraphics[width=0.8\linewidth]{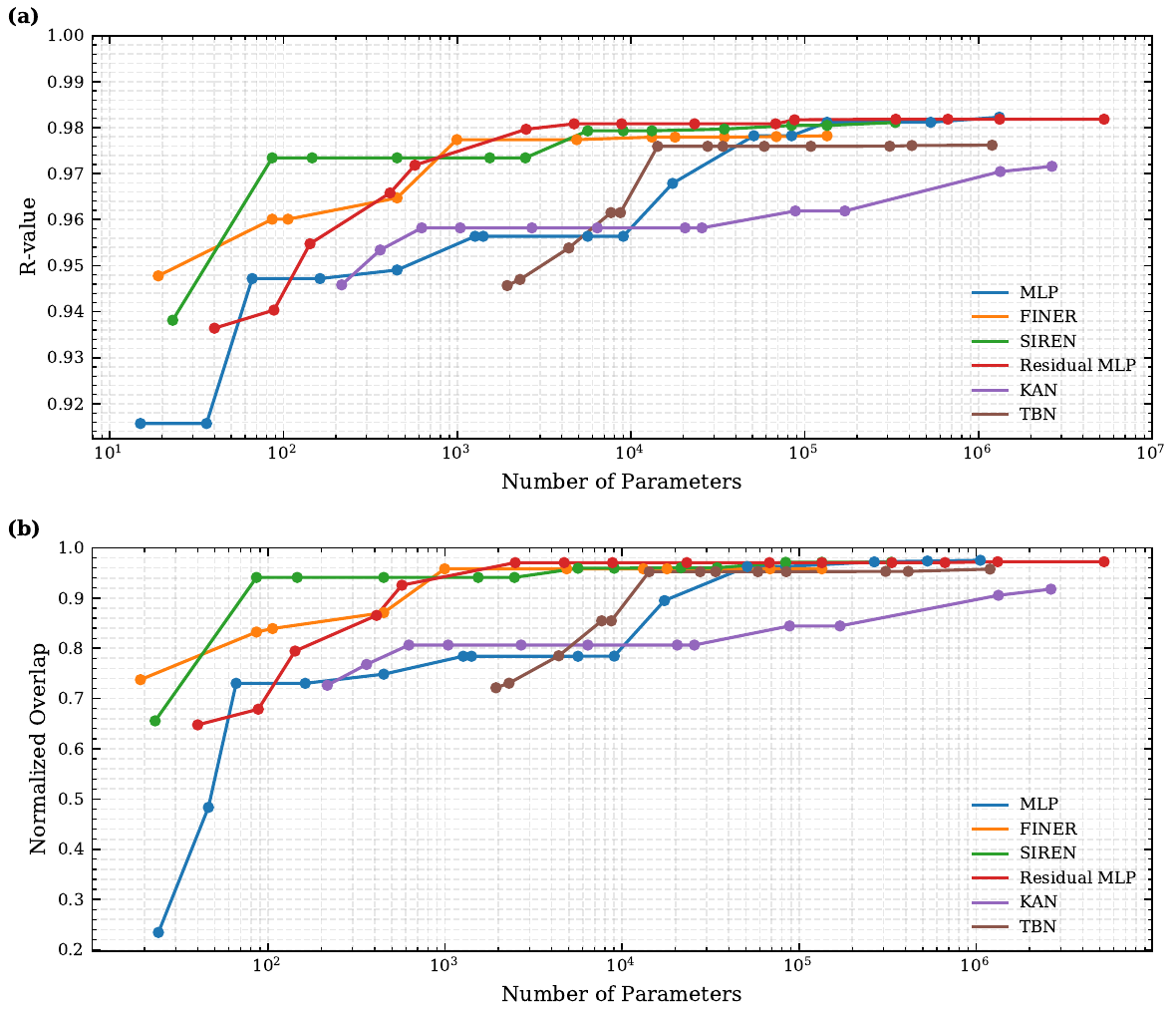}

    \caption{R-value (a) and normalized overlap (b) versus number of parameters for the best performance of each NN architecture predicting on the MoTe$_2$ $6\times 6$ $\bsl{k}$  point mesh.
    }
    \label{fig:MoTe2_paramater_eff}
\end{figure}

Finally, we examine the parameter efficiency plots in \cref{fig:MoTe2_paramater_eff}. We see that the SIREN, FINER, TBN, and residual MLP are able to achieve most of their performance with less than $10^4$ parameters, while the MLP and KAN need far more parameters to reach their highest level of performance.

\begin{table}[htb]
    \centering

    \setlength{\tabcolsep}{8pt}

    \renewcommand{\arraystretch}{1.20}

    \begin{tabular}{l|c|c|c|r|l}
        \toprule
        \textbf{Architecture} & $R_\text{value}$ & $R_\text{overlap}$ & \textbf{Energy (meV)} & \textbf{Parameters} & \textbf{Hyperparameters}\\
        \midrule
        MLP$^{a}$ & $0.9823$ & $0.9737$ & $2123.9425$ & $1,318,914$ & $h=512$, $d=6$ \\
        Residual MLP$^{a}$ & $0.9818$ & $0.9707$ & $2097.7711$ & $334,210$ & $h=128$, $d=10$ \\
        KAN$^{a}$ & $0.9716$ & $0.9178$ & $2135.0411$ & $2,647,040$ & $h=256$, $d=5$, $G=5$, $w_b=1.0$, $w_s=5.0$ \\
        SIREN$^{a,\dagger}$ & $0.9811$ & $0.9709$ & $2133.2267$ & $331,778$ & $h=256$, $d=5$, $\omega_0=3.0$, $\omega_1=5.0$ \\
        FINER$^{a}$ & $0.9782$ & $0.9526$ & $2139.4676$ & $134,402$ & $h=256$, $d=2$, $\omega_0=1.0$, $\omega_1=1.0$ \\
        TBN$^{b}$ & $0.9762$ & $0.9574$ & $2130.3253$ & $1,198,466$ & $h=128$, $d=6$, $N_f=4$ $N_h=1$, $\mathrm{drop}=0.00$ \\
        \bottomrule
            \multicolumn{6}{l}{$^{\dagger}$ Trained on three $12-\bsl{k}$ meshes} \\
        \multicolumn{6}{l}{$^{a}$  $(\alpha_{\text{PSD}_Q},\alpha_{\text{PSD}_G},\alpha_{E_{\text{dist}}})=(1,1,0)$ }\\
        \multicolumn{6}{l}{$^{b}$  $(\alpha_{\text{PSD}_Q},\alpha_{\text{PSD}_G},\alpha_{E_{\text{dist}}})=(0,1,0)$} \\
    \end{tabular}
    \caption{Peak performance for each architecture training on the one-band projected twisted bilayer MoTe$_2$ model (see \cref{eq:ED_hamiltonian_app}) for the 
    interpolation task, evaluated on the $6\times 6$ $\bsl{k}$ mesh. 
    Each network is given a set of NN hyperparameters, as described in \cref{app:Architectures}. Not all losses and constraints are applied for each model, and they are noted as such. Additionally, architectures marked $^\dagger$ were trained on three $12-\bsl{k}$ meshes, while the rest are trained on a single $3\times 6$ mesh.}
    \label{tab:MoTe2_architectures} 
\end{table}

\subsubsection{Mesh Comparison}
\label{sec:mesh_comparison}
We compare the single-mesh $(3\times6)$ and three-mesh ($3\times4$, $2\times 6$, tilted $6\times 2$) training in \cref{tab:MoTe2_mesh_comparison}. From this result, we can see that both methods yield similar results, with marginal improvements from a three-mesh training for the SIREN with 
a best R-value of $0.9805$ vs. $0.9811$. The  MLP, Residual MLP, KAN, FINER and TBN, on the other hand, all have a slight decrease between the two. This similarity suggests that the coverage of the $3\times 6$ mesh is similar to that of varied meshes over different grid shapes for the purpose of interpolation training and, as smaller meshes are computationally less expensive, could provide efficiency gains for the full training process.

Overall, we select residual MLP as the model with best performance, taking into account both accuracy and energy.

\begin{table}[htb]

    \centering

    \setlength{\tabcolsep}{8pt}

    \renewcommand{\arraystretch}{1.20}

    \begin{tabular}{l|ccc|ccc}

        \toprule

& \multicolumn{3}{c|}{\textbf{Single-mesh ($3\times6$)}} & \multicolumn{3}{c}{\textbf{Three-mesh ($3\times4$, $2\times 6$, tilted $6\times 2$)}} \\
        \textbf{Architecture} & $R_\text{value}$ & $R_\text{overlap}$ & $E$ (meV) & $R_\text{value}$ & $R_\text{overlap}$ & $E$ (meV) \\
        \midrule
        MLP             & $0.9823$ & $0.9737$ & $2123.942$ & $0.9799$ & $0.9519$ & $2115.189$ \\
        Residual MLP    & $0.9818$ & $0.9707$ & $2097.771$ & $0.9817$ & $0.9663$ & $2116.083$ \\
        KAN             & $0.9716$ & $0.9178$ & $2135.041$ & $0.9651$ & $0.8710$ & $2145.995$ \\
        SIREN           & $0.9805$ & $0.9681$ & $2131.023$ & $0.9811$ & $0.9709$ & $2133.227$ \\
        FINER           & $0.9782$ & $0.9531$ & $2139.966$ & $0.9782$ & $0.9526$ & $2139.468$ \\
        TBN             & $0.9762$ & $0.9574$ & $2130.325$ & $0.9761$ & $0.9522$ & $2139.126$ \\
        \bottomrule
    \end{tabular}
    \caption{Comparison of single-mesh and three-mesh interpolation 
    on the twisted MoTe$_2$ model (see \cref{eq:ED_hamiltonian_app}), predicting the $6\times 6$ $\bsl k$ point mesh. Here, the best $R\utext{value}$ is chosen for each model, to demonstrate the best possible performance, and we provide the corresponding $R\utext{overlap}$ and the predicted energy on $6\times 6$. }
    \label{tab:MoTe2_mesh_comparison}
\end{table}

\subsection{Variational  Optimization}

In this section, we develop a method to variationally optimize our NN. Specifically, we note that the energies predicted by the interpolation-trained NNs (\cref{sec:MoTe2_interpolation_training}) are far higher than the ED ground state on $6\times 6$, and hence wish to improve the accuracy of their prediction.

To do so, we first initialize our NN pipeline with weights given by doing a $4$-mesh interpolation training. 
Specifically, we run our NNs for $N_{epochs}=5000$ epochs, with a cosine scheduled learning rate from $\eta =10^{-3}$ to $\eta/100$. 
Additionally, we use a warmup fraction of $N_w=500$ and ramp $r=0.7$. 
During this process, we use the reconstruction loss on 4 meshes (conventional $2\times 6$, $3\times 4$, $3\times 6$, and tilted $6\times 2$), choose $\alpha_{\text{PSD}_Q}=\alpha_{\text{PSD}_G} = 1$ on the 4 meshes and an auxiliary mesh $6\times 6$, and set $\alpha_{E_\text{dist}} = 0$.

After the interpolation training, the NN is initialized.
With an initialized NN, we then perform our variational optimization. This method works on the basis that, as our NN can make predictions an arbitrary $\bsl{k}$ mesh, we may train the NN to make accurate predictions on a set of meshes by variationally decreasing energy over each mesh while also enforcing the ${}^2Q$ and ${}^2G$ positivity conditions through $\L_{\text{PSD}_{Q/G}}$. 
The benefit of this method is two-fold: it forces the NN to care about more $2$-RDMs when making predictions, and it provides an early stopping mechanism.

We first define a loss function using the PSD  loss  of $\prescript{2}{}{Q}$ and $\prescript{2}{}{G}$, $\mathcal{L}_{PSD_{Q/G}}$ from \cref{eq:lambda_Q_G}, and the energy loss minimization loss, $\mathcal{L}_{E_{min}}$ from \cref{eqn:energy_min}. 
The energy minimization loss is normalized per site as energy scales with $N$. 
Hence, the full loss function is given by
 \begin{equation}
    \mathcal{L}(\theta) = \alpha_{PSD_{Q}}\mathcal{L}_{PSD_{Q}}(\theta) + \alpha_{PSD_{G}}\mathcal{L}_{PSD_{G}}(\theta)   + \sum_i \alpha_{E_{min}}^i \frac{\mathcal{L}^i_{{E}_{min}}(\theta)}{N_s^i}
 \end{equation}
where we sum over each mesh in our energy minimization set. 
We note that $\mathcal{L}_{PSD_{Q}}$ and $\mathcal{L}_{PSD_{G}}$ are evaluated on all optimization target meshes and auxiliary meshes, while $\mathcal{L}^i_{{E}_{min}}$ only involves the optimization target meshes.
Specifically, $\mathcal{L}^i_{E_{min}}$ refers to the loss on the $ith$ target mesh, and $N_s^i$ refers to the number of sites in that mesh. For each target mesh, we may choose an $\alpha_{E_{min}}^i$ to weight the loss. Optimizing the energy on smaller target meshes is  easier than on larger targe tmeshes, so we define a weight on the target meshes which scales linearly with the number of sites to encourage a more uniform convergence. Specifically, we take
\begin{equation}
    \alpha_{E_{min}}^i = 1 + \frac{N_s^{i}-12}{24} (\kappa-1)
\end{equation}
and we examine the impact of different $\kappa$ in training. The benefit of this equation is that  our smallest target meshes, with 12 $\bsl{k}$ points have, have a weight of $1$, while further weights linearly increase and the weight on our $6\times 6$ mesh is $\kappa$.  We take  $\alpha_{PSD_{Q}}=\alpha_{PSD_{G}}=10^6$.

We now establish an early stopping mechanism. Because our NN  predicts and minimizes 2-RDMs over multiple mesh sizes, we may consider the energy predictions on sets of meshes the NN does not minimize energy on. This provides us a mechanism to stop training: when some validation set begins to increase it's predicted energy dramatically, we have entered the overfitting regime.  
In order to implement this method, we choose a validation set of meshes and use the prediction of energy as the validation loss.

For this work, the optimization target meshes are chosen to be $3\times 4$, $3\times 5$, $3\times 6$, $4\times 6$, $5\times 6$ and $6\times 6$ meshes, and we optionally also include a tilted $12\times 4$ mesh with $M=\mat{1 & 1 \\ 0 & 1}$ in   \cref{eqn:tilted_convention}. 
For both the auxiliary meshes, we choose $2\times 6$ and tilted $6\times 2$, and we use the energy prediction on $2\times 6$ as the validation loss.

To examine the effectiveness of our method, we optimize our interpolation-initialized NNs for  $100,000$ epochs. For the first $60,000$ epochs, we cosine anneal the learning rate from $10^{-5}$ to $10^{-7}$. Additionally, for this set of epochs, we using a ramp ratio $r=0.05$, leading to a gradual increase in loss over $3,000$ epochs as defined in \cref{eqn:ramp}.  For the next $40,000$ epochs, we continue to perform the minimization with a learning rate of $10^{-7}$. We perform early stopping when the validation loss (predicted energy on the conventional $2\times 6$ mesh) reaches $2$ meV above its minimum, where the minimum is taken after the initial ramp phase to prevent any minima spikes. 

We perform this examination over two seeds for randomization and take $\kappa\in\{1.5, 2.0,2.5,3.0\}$, and use the early stopping criteria above.
We provide the results of the prediction on a $6\times 6$ mesh with and without the tilted $12\times 4$ mesh being the target mesh included in \cref{tab:variational_6x6}. As can be seen from these results, our variational optimization continually outperforms the pure interpolation, leading to energies that are closer to the ED ground truth while also increasing the $R\utext{value}$ and $R\utext{overlap}$ scores. Our peak performance occurs for in the seed 12 trial with a $\kappa=3.0$ and no $12\times 4$ mesh interpolation. In this trial, our NN achieved an $R\utext{value}=0.98964$ and $R\utext{overlap}=0.98935$ when evaluated on the $6\times 6$ mesh. Additionally, its prediction on this mesh is only $0.1139$ meV below the ED ground state. To get a full picture of this network, we evaluate its predictions on other meshes and show them in \cref{tab:multiple_meshes}.

\begin{table}[t]
\centering
\label{tab:variational_6x6}
\begin{tabular}{llrrrrrr}
\toprule
 &  & \multicolumn{3}{c}{Seed 11} & \multicolumn{3}{c}{Seed 12} \\
\cmidrule(lr){3-5}\cmidrule(lr){6-8}
Setup & $\kappa$ & $E_{\mathrm{pred}}$ (meV) & $R\utext{value}$ & $R\utext{overlap}$ & $E_{\mathrm{pred}}$ (meV) & $R\utext{value}$ & $R\utext{overlap}$ \\
\midrule
with $12\times4$ & 1.5 & 2065.9308 & 0.97777 & 0.93622 & 2037.6909 & 0.98501 & 0.97205 \\
 & 2.0 & 2063.8766 & 0.97800 & 0.93719 & 2023.2591 & 0.98878 & 0.98674 \\
 & 2.5 & 2062.0672 & 0.97820 & 0.93810 & 2049.9095 & 0.98270 & 0.96632 \\
 & 3.0 & 2060.8860 & 0.97839 & 0.93892 & 2048.4330 & 0.98289 & 0.96678 \\
 & 3.5 & 2059.0198 & 0.97859 & 0.93967 & 2022.9335 & 0.98985 & 0.98706 \\
 & 4.0 & 2063.1764 & 0.97799 & 0.93704 &  2022.1637
 & 0.98897 & 0.98799\\
\midrule
without $12\times4$ & 1.5 & 2038.0997 & 0.98296 & 0.96704 & 2022.4649 & 0.98906 & 0.98686 \\
 & 2.0 & 2035.6772 & 0.98343 & 0.96878 & 2021.1900 & 0.98940 & 0.98825 \\
 & 2.5 & 2033.5800 & 0.98378 & 0.96992 & 2020.4463 & 0.98962 & 0.98925 \\
 & 3.0 & 2031.9786 & 0.98406 & 0.97082 & 2020.3141 & 0.98964 & 0.98935 \\
\bottomrule
\end{tabular}
\caption{$R\utext{value}$ and $R\utext{overlap}$ for prediction on the $6\times 6$ mesh during the variational minimization on tMoTe$_2$ (see \cref{eq:ED_hamiltonian_app}), with an interpolation-initialized NN. We conduct trials with and without treating the tilted $12\times 4$ mesh as a target mesh, and present the energies and accuracies predicted on a $6\times 6$ mesh. Trials were conducted until the energy predicted for a validation $2\times 6$ $\bsl{k}$ point mesh went $2$ meV above its minimum predicted energy. }
\end{table}

\begin{table}[t]
\centering
\label{tab:single_no12x4_seed12_k3}
\begin{tabular}{|c|c|c|c|c|c|c|}
\toprule
Mesh & $E_{{pred}}$ (meV) & $E_{pred}-E_{ED}$ (meV) & $R\utext{value}$ & $R\utext{overlap}$ & $\lambda_{\min}^Q$ & $\lambda_{\min}^G$ \\
\midrule
2$\times$6 & 629.6170 & 13.509 & 0.964699 & 0.938740 & $-1.663\cdot10^{-4}$ & $4.068\cdot10^{-5}$ \\
6$\times$2 & 653.3514 & 13.551 & 0.962043 & 0.955867 & $-1.660\cdot10^{-4}$ & $3.809\cdot10^{-5}$ \\
\midrule
3$\times$4 & 636.7304 & $-1.0086$ & 0.971119 & 0.938488 & $-2.091\cdot10^{-4}$ & $-6.341\cdot10^{-5}$ \\
3$\times$5 & 809.7487 & $-0.3663$ & 0.974450 & 0.935629 & $-2.086\cdot10^{-4}$ & $-1.528\cdot10^{-4}$ \\
3$\times$6 & 979.0046 & $-0.4234$ & 0.978241 & 0.943699 & $-2.121\cdot10^{-4}$ & $-3.632\cdot10^{-4}$ \\

4$\times$6 & 1333.3495 & 0.0385 & 0.984423 & 0.982603 & $-2.191\cdot10^{-4}$ & $-4.784\cdot10^{-4}$ \\
5$\times$6 & 1678.3400 & $-0.1340$ & 0.988221 & 0.991489 & $-3.304\cdot10^{-4}$ & $-1.283\cdot10^{-3}$ \\
6$\times$6 & 2020.3141 & $-0.1039$ & 0.989639 & 0.989350 & $-4.599\cdot10^{-4}$ & $-2.326\cdot10^{-3}$ \\
\bottomrule
\end{tabular}
\caption{
\label{tab:multiple_meshes}
Per-mesh evaluation for the NN variationally optimized on tMoTe$_2$ (see \cref{eq:ED_hamiltonian_app}) with the lowest predicted energy for the $6\times 6$ 2-RDM.}

\end{table}

\subsection{Boundary-Point Semidefinite Programming}

To compare with the neural network results, we also perform the BPSDP.
We first briefly review the BPSDP algorithm described in \refcite{Mazziotti_2011_BPSDP}.
The general algorithm is a first-order primal--dual method for solving SDP problems of the form
\begin{equation}
    E_{\mathrm{SDP}}
    =
    \min_{x}
    \ c^T x ,\text{ subject to }
    Ax=b \text{ and }
    M_\ell(x) \succeq 0 \quad \forall l,
\end{equation}
where $x$ denotes the independent variables of the reduced density matrices,
$c$ is the energy vector, $Ax=b$ represents the linear equality constraints, and $M_\ell(x)$ are the matrix blocks constrained to be positive semidefinite with $l$ ranging over different kinds.
The corresponding dual problem is
\begin{equation}
    \max_{y,z} \ b^T y ,\text{ subject to }
    c-A^T y-z=0 \text{ and }
    M_\ell(z)\succeq 0 ,
\end{equation}
where $y$ are called the dual variables which are also real, $y$ has the same length as $b$, and $z$ is a real vector of the size $x$.
The BPSDP algorithm starts with an initial $x$, $y$ and $z$, and iteratively updates them until the primal error $\|Ax-b\|$, dual error
$\|c-A^T y-z\|$, and primal--dual energy difference
\begin{equation}
    \sigma(E)=\left|c^T x-b^T y\right|
\end{equation}
are sufficiently small.

We adapted the algorithm proposed in \refcite{Mazziotti_2011_BPSDP}  by incorporating the translational invariance.
To be concrete, let us consider our case with only $(2,2)$ condition.
Due to translationally invariant, $x$ represents all the independent real parameters in $\tilde{D}$, $\tilde{Q} $ and $\tilde{G}$ tensors, and thus we can choose $M_1(x) = \tilde{D}$, $M_2(x) = \tilde{Q}$ and $M_3(x) = \tilde{G}$, as all of them need to be PSD.
Here the independence should also incorporates the fermionic antisymmetry in $D$ and $Q$.
The translational invariance also allows us to enforce the PSD constraints as the PSD constraints on $\tilde{D}_{\bsl{k},\bsl{k}',\bsl{q}}$, $\tilde{Q}_{\bsl{k},\bsl{k}',\bsl{q}}$ and $\tilde{G}_{\bsl{k},\bsl{k}',\bsl{q}}$ at each fixed $\bsl{q}$.
$c$ is defined by $\epsilon_{\bsl{k}}$ and $V_{\bsl{k},\bsl{k}',\bsl{q}}$ in \cref{eq:ED_hamiltonian_app} such that
\eq{
\frac{1}{N -1}\sum_{\bsl{k},\bsl{q}}\epsilon_{\bsl{k}}\tilde{D}_{\bsl{k},\bsl{k},\bsl{q}}(x)  + \sum_{\bsl{k},\bsl{k}',\bsl{q}}V_{\bsl{k},\bsl{k}',\bsl{q}}\tilde{D}_{\bsl{k},\bsl{k}',\bsl{q}} = c^T x
}
for any $x$.
Then the real matrix $A$ and real vector $b$ are defined such that $Ax=b$ reproduces all components in \cref{eq:D-G_relation,eq:D-Q_relation}.
The length of $b$ equals to the number of constraints.
The translationally invariance can also be implemented for the $T1$ and $T2$ conditions similarly.

We first perform the BPSDP with the full $(2,2)$ conditions on the conventional $3\times 4$, $3\times 5$, $3\times 6$, $4\times 6$, $5\times 6$ and $6\times 6$, and summarize the results in \cref{tab:BP22-results}.
We note that the constraint here takes into account of all elements in \cref{eq:D-G_relation,eq:D-Q_relation}.
The improvement after adding $T1$ is shown in \cref{tab:BP22-T1-results}, where the independent components in \cref{eq:T1_components} are included in $Ax-b$.
We further add the ordinary $T2$ condition in \cref{tab:BP22-T1-T2-results}, which adds the independent components in \cref{eq:T2_components} into $Ax-b$.

\begin{table}[h]
\centering
\begin{tabular}{|cc|c|c|c|c|c|c|c|}
\hline
$L_1$ & $L_2$ & $N_{para}$ & $R_{\text{value}}$ & $R_{\text{overlap}}$ & \
$E/\mathrm{meV}$ & $\sigma(E)/(10^{-5}\,\mathrm{meV})$ & \
$\|Ax-b\|/10^{-5}$ & $\|c-A^T y-z\|/10^{-5}$ \\
\hline
3 & 4 & 2460 & 0.967757 & 0.951361 & 635.949 & 4.911 & 9.900 & 9.750 \\
3 & 5 & 4845 & 0.965567 & 0.936443 & 807.348 & 8.084 & 9.962 & 9.935 \\
3 & 6 & 8442 & 0.965313 & 0.934521 & 975.971 & 8.722 & 9.829 & 9.983 \\
4 & 6 & 20208 & 0.968313 & 0.897407 & 1325.605 & 17.586 & 9.979 & \
9.968 \\
5 & 6 & 39630 & 0.973158 & 0.905827 & 1668.140 & 18.831 & 9.997 & \
9.579 \\
6 & 6 & 68760 & 0.977307 & 0.912077 & 2007.788 & 27.072 & 9.987 & \
8.869 \\
\hline
\end{tabular}
\caption{Numerical results for the BPSDP with the full $(2,2)$ conditions on the conventional meshes.
$L_1\times L_2$ label the mesh, and $N_{para}$ labels the number of independent real parameter needed.
$E$ is the mean value of the energies obtained from the primal and dual SDP problems.
$\sigma(E)$, $\|Ax-b\|$ and 
$\|c-A^T y-z\|$ are the absolute difference between energies obtained from the primal and dual SDP problems, the primal error, and the dual error, respectively.
}
\label{tab:BP22-results}
\end{table}

\begin{table}[h]
\centering
\begin{tabular}{|cc|c|c|c|c|c|c|c|}
\hline
$L_1$ & $L_2$ & $N_{para}$ & $R_{\text{value}}$ & $R_{\text{overlap}}$ & \
$E/\mathrm{meV}$ & $\sigma(E)/(10^{-5}\,\mathrm{meV})$ & \
$\|Ax-b\|/10^{-5}$ & $\|c-A^T y-z\|/10^{-5}$ \\
\hline
3 & 4 & 6496 & 0.968021 & 0.947426 & 635.974 & 20.482 & 9.984 & 5.271 \
\\
3 & 5 & 18650 & 0.965948 & 0.934837 & 807.382 & 34.583 & 9.985 & \
9.968 \\
3 & 6 & 45450 & 0.966702 & 0.935091 & 976.024 & 57.756 & 9.990 & \
8.835 \\
4 & 6 & 190904 & 0.969623 & 0.904283 & 1325.674 & 95.238 & 8.322 & \
9.998 \\
5 & 6 & 589090 & 0.974337 & 0.91211 & 1668.230 & 101.943 & 9.994 & \
8.823 \\
6 & 6 & 1484892 & 0.978123 & 0.917177 & 2007.910 & 154.068 & 13.035 & 
9.877 \\
\hline
\end{tabular}
\caption{Numerical results for the BPSDP with the full $(2,2)$ conditions and the $T1$ condition on the conventional meshes.
$L_1\times L_2$ label the mesh, and $N_{para}$ labels the number of independent real parameter needed.
$E$ is the mean value of the energies obtained from the primal and dual SDP problems.
$\sigma(E)$, $\|Ax-b\|$ and 
$\|c-A^T y-z\|$ are the absolute difference between energies obtained from the primal and dual SDP problems, the primal error, and the dual error, respectively.
}
\label{tab:BP22-T1-results}
\end{table}

\begin{table}[h]
\centering
\begin{tabular}{|cc|c|c|c|c|c|c|c|}
\hline
$L_1$ & $L_2$ & $N_{para}$ & $R_{\text{value}}$ & $R_{\text{overlap}}$ & \
$E/\mathrm{meV}$ & $\sigma(E)/(10^{-5}\,\mathrm{meV})$ & \
$\|Ax-b\|/10^{-5}$ & $\|c-A^T y-z\|/10^{-5}$ \\
\hline
3 & 4 & 58768 & 0.979828 & 0.953776 & 637.235 & 86.822 & 8.681 & \
9.993 \\
3 & 5 & 184025 & 0.982268 & 0.961272 & 809.332 & 22.032 & 9.998 & \
9.839 \\
3 & 6 & 466812 & 0.983241 & 0.962182 & 978.451 & 113.126 & 9.999 & \
9.633 \\
4 & 6 & 2019128 & 0.985795 & 0.962212 & 1329.720 & 128.516 & 9.998 & \
8.964 \\
5 & 6 & 6265840 & 0.987974 & 0.964069 & 1673.749 & 182.811 & 9.999 & \
9.841 \\
6 & 6 & 15773292 & 0.989446 & 0.963958 & 2014.858 & 300.347 & 9.998 & \
9.603 \\
\hline
\end{tabular}
\caption{Numerical results for the BPSDP with the full $(2,2)$ conditions, the $T1$ condition and the ordinary $T2$ conditions on the conventional meshes.
$L_1\times L_2$ label the mesh, and $N_{para}$ labels the number of independent real parameter needed.
$E$ is the mean value of the energies obtained from the primal and dual SDP problems.
$\sigma(E)$, $\|Ax-b\|$ and 
$\|c-A^T y-z\|$ are the absolute difference between energies obtained from the primal and dual SDP problems, the primal error, and the dual error, respectively.
}
\label{tab:BP22-T1-T2-results}
\end{table}

\end{document}